\documentclass[reprint,twocolumn,floatfix]{revtex4-2}
\usepackage[english]{babel}
\usepackage[utf8]{inputenc}

\usepackage[colorlinks=true,citecolor=blue,linkcolor=blue,urlcolor=blue,pdfencoding=auto]{hyperref}
\usepackage[sort&compress]{natbib}

\usepackage{graphicx,times}
\usepackage{color} 

\usepackage{amssymb}
\usepackage{cancel}
\usepackage{soul}

\usepackage{subeqnarray}
\usepackage{diagbox}
\usepackage{makecell}
\usepackage{amsmath}
\usepackage{mathtools}
\usepackage{empheq}
\usepackage{bbding}

\usepackage[dvipsnames]{xcolor}
\usepackage{tikz}
\usepackage{bbm}

\definecolor{aprcolor}{rgb}{0.7,0.1,0.2}

\newcommand{\mrm}[1]{\mathrm{#1}}
\newcommand{\msf}[1]{\mathsf{#1}}
\newcommand{\mcl}[1]{\mathcal{#1}}
\newcommand{\bsb}[1]{\boldsymbol{#1}}
\newcommand{\mbf}[1]{\mathbf{#1}}

\DeclareSymbolFont{greekletters}{OML}{FiraSans}{m}{n}

\newcommand{\dd}{{\mrm{d}}}
\newcommand{\Omeps}{{\Omega}{\epsilon}}

\DeclareMathOperator{\msOmega}{\msf{\Omega}}

\DeclareMathOperator{\mGamma}{\msf{\Gamma}}
\DeclareMathOperator{\mSigma}{\msf{\Sigma}}
\DeclareMathOperator{\mDelta}{\msf{\Delta}}

\DeclareMathOperator{\bU}{\mbf{U}}
\DeclareMathOperator{\bV}{\mbf{V}}
\DeclareMathOperator{\bW}{\mbf{W}}
\DeclareMathOperator{\bw}{\mbf{w}}

\DeclareMathOperator{\bef}{\mbf{f}}

\DeclareMathOperator{\bphi}{{\bsb{\phi}}}
\DeclareMathOperator{\bPhi}{{\bsb{\Phi}}}
\DeclareMathOperator{\bq}{{\mbf{q}}}
\DeclareMathOperator{\bQ}{{\mbf{Q}}}

\DeclareMathOperator{\bI}{\mbf{I}}

\DeclareMathOperator{\bz}{\boldsymbol{z}}

\DeclareMathOperator{\be}{\mbf{e}}
\DeclareMathOperator{\br}{\mbf{r}}

\DeclareMathOperator{\bzeta}{\bsb{\zeta}}

\DeclareMathOperator{\mcL}{\mcl{L}}
\DeclareMathOperator{\mcT}{\mcl{T}}

\DeclareMathOperator{\bmW}{{\bsb{\mathcal{W}}}}

\DeclareMathOperator{\msc}{\msf{c}}
\DeclareMathOperator{\msl}{\msf{l}}

\DeclareMathOperator{\mcM}{\mcl{M}}

\DeclareMathOperator{\msA}{\msf{A}}
\DeclareMathOperator{\msB}{\msf{B}}
\DeclareMathOperator{\msC}{\msf{C}}
\DeclareMathOperator{\msD}{\msf{D}}
\DeclareMathOperator{\msE}{\msf{E}}
\DeclareMathOperator{\msK}{\msf{K}}

\DeclareMathOperator{\msM}{\msf{M}}
\DeclareMathOperator{\msN}{\msf{N}}
\DeclareMathOperator{\msT}{\msf{T}}
\DeclareMathOperator{\msY}{\msf{Y}}

\DeclareMathOperator{\msZ}{\msf{Z}}

\DeclareMathOperator{\msF}{\msf{F}}
\DeclareMathOperator{\msL}{\msf{L}}
\DeclareMathOperator{\msG}{\msf{G}}
\DeclareMathOperator{\msJ}{\msf{J}}
\DeclareMathOperator{\msU}{\msf{U}}

\newcommand{\mone}{\mathbbm{1}}

\newcommand{\dQ}{{\mathrm{d}}{\bQ}}
\DeclareMathOperator{\dPhi}{{\mathrm{d}}{\bPhi}}
\newcommand{\TL}{{\text{TL}}}

\usepackage{svg}

\begin{document}
	\title{Exact quantization of nonreciprocal quasi-lumped electrical networks}
	\author{A. Parra-Rodriguez}
	\email{adrian.parra.rodriguez@gmail.com}
	\affiliation{Institute of Fundamental Physics IFF-CSIC, Calle Serrano 113b, 28006 Madrid, Spain}
	\author{I. L. Egusquiza}
	\email{inigo.egusquiza@ehu.es}
	\affiliation{Department of Physics, University of the Basque Country UPV/EHU, Apartado 644, 48080 Bilbao, Spain}
	\affiliation{EHU Quantum Centre, University of the Basque Country UPV/EHU, Apartado 644, 48080 Bilbao, Spain}
	
	\begin{abstract}
		Following a consistent geometrical description previously introduced in Ref.~\cite{ParraRodriguez:2024a}, we present an exact method for obtaining canonically quantizable Hamiltonian descriptions of nonlinear, nonreciprocal quasi-lumped electrical networks. We identify and classify  singularities arising in the quest for Hamiltonian descriptions of general quasi-lumped element networks via the Faddeev-Jackiw technique.  We offer systematic solutions to cases previously considered singular--a major challenge in the context of canonical circuit quantization. The solution relies on the correct identification of the reduced classical circuit-state manifold, i.e., a mix of flux and charge fields and functions. Starting from the geometrical description of the transmission line, we provide a complete program including lines coupled to one-port lumped-element networks, as well as multiple lines connected to multiport nonreciprocal lumped-element networks, with intrinsic ultraviolet cutoff. On the way we naturally extend the canonical quantization of transmission lines coupled through frequency-dependent, nonreciprocal linear systems, such as practical circulators. Additionally, we demonstrate how our method seamlessly facilitates the characterization of general nonreciprocal, dissipative linear environments. This is achieved by extending the Caldeira-Leggett formalism, using continuous limits of series of immittance matrices. We provide a tool in the analysis and design of electrical circuits and of special interest in the context of canonical quantization of superconducting networks. For instance, this work will provide a solid ground for a precise non-divergent input-output theory in the presence of nonreciprocal devices, e.g., within (chiral) waveguide QED platforms.
	\end{abstract}

	\maketitle
	
	\section{Introduction}
	\label{sec:introduction}
	The lumped element model of Maxwell's equations, also known as the quasi-static approximation, has been extremely successful in describing many aspects of the low-energy dynamics of electrical circuits~\cite{Jackson:1999,Feynman:2010}, including the rapidly growing field of superconducting quantum circuits~\cite{CaldeiraLeggett:1981,Caldeira:1983,YurkeDenker:1984,Devoret:2013,Blais:2021}. In this process, the distributed partial differential field equations (infinite-dimensional state space) are reduced to ordinary differential equations (finite-dimensional state space), with an intrinsic high-frequency cutoff embedded in this mapping~\cite{ParraRodriguezPhD:2021}. Naturally, this description cannot capture all the phenomenology of electromagnetic waves guided by matter. Nonetheless, the \emph{transmission line} approximation, which also involves quasi-staticity, simplifies Maxwell's equations to a set of far simpler partial differential equations in 1+1 dimensions. Indeed, it has proven essential for the quantum description of modern-day superconducting electrical circuits to consider transmission lines in conjunction with lumped elements~\cite{YurkeDenker:1984,Janhsen:1992,Blais:2004,Minev:2021,Blais:2021}, also known as the \emph{quasi-lumped} element model.
	
	In a previous work \cite{ParraRodriguez:2024a}, we put forward a geometric construction to implement the crucial Kirchhoff constraints, along with those arising from ideal lineal elements such as gyrators~\cite{Tellegen:1948a} or circulators~\cite{Pozar:2009} (nonreciprocal elements~\cite{Caloz:2018}), transformers~\cite{Belevitch:1950}, and linear
	resistors. This allows for Hamiltonian descriptions of lumped element circuits, supplemented by a Rayleigh dissipation function. When setting aside the dissipative elements, canonical quantization would naturally follow. We also put forward a topological ansatz for one-port reactive and source elements, that allows the prediction of the final topology of configuration and phase space for circuits constituted of those types of element.
	
	In this article, we further extend that approach, providing classical Lagrangian and (canonically-quantizable) Hamiltonian descriptions of ideal circuits with transmission lines (TLs). These circuits are coupled to lumped element circuits comprising linear or nonlinear (NL) capacitors and inductors, voltage and current sources, transformers, and nonreciprocal (NR) devices. By combining all the aforementioned linear elements, our method can treat the most general linear blackbox devices exhibiting NR behavior \cite{Kamal:2011,Viola:2014,Kerckhoff:2015,Sliwa:2015,Chapman:2017,Barzanjeh:2017,Rosenthal:2017}. These devices act as couplers between input/output waveguides, the aforementioned transmission lines, and nonlinear degrees of freedom, such as Josephson \cite{Josephson:1962} or phase-slip junctions \cite{Mooij:2006}.
	
	Furthermore, building on our previous work \cite{ParraRodriguez:2018, ParraRodriguezPhD:2021}, our exact models, by design, require no renormalization and exhibit intrinsic high-energy cutoffs \cite{Paladino:2003}, a mechanism recently observed in \cite{Ao:2023}. The approach presented here is simpler and more direct than our previous proposal, wherein a doubled configuration space and a subsequent reduction of variables were utilized \cite{ParraRodriguez:2022,Egusquiza:2022}. This enhanced method not only simplifies the process but also broadens the range of circuits from which Hamiltonian dynamics with canonical coordinates can be constructed, thus paving the way for the systematic derivation of canonically-quantized Hamiltonian models.
	
	The manuscript is structured as follows. In section~\ref{sec:geom-descr-class} we briefly revisit the geometrical approach for constructing Hamiltonian descriptions of electrical circuits in Ref.~\cite{ParraRodriguez:2024a}, setting up some notation and emphasising the conceptual aspects. Readers already familiar with the method may want to skip the main part of the section and quickly review the Subsec.~\ref{subsec:summary_red_method_LE} with an even more direct step-wise reduction algorithm. Subsequently, three illustrative circuit examples are explicitly worked out. We also present a summary of the topological ansatz, and an example of its application and power, in Subsec.~\ref{subsec:micro-topological}. We then apply the geometrical description to transmission lines in section \ref{sec:TLs-geom-description}, initially examining a discrete description before proceeding to the continuum limit, with special attention to certain subtleties. Given that these concepts are newly introduced, we carefully examine TLs coupled to one-port lumped-element networks (reciprocal circuits) and explicitly show how alternative discretizations yield the same results. These subtleties are required to check the consistency of the framework, but not for its application once established, so the reader interested only in applications might prefer to skip the initial part of the section and jump straight to the summary in Subsec.~\ref{subsec:summary_TLs} (in conjunction with the previous summary of Subsec.~\ref{subsec:summary_red_method_LE}). We analyze the impact of the topological ansatz of Subsec.~\ref{subsec:micro-topological} on the framework, in particular in regard to different discretizations and line configurations, in Subsec.~\ref{subsec:TL-topological}. We showcase the potential of this novel technique in multi-line circuits connected to nonreciprocal multiport lumped-element networks in Sec.~\ref{sec:TLs-multiport-networks}, particularly addressing the quantization of TLs connected through frequency-dependent NR devices (e.g., circulators). We show  that the resulting theory is finite, in that so are Lamb shifts, for example. In Sec.~\ref{sec:quasi_lumped_circuit_examples}, we illustrate the general methodology with a pair of canonical circuit examples: TLs coupled through reciprocal and nonreciprocal multiport linear systems to Josephson-junction circuits. Moving beyond the quasi-lumped element circuit class,  Sec.\ref{sec:C-L_models} demonstrates how to apply the continuous limit to (NR) dissipative circuits within the Caldeira-Leggett formalism (dissipative NR black-box approach), employing the new geometrical method. This expansion significantly broadens the scope of the subject matter, building upon and extending the principles established in classic literature on the topic~\cite{Caldeira:1983, YurkeDenker:1984,Devoret:1997,Vool:2017,ParraRodriguez:2019,ParraRodriguez:2022b}. We conclude with a final summary and offer some perspectives on future applications and developments.
	\section{Geometrical description of classical lumped-element electrical circuits: a short review}
	\label{sec:geom-descr-class}
	
	In traditional lumped-element circuit theory \cite{Guillemin:1953}, the electrical state is determined by the values of voltage drop $v^b$ and
	intensity $i^b$ in each port/branch $b$. As our main concern lies with capacitive and inductive elements, it is more convenient to describe the state redundantly by using branch fluxes $\phi^b$ and branch charges $q^b$, where $\dot{\phi}^b=v^b$,
	$\dot{q}^b=i^b$, and $\dot{f}\equiv \partial_t f$ apply to both these and subsequent equations.  We collect all charge and flux variables to form an initial manifold $\mcl{M}_{2B}=\mathbbm{R}^{2B-k}\times (S^1)^k$, where we allow for compact variables  to  represent Josephson ~\cite{Josephson:1962,Devoret:2021} (fluxes) or phase-slip ~\cite{Mooij:2006,Astafiev:2012} (charges) junctions, the dual nonlinear elements in superconducting quantum circuits. 
	
	Intuitively, charges and fluxes should be conjugate variables. However, constraints exist. Firstly, the Kirchhoff current (KCL) and voltage (KVL) laws, $\sum_{b\in \mcl{N}} i^b=0$ and $\sum_{b\in \mcl{P}}  v^b=0$, respectively, apply, with $b\in \mcl{N}$ $(\mcl{P})$ denoting branches flowing into node $\mcl{N}$ (present in the loop $\mcl{P}$). These conditions apply no matter the actual dynamics, and we therefore express them as geometric conditions,
	$\sum_{b\in \mcl{N}} \dd q^b=0$ and $\sum_{b\in \mcl{P}} \dd \phi^b=0$, using differential forms. The reason to do this is that these are conditions on the tangents of trajectories in $\mathcal{M}_{2B}$, therefore conditions on tangent vectors, that are here expressed in a dual form. In a manner most intuitive to physicists, the infinitesimal change in $q^b$, say, expressed by $\dd q^b$, is proportional to $\dd t$, and Kirchhoff's laws apply no matter the unit of time or parameterization. Thus, here and in what follows, we shall use this language to express these and other constraints. We present a lightning summary of relevant mathematical terms in Sec.~\ref{sec_diff_forms_app}. Secondly, and in
	order to provide canonical descriptions for linear time-independent (LTI)
	passive multiport devices, one must consider ideal NR elements (e.g., gyrators and circulators) and
	transformers. From the perspective we put forward, these are 
	additional constraints on branch voltage and intensity sets~\cite{Foster:1924,Cauer:1926,Tellegen:1948,Belevitch:1950,Newcomb:1966,Solgun:2015}. Such multiport descriptions are also known in the superconducting technology jargon as \textit{blackboxes}~\cite{Nigg:2012,Solgun:2014,Solgun:2015}, given that they do not inform about the internal microscopic dynamics.
	
	All in all, these constraints, which are imposed on branch voltages and currents, are collectively represented by a matrix $\msF$ that annihilates the state vector differential $\dd{\bzeta}^T=(\dd{\bphi}^T, \dd{\bq}^T)$, i.e.,  $\msF\dd{\bzeta}=0$. Linear as they are, these constraints are integrable into fluxes and charges because the lumped approximation demands that the space of states be a product of one-dimensional manifolds \cite{ParraRodriguez:2024a}. Graph algebra, as applied to electrical circuits since the proposal of Weyl \cite{Weyl:1923}, tells
	us that if the constraints are simply of Kirchhoff type, then they can be
	satisfied by expressing branch fluxes and charges in terms of node fluxes and
	loop charges. Further reduction is necessary in the presence of
	gyrators (we will use the gyrator as a representative of the class of ideal NR elements throughout the manuscript) and transformers, determining the minimal
	set of flux and charge variables required to express all the allowed
	configurations. Observe that in this procedure several issues may arise. For example, additional reductions may be possible for a specific circuit topology, such as the archetypal reduction of two linear capacitors or inductors, in series or parallel, to their equivalent components. Indeed, for general nonlinear circuits, one might encounter an imbalance of charge and flux variables, thus apparently precluding a Hamiltonian description~\cite{Rymarz:2023}.
	\begin{figure}[h!]
		\centering
		\includegraphics[width=1\columnwidth ]{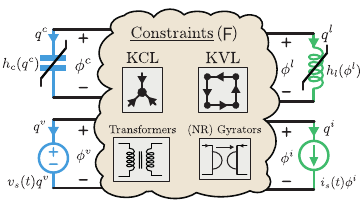}
		\caption{
			Outside of the cloud, the set of dynamical lumped elements that are characterized by energy functions (energetic constitutive equations) such as (nonlinear) inductors ($\mcl{L}$) and capacitors ($\mcl{C}$), voltage sources ($\mcl{V}$), and current sources ($\mcl{I}$). Within the cloud, connections among these elements  are implemented by constraints such as Kirchhoff's laws (KCL and KVL), and two more general ones, ideal transformers ($\mcl{T}_F$), recently known as energy-participation ratios \cite{Nigg:2012,Minev:2021b,Ciani:2023}, and gyrators ($\mcl{G}$), both of  which  can be used to express more general constraints between fluxes and charges effectively breaking time-reversal symmetry, see~\cite{ParraRodriguez:2024a}. As to gyrators, we depict the canonical two-port nonreciprocal element to represent its class.}
		\label{fig:Elements_noR}
	\end{figure}
	
	The key to understanding this issue relies on recognising that charge-flux conjugation presents itself differently between capacitive and voltage source branches, on the one hand, and inductive and current source ones, on the other. Thus, we
	\cite{ParraRodriguez:2024a} proposed to codify these different forms of conjugation in the following closed two-form on the product manifold of branch flux and charge variables,
	\begin{align}
		\label{eq:twoform2b}
		\begin{aligned}
			\omega_{2B}=&\,\, \frac{1}{2}
			\left[\sum_{l\in\mathcal{L}}\dd\phi^{{l}}\wedge\dd q^l+
			\sum_{c\in\mathcal{C}}\dd q^c\wedge\dd \phi^{{c}}\right]\\
			&+\frac{1}{2} \left[\sum_{i\in\mathcal{I}}\dd \phi^{{i}}\wedge\dd q^i +
			\sum_{v\in\mathcal{V}}
			\dd q^v\wedge\dd \phi^{{v}}\right]\,.
		\end{aligned}
	\end{align}
	
	Here, if necessary, we incur in the standard abuse of notation $\dd \alpha$ for
	the winding one form in an $S^1$ variable (as in the case of flux variable associated with a superconducting Josephson island~\cite{Devoret:2021}), by writing it in terms of the angle $\alpha$ in one coordinate patch. We have classified all branches in  one of the following categories, all of them ideal: linear and nonlinear capacitors ($\mcl{C}$) and inductors ($\mcl{L}$),  voltage  ($\mcl{V}$) and current ($\mcl{I}$) sources,   and transformer ($\mcl{T}_F$) and gyrator  ($\mathcal{G}$) branches. Only the reactive ($\mathcal{C}$ and $\mathcal{L}$) and source ($\mathcal{I}$  and $\mathcal{V}$) branches do actually present conjugate pairs of variables  from the dynamical perspective, while the $\mcl{T}_F$ and $\mathcal{G}$ sets will be constraints.  In parallel independent work Osborne et al also identified this object, albeit only in the presence of KCL and KVL constraints~\cite{Osborne_v2:2024}.
	
	We also associate a total energy function 
	\begin{align}
		H=\sum_{l\in\mathcal{L}}h_l(\phi^l)+\sum_{c\in\mathcal{C}}h_c(q^c)\label{eq:H_method}
	\end{align}
	for all capacitive and inductive (passive, storing energy) elements, because their constitutive equations 
	\begin{align}
		\label{eq:conscl}
		\dot{\phi}^c= v^c= \frac{\partial h_c}{\partial q^c}\,,\qquad
		\dot{q}^l = i^l = \frac{\partial h_l}{\partial \phi^l}\,,
	\end{align}
	do resemble Hamiltonian equations. In fact, it must be here noted that, if uncoupled, they seem to  come from a two-form twice that of the defined $\omega_{2B}$ (observe that the remaining two equations would be trivial $\dot{q}^c=\dot{\phi}^l=0$). An extra energy term 
	\begin{align}
		H_d(t)=\sum_{v\in \mcl{V}} q^v v_v(t)+\sum_{i\in \mcl{I}}\phi^i i_{i}(t)\label{eq:Hd_method}
	\end{align}
	will capture the voltage $v_v(t)$ and current $i_i(t)$ sources, see further details in~\cite{ParraRodriguez:2024a}. In the following, we will denote the total energy contributions by $H_T=H+H_d(t)$, combining the passive and active components.
	
	The constraints derived from   $\mcl{T}_F$ and
	$\mathcal{G}$, together with the Kirchhoff constraints, are a set of
	differential equations for the branch fluxes and charges. As a matter
	of fact, the scale of time is irrelevant for these constraints, that
	are autonomous. Thus, for our
	purposes it is best to consider them as an external Pfaff system, i.e., a set of
	linear homogeneous equations for differential forms on the manifold of
	states. This system, as stated, is integrable, which means that we can
	find an integrable submanifold of  $\mcl{M}_{2B}$, such that
	its cotangent space automatically satisfies the Pfaff system. In
	circuit theory the integrable submanifold is traditionally
	parameterized in terms of loop charges and flux nodes, if the only
	constraints are Kirchhoff, and generalizations thereof in the presence
	of additional gyrator and transformer constraints.
	
	Thus, we restrict the energy function to the relevant submanifold, expressing it in terms of a good coordinatization thereof, and we eliminate terms from $\omega_{2B}$ to satisfy the differential equations, i.e., we restrict $\omega_{2B}$ to the integral manifold $\mcM$, 
	\begin{align}
		\omega=\imath^* \omega_{2B}=\tfrac{1}{2}\omega_{\alpha\beta} \dd z^\alpha\wedge \dd z^\beta,\label{eq:twoform_restricted}
	\end{align}
	where $\imath^*$ represents the pull-back under the immersion map $\imath:\mcM\rightarrow\mcM_{2B}$. In our case, where the set of constraints is linear, this is tantamount to obtaining the kernel of the constraint matrix $\msF$, which can be easily done by Gaussian elimination. Now, in the absence of
	dissipation, the equations of motion that would follow from standard
	circuit theory, i.e., those that follow from the constitutive equations
	(\ref{eq:conscl}) when the constraints are imposed, are the same as the Euler--Lagrange equations of motion for the
	Lagrangian 
	\begin{align}
		\label{eq:L_method}
		L= L_{\omega}-H
		-S_\alpha(t){z}^\alpha,
	\end{align}
	where $L_{\omega}=\tfrac{1}{2}\omega_{\alpha\beta}z^\alpha\dot{z}^\beta$, with $z^\alpha$ coordinates in the restricted submanifold
	$\mathcal{M}$, and $\omega_{\alpha\beta}$ the components of the restricted two-form in that system of coordinates. $S_\alpha(t)$ is also constructed by restricting the driving term $H_d(t)$ to $\mathcal{M}$. Observe that generically
	the $z^\alpha$ coordinates need not be strictly node fluxes or loop charges. In fact, in the presence of NR elements, with characteristic resistance/conductance parameter, they will be a mixture of those. Notice that in determining the integral manifolds of the constraint and their coordinates there is a freedom
	in some integration constants. These will only be relevant for the dynamics derived from Eq. (\ref{eq:L_method}) through the energy function $H$, and, as such, can generically be chosen so the origin of coordinates is an energy minimum. 
	
	The task of constructing a Hamiltonian description of the circuit has
	not been yet completed at this point in general. The issue is that the
	two-form $\omega$ might be degenerate,
	i.e., $\mathrm{det}(\omega_{\alpha\beta})=0$. To give but one simple example, this would happen in presence of a capacitor only loop. One way forward, therefore, is to follow the method of Faddeev and Jackiw
	\cite{Faddeev:1969,Faddeev:1988,Jackiw:1993,Toms:2015}, as made explict in
	\cite{ParraRodriguez:2024a}. By construction, the rank
	of the two-form is homogeneous on the restricted submanifold. We thus
	search for  the zero modes of
	the two-form $\omega$, a set of vectors
	$W=\left\{\boldsymbol{W}_I\right\}_{I=1}^{|W|}$ with \[\boldsymbol{W}_I=
	W^\alpha_I\frac{\partial}{\partial z^\alpha}\,,\] 
	such that ${W}_I^\alpha\omega_{\alpha\beta}=0$.
	Here $|W|=\mathrm{dim}\left[\mathrm{ker}(\omega)\right]$ informs us of
	how many coordinates are, in a sense, superfluous. We now have to use
	these vectors to identify a coordinate system in which we have
	discarded or separated out superfluous coordinates, or, alternatively,
	to further reduce the space of states by constraining to a smaller
	manifold that is symplectic.  These vectors, for circuits, commute, and therefore can be integrated. In other words, each zero-mode vector $\boldsymbol{W}_I$ defines locally a zero-mode coordinate $w^I$, and there exists locally a set of coordinates $\xi^\mu$ that complement the zero-mode ones. The two-form is then explicitly $\omega=f_{\mu \nu}\dd \xi^\mu\wedge\dd \xi^\nu$. The original variables are expressed in terms of this new set, $z^\alpha(\xi^\mu,w^I)$. Now, we apply these vectors to the total Hamiltonian, and we obtain a new set of constraints
	\begin{equation}
		\label{eq:newcosntraints}
		\boldsymbol{W}_I(H_T)=W_I^\alpha\frac{\partial H_T}{\partial z^\alpha}=\frac{\partial H_T(\xi^\mu,w^I)}{\partial w^I}=0,
	\end{equation}
	required for consistency of the equations of motion $\omega_{\alpha\beta}\dot{z}^\beta=\partial H_T/\partial z^\alpha$, as can be seen by contracting from the left with $W_I^\alpha$.
	
	There is a wide spectrum of phenomenology in general at this point of the method. For the case of interest to us there are three possibilities, that can appear simultaneously in a particular example. The simplest case is that  coordinates $w^I$ can be solved in terms of $\xi^\mu$ and time $t$ (through the driving functions $S_\alpha$), see an example below in Subsec.~\ref{subsec:LE_examples}. Then the manifold of states is accordingly reduced in dimension, and the reduction is also applied to the two-form and the Hamiltonian explicitly. A bit more involved is the case in which the solution of the new constraints has to be parametric since we might run into a serious complication, namely that the restricted two-form is no longer
	of homogeneous rank. These cases have to be separately
	studied, see further comments in~\cite{ParraRodriguez:2024a}. Finally, and of particular relevance to transmission lines, as we shall see presently, we can find that some constraints are identically satisfied, in which case we say that we are in the presence of a \emph{gauge freedom}. We postpone their explicit
	treatment to later examples.
	
	Setting aside the subtleties associated with non-homogeneity and gauge
	constraints for the time being, we have obtained a description of the
	classical system in terms of a non-degenerate two-form and an energy
	function. That is precisely the classical Hamiltonian description we were
	searching for. Our central motivation is the quantum mechanical
	description of superconducting circuits, and  this classical
	Hamiltonian description is the starting point for canonical
	quantization. In most cases of interest this will be
	straightforward. Nonetheless we cannot fail to mention two possible
	hurdles.  First, the existence of \emph{global} Darboux coordinates is not guaranteed if either non
	trivial topology or inhomogeneity of the two-form are present. Second,
	there will in general exist inequivalent quantizations, both because of ordering issues or because of topological properties of the final manifold. Thus there are situations for which canonical quantization,
	which has been the workhorse of circuit QED, might not be the most
	favoured method. The Faddeev--Jackiw method we put forward has, in this
	respect, the advantage that alternatives exist, such as path integral
	quantization (see for instance \cite{Toms:2015}).
	
	The abstract geometrical formulation summarized here is most convenient for conceptualization and for the statement of general properties of all circuits, such as the integrability of the Kirchhoff constraints. When it comes to actual computations, however, there are two preferred approaches, depending on the objective. First, the reduction process is fully algoritmizable, and to that purpose explicit matrix notations will be most adequate, see Appendices in Ref.~\cite{ParraRodriguez:2024a}. Second, and this is the approach that we follow in the rest of the paper, exterior algebra is much more useful to deal with circuits with strong regularities, such as the discrete presentations of TLs that we shall examine presently. We include a lightning review of the relevant tools in Appendix \ref{sec_diff_forms_app}.
	\subsection{Summary of the reduction method}\label{subsec:summary_red_method_LE}
	To summarize, the complete reduction method to obtain canonically-quantized Hamiltonians, when possible (see obstructions above and in~\cite{ParraRodriguez:2024a}), requires the following five steps:
	\begin{enumerate}
		\item Construct $\omega_{2B}$ using the flux and charge branch variables of the circuit elements, Eq. \eqref{eq:twoform2b}.
		\item Implement the immersion map $\imath^*$, Eq. \eqref{eq:twoform_restricted}, to reduce the linear constraints to arrive to $\omega$, e.g., by gaussian elimination. For circuits without transformers/nonreciprocal elements, the set of node fluxes and loop charges is a basis for the circuit state.
		\item Change to a coordinate system ($z^\alpha \rightarrow\{\xi^\mu\}\cup \{w^I\}$) in which the kernel of the two-form and its complementary subspaces are separated, i.e., $\omega=f_{\mu\nu}\dd \xi^\mu\wedge \dd \xi^\nu$. 
		\item The equations $\bW_I(H_T)=\partial H_T/\partial w^I=0$ are constraints, i.e., equations for  $w^I$ in terms of  $\xi^\mu$ and $t$ (from the source terms). If the solutions for all $w^I$ are smooth then we have arrived at a  reduced Hamiltonian dynamics, with variables $\xi^\mu$ and Poisson bracket $\{\xi^\mu,\xi^\nu\}=f^{\mu\nu}=(f_{\mu\nu})^{-1}$. Apply (if needed) standard sequences of transformations to take the matrix $f_{\mu\nu}$ to canonical (symplectic) form.
		\item This well-behaved classical Hamiltonian dynamics is promoted to quantum dynamics by canonical quantization, i.e.,  the conjugate variables map to quantum operators acting either in $L^2(S^1)$ (compact) or  $L^2(\mathbb{R})$ (extended), depending on whether superconducting islands (and/or phase-slips) are present or not, respectively.
	\end{enumerate}
	
	Before we enter the new geometrical description of transmission lines, let us illustrate the method, as initially introduced in Ref.~\cite{ParraRodriguez:2024a} and reexpounded in this section, by computing the Hamiltonian of three circuits: (i) a capacitively-shunted Josephson junction coupled to an LC oscillator driven by a voltage source (the basic circuit for charge qubits~\cite{Buettiker:1987,Koch:2007}), (ii) an LC network characterized by an admittance response matrix within the blackbox approach~\cite{Nigg:2012,Solgun:2015,ParraRodriguez:2024a}, and (iii) a nonreciprocal circuit used to stabilize GKP states~\cite{Rymarz:2021}. Notice that these  examples, in particular, could be analysed with the node-flux method~\cite{Devoret:1997,Solgun:2015,ParraRodriguez:2019}. We stress that all existing methodologies for the node-flux method can be incorporated in the geometric perspective.
	\subsection{Lumped-element circuit examples}\label{subsec:LE_examples}
	\subsubsection{Josephson junction coupled to a driven LC oscillator}
\begin{figure}[ht!]
		\centering
		\includegraphics[width=1\columnwidth ]{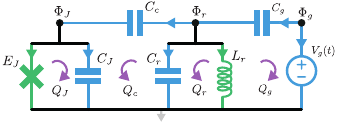}
		\caption{Lumped element circuit example: an LC oscillator is capacitively coupled to both a Josephson junction and a voltage source.  Given that in this circuit there are only Kirchhoff's constraints, node fluxes and loop charges are a good basis for all the branch fluxes and charges.}
		\label{fig:CJJ_Cc_LC_Vg}
	\end{figure}

	Let us now follow the procedure previously explained with the derivation of the Hamiltonian for the circuit in Fig.~\ref{fig:CJJ_Cc_LC_Vg}, comprising a C-shunted Josephson junction capacitively coupled to a driven LC harmonic oscillator.
	
	Given that the circuit does not contain ideal transformer/nonreciprocal constraints, a basis of node fluxes $\bPhi^T=(\Phi_J,\Phi_r,\Phi_g)$ and loop charges $\bsb{Q}^T=(Q_J,Q_c,Q_r,Q_g)$ ($\bz^T=(\bsb{Q}^T,\bPhi^T)$) can describe all the branch variables $\bzeta^T=(\bsb{q}^T,\bphi^T)$. The relation derives from integrating   $\dd{\bzeta}=\msK \dd{\bz}$. The matrix $\msK$ generates the kernel of the  linear constraint matrix $\msF=\msF_{\text{Kir}}$, see \cite{ParraRodriguez:2024a} for more details. In fact the flux-node analysis method provides us directly with this relation. For instance, with the chosen directions in Fig.~\ref{fig:CJJ_Cc_LC_Vg}, we have $q_J=-Q_J$, $q_{C_J}=Q_J-Q_c$, etc. Following the method,  write the two-form (\ref{eq:twoform_restricted}) for the circuit starting from (\ref{eq:twoform2b}), express the branch variables explicitly in terms of node fluxes and loop charges, and then the linear Kirchhoff's constraints are automatically conserved, with
	\begin{align}
		\omega=&\,\frac{1}{2}\left[\dd Q_J\wedge \dd \Phi_J+(\dd Q_J+\dd Q_c)\wedge \dd \Phi_J\right.\nonumber\\
		&\left.+\dd Q_c\wedge (\dd \Phi_r-\dd \Phi_J)+(\dd Q_g+\dd Q_r)\wedge \dd \Phi_r\right.\nonumber\\
		&\left.+\dd Q_g\wedge(\dd \Phi_r-\dd \Phi_g)+\dd Q_g\wedge\dd \Phi_g\right]\nonumber\\
		=&\,\dd Q_J\wedge \dd \Phi_J+(\dd Q_r+\dd Q_g)\wedge\dd \Phi_r.
	\end{align}
	This  implements in one go steps 1 and 2 of the previous summary of the method. In this flux-charge basis, thus, the  Lagrangian (\ref{eq:L_method}) reads
	\begin{align}
		L=Q_J\dot{\Phi}_J+(Q_r+ Q_g)\dot{\Phi}_r-H-H_d(t),
	\end{align}
	\begin{align}
		\begin{aligned}
			H=&\,\frac{(Q_J+Q_c)^2}{2C_J}+\frac{Q_c^2}{C_c}+\frac{(Q_r-Q_c)^2}{2C_r}+\frac{Q_g^2}{C_g}\\
			&-E_J\cos(2\pi\Phi_J/\Phi_Q)+\frac{\Phi_r^2}{2 L_r},
		\end{aligned}
	\end{align}
	and the drive term $H_d(t)=Q_g V_g(t)$, to form $H_T=H+H_d(t)$. Here $\Phi_Q=h/2e$ is the superconducting flux quantum. Observe that in this first step $\Phi_g$ has already disappeared, i.e. it is already one the $w^I$ variables mentioned above. However, there is still an excess of charge variables which are constants of motion. In fact, as two independent flux one-forms ($\dd\Phi_J$ and $\dd\Phi_r$) appear in $\omega$, there can be only two independent charges in the reduced two-form. It is immediate to observe that $\dd Q_J$ and the combination $\dd Q_r+\dd Q_g= \dd\tilde{Q}_r$ are those two.  Thus the two-form now reads $\omega=\dd Q_J\wedge \dd \Phi_J+\dd\tilde{Q}_r\wedge\dd \Phi_r$, i.e., $\xi^\mu=\{Q_J,\Phi_J,\tilde{Q}_r,\Phi_r\}$ and $w^I=\{\Phi_g,Q_g,Q_c\}$. Observe that the kernel of $\omega$ has dimension 3, with a basis of vectors
	\begin{align}
		\bsb{W}_{Q_g}=\partial_{Q_g}, \quad \bsb{W}_{Q_c}=\partial_{Q_c},\quad \bsb{W}_{\Phi_g}=\partial_{\Phi_g},
	\end{align}
	and that the two-form is in canonical form ($f^{\mu\nu}$ is the  symplectic matrix). As $\Phi_g$ does not appear in the transformed Hamiltonian (recall the introduction of the charge $\tilde{Q}_r$), it represents a gauge freedom that can be directly eliminated from the description. Continuing with step 4, using the two other vectors, we obtain two linear equations for the ``charge-type" loops (here, the voltage source behaves as a capacitor in this respect, hence the color choice in  Fig.~\ref{fig:Elements_noR}), 
	\begin{align}
		\bsb{W}_{Q_c}(H_T)&= \frac{Q_J+Q_c}{C_J} + \frac{Q_c}{C_c} -\frac{\tilde{Q}_r-Q_g-Q_c}{C_r}=0,\nonumber\\
		\bsb{W}_{Q_g}(H_T)&= V_g + \frac{Q_g}{C_g} -\frac{\tilde{Q}_r-Q_g-Q_c}{C_r}=0,\nonumber
	\end{align}
	which can be used to solve $Q_c$ and $Q_g$ in terms of $Q_J$, $\tilde{Q}_r$ and $V_g$. Finally, introducing the solutions in the total energy term $H_T$ we obtain the Hamiltonian dynamics for the pair of conjugated variables $\{\Phi_J,Q_J\}=\{\Phi_r,\tilde{Q}_r\}=1$ governed by 
	\begin{align}
		\begin{aligned}
			H_T=&\,\frac{Q_J^2}{2 \tilde{C}_J}+\frac{\tilde{Q}_r^2}{2 \tilde{C}_r}+\frac{Q_J \tilde{Q}_r}{2\tilde{C}_{Jr}}+U(\Phi_J,\Phi_r)\\
			&+(r_{Jg}Q_J+r_{Jg}\tilde{Q}_r) V_g(t),
		\end{aligned}   
	\end{align}
	and $\tilde{C}_J=C_*^2/(C_c+C_r+C_g)$, $\tilde{C}_r=C_*^2/(C_c+C_J)$,$\tilde{C}_r=C_*^2/(2C_c)$, $r_{Jg}=C_c C_g/C_*^2$, $r_{rg}=C_g(C_c+C_J)/C_*^2$, where $C_*^2=(C_J(C_g+C_r)+C_c(C_g+C_J+C_r))$. Having reached this classical Hamiltonian, canonical quantization follows by promoting the classical variables to quantum operators. The harmonic pair  will map $\{\Phi_r,Q_r\}=1\rightarrow[\hat{\Phi}_r,\hat{Q}_r]=i\hbar$. On the other hand, we recall that as the Josephson junction forms a superconducting island, its phase variable lives in a circle $\varphi_J=2\pi\Phi_J/\Phi_Q\in S^1(2\pi)$, and thus, its conjugate quantized Cooper-pair number operator $\hat{n}_J=\hat{Q}_J/2e$ has an integer spectrum~\cite{Devoret:2021}. Together they will have the standard number-phase commutation relations $[\hat{n}_J,e^{i\hat{\varphi}_J}]=e^{i\hat{\varphi}_J}$. Additionally, recall that a possible charge offset of the superconducting island, giving rise to inequivalent quantizations of the classical C-shunted Josephson  Hamiltonian, must be taken into account~\cite{Egusquiza:2022b}. In other words, there are different $\hat{n}_J$ operators, associated to different boundary conditions/monodromy and parameterized by the gate charge number $n_g\in[0,1)$, with respective spectra $\sigma(\hat{n}_J^{(n_g)})=\left\{ (n-n_g): n\in\mathbb{Z}\right\}$. However, as is well known, such a charge bias (which cannot be trivially \emph{gauged} away for the Josephson variable) can be modified with a constant external voltage~\cite{Buettiker:1987,Bouchiat:1998} (for instance, the dc signal of $V_g(t)$ could play this role).
	
	\subsubsection{Blackbox description of an LC network}
	
\begin{figure}[ht!]
		\centering  \includegraphics[width=.9\columnwidth ]{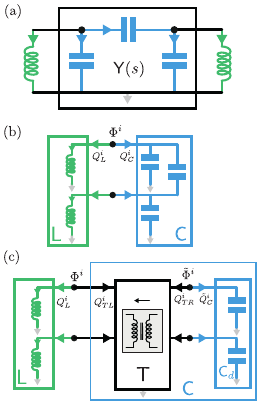}
		\caption{(a) A 2 active-node  lumped-element $LC$ network is (b) deformed into two  multiport capacitive and inductive linear elements connected in parallel, and  characterized by $2\times 2$ (full-rank) $\msC$ and $\msL$ matrices, respectively. (c) Following standard circuit theory, the non-diagonal (possibly, even singular) capacitance matrix can be further decomposed in two capacitors connected to an ideal transfomer~\cite{Newcomb:1966}.}
		\label{fig:LC_network}
	\end{figure}

	In the previous example we have shown how to derive the quantum Hamiltonian for a Josephson junction coupled to a harmonic oscillator using an specific lumped-element representation and reducing all (Kirchhoff's) constraints. We now provide another lumped-element example, an $LC$ network, and show three different approaches to obtain equivalent Hamiltonian descriptions, see Fig.~\ref{fig:LC_network}. We demonstrate that the common reductions and techniques previously developed are also effective in this context, and we illustrate the use of ideal transformer constraints.
	
	In the analysis of Fig.~  \ref{fig:LC_network}(a), and following the steps summarized above, we would start with 10 branch variables (a flux and charge variable per branch), and perform the reduction as in the previous example. From inspection, one sees that in the nodal analysis there are two independent active nodal fluxes and three loop charges. Indeed, there must be a constraint between those loop charges, so that they can all be expressed in terms of two (always invertible for linear systems). 
	
	Using both well-established circuit analysis and our perspective we shall reach exactly the same conclusion. Namely, this comes about by  understanding the network as two multiport $L$ and $C$ networks connected in parallel, see Fig.~\ref{fig:LC_network}(b). Each sub-network can be characterized by a matrix ($\msC$ and $\msL$) with  dimension the number of its outer port branches ($n$ ports, meaning $2n$ variables per network), ignoring the inner loops and nodes (and thus applying directly a variable reduction). For full-rank matrices, the multiport elements will contribute with energy terms 
	\begin{align}
		h_L(\bPhi_L)&=\frac{1}{2}\bPhi_L^T\msL^{-1}\bPhi_L,\\
		h_C(\bQ_C)&=\frac{1}{2}\bQ_C^T\msC^{-1}\bQ_C,
	\end{align}
	where $\bPhi_L=(\Phi_{1},\Phi_{2})^T$ and $\bQ_C=(Q_{C}^1,Q_{C}^2)$. Observe that we are using a common ground plane for all the port branches, and thus, one terminal suffices to characterize the port. It can be easily verified that for the full-rank multiport purely capacitive device, there is a contribution to the pre-canonical two-form given by $\omega_C=\dQ_C^T\wedge\dPhi_C/2$, and analogously for the inductive multiport element. Thus, from this starting point we have the total pre-canonical two-form 
	\begin{align}\label{eq:multiport_canonical}
		\omega_{2B}=\frac{1}{2}\left(\dPhi_L^T\wedge\dQ_L+\dQ_C^T\wedge\dPhi_C\right)
	\end{align}
	where $\dd{\bsb{a}}^T\wedge\dd{\bsb{b}}\equiv\sum_i \dd{a^i}\wedge\dd{b^i}$. Now, the map immersion is performed by reducing the Kirchhoff's constraints, $\dPhi_L=\dPhi_C$ and $\dQ_L=-\dQ_C$, such that
	\begin{align}
		\omega=\dQ_C^T\wedge\dPhi_L,
	\end{align}
	which is already in canonical form.  The computed Lagrangian is thus, 
	\begin{align}
		L=\bQ_C^T\dot{\bPhi}_L-H,\label{eq:L_LC_network}
	\end{align}
	with the energy term (now a  Hamiltonian function) 
	\begin{align}
		H=\frac{1}{2}\bQ_C^T\msC^{-1}\bQ_C+\frac{1}{2}\bPhi_L^T\msL^{-1}\bPhi_L,\label{eq:H_LC_network}
	\end{align}
	with conjugated variables $\{\bPhi_L,\bQ_C^T\}=\mone$.
	
	It is educational to note that this Hamiltonian dynamics can also be derived from an alternate equivalent circuit featuring two capacitors and an ideal Belevitch transformer~\cite{Newcomb:1966,Solgun:2015}, also known under the name of energy participation ratios in the cQED community~\cite{Ciani:2023,Minev:2021b}, see Fig.~  \ref{fig:LC_network}(c). For instance, consider the admittance response of the two-port capacitance network seen by the inductances and characterized by matrix $\msY(s)=s \msC$ in the Laplace space, with $s\in \mathbb{C}$. The orthogonal matrix $\msT$ diagonalizes $\msC=\msf{T}^T\msC_d\msf{T}$, and has as entries the turn-ratios of an ideal Belevitch transfomer~\cite{Belevitch:1950}, such that to the usual Kirchoff's constraints 
	\begin{align}
		\begin{aligned}
			\dPhi_{TL}&=\dPhi_L,\quad
			\dQ_{TL}=-\dQ_L,\\
			\dd{\tilde{\bPhi}_C}&=\dPhi_{TR},\quad
			\dQ_{TR}=-\dd{\tilde{\bQ}_{C}},    
		\end{aligned}
	\end{align}
	where $\dd{\tilde{\bPhi}_C}=(\dd{\tilde{\Phi}^1},\dd{\tilde{\Phi}^2})^T$, we must add the transfomer relations, readily 
	\begin{align}
		\dQ_{TL}&=-\msT^T\dQ_{TR},\\
		\dPhi_{TR}&=\msT\dPhi_{TL}.
	\end{align}
	Observe that the admittance response matrix is easily recovered from the above two equations and the admittance matrix of the capacitance network, i.e., $\bI_{TL}(s)=\msT^T(s\msC_d)\msT \bV_{TL}(s)$. 
	Thus, following the method (step 2), we implement the immersion map to obtain 
	\begin{align}
		\omega&=\imath^*\left(\frac{1}{2}\dPhi_L^T\wedge\dQ_L+\frac{1}{2}\dd{\tilde{\bQ}_C}^T\wedge\dd{\tilde{\bPhi}_C}\right)\nonumber\\
		&=(\msT^T\dd{\tilde{\bQ}_C})^T\wedge\dPhi_L^T,
	\end{align}
	such that we can immediately write the Lagrangian (\ref{eq:L_method}) as
	\begin{align}
		L=(\msT^T\tilde{\bQ}_C)^T\dot{\bPhi}_L-H,\label{eq:L_LC_network_T}
	\end{align}
	with the energy term  
	\begin{align}
		H&=\frac{1}{2}\tilde{\bQ}_C^T\msC_d^{-1}\tilde{\bQ}_C+\frac{1}{2}\bPhi_L^T\msL^{-1}\bPhi_L.\label{eq:H_LC_network_T}
	\end{align}
	Notice that the above Lagrangian and energy function are equivalent to Lagrangian (\ref{eq:L_LC_network}) and Hamiltonian (\ref{eq:H_LC_network}) under the change of coordinates $\bQ_C=\msT^T\tilde{\bQ}_C$ (equivalent to steps 3 and 4). Canonical quantization follows in the standard way for harmonic oscillators.
	
	In fact, the central observation leading to Eq. \eqref{eq:multiport_canonical} can be extended  to  a more general situation. Consider a general purely capacitive multiport element. Its constitutive equation, in the spirit of Chua \cite{Chua:1980}, is a set of  independent relations between its port charge and its port voltage variables, $\bef(\bQ,\bV)=0$. There is a  particular subclass of interest, when the port voltages can all be expressed in terms of the port charges. In this situation  the constitutive equation is rephrased to $\dot{\bPhi}_C=\nabla_{\bQ_C} H(\bQ_C)$, with $\Phi_C^i$ and $Q_C^i$ the $i$-th port's flux and charge variables. I.e., in this case the constitutive relations have an energy interpretation. Then, exactly as in our starting point, we associate   $\omega_C=\dQ_C^T\wedge\dPhi_C/2$ as the contribution to the two-form, and analogously for purely inductive multiport elements.
	
	Observe that the linear case above, with an admittance description, corresponds to the port currents being expressed in terms of the voltages. If the $\msC$ matrix is full rank, however, we are in the previous case. We now see the role of the transformers in the linear case: they allow us to reexpress the constitutive relations so as to extend the construction of the two-form and the energy function even for $\msC$ not being full rank.
	
	\subsubsection{Nonreciprocal circuit for GKP states}
\begin{figure}[ht!]
		\centering  \includegraphics[width=1.0\columnwidth ]{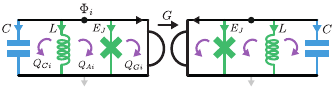}
		\caption{The nonreciprocal circuit comprising capacitors, inductors, pure Josephson elements, and a gyrator introduced in Ref.~\cite{Rymarz:2021} to stabilize GKP states.}
		\label{fig:NR_GKP_circuit}
	\end{figure}

	Finally, let us give an example of a lumped-element nonreciprocal circuit introduced in \cite{Rymarz:2021} with the objective to encode GKP states in the hardware of superconducting circuits, see Fig.~\ref{fig:NR_GKP_circuit}. We can write the two-form (\ref{eq:twoform_restricted}), by first reducing the Kirchhoff constraints (using the set of loop charges and active node fluxes), and then integrating out the nonreciprocal constraints (steps 1 and 2), readily  
	\begin{align}
		\omega=&\,\frac{1}{2}\imath^*\left(\dQ_C^T\wedge\dPhi_C+\dPhi_L^T\wedge\dQ_L+\dPhi_J^T\wedge\dQ_J\right)\nonumber\\
		=&\,\frac{1}{2}\dQ_C^T\wedge\dPhi - \frac{1}{2}\dPhi^T\wedge(\dQ_C+\dQ_A)\nonumber\\
		&+\frac{1}{2}\dPhi^T\wedge(\dQ_A-\dQ_G)\nonumber\\
		=&\,(\dQ_C+\tfrac{\msY}{2}\dPhi)^T\wedge\dPhi,
	\end{align}
	where in the last step we have implemented the reduction of the gyrator branch charges through the dimensionful constraint $\dQ_G=\msY\dPhi$, with $\msY=G i\sigma_y$ the skew-symmetric matrix characterizing the ideal nonreciprocal element. Here, $\sigma_y$ denotes the second Pauli matrix, and $G$ is a conductance parameter.
	Following the method, we can then write the Lagrangian  
	\begin{align}
		L=(\bQ_C+\tfrac{\msY}{2}\bPhi)^T\dot{\bPhi}-H
	\end{align}
	where the energy function is defined as
	\begin{align}
		H=\frac{1}{2}\bQ_C^T\msC^{-1}\bQ_C+\frac{1}{2}\bPhi^T\msL^{-1}\bPhi+U_J(\bPhi),
	\end{align}
	with $\msC=\text{diag}(C,C)$  $\msL=\text{diag}(L,L)$, and $U_J(\bPhi)=-\sum_i E_{Ji}\cos(2\pi\Phi_i/\Phi_Q)$. Now, a shift of the capacitor charges $\tilde{\bQ}_C=\bQ_C+\tfrac{\msY}{2}\bPhi$ allows us to write a canonical two-form, $\omega=\dd{\tilde{\bQ}_C}\wedge\dPhi\equiv f_{\mu\nu}\dd \xi^\mu\wedge\dd \xi^\nu$ such that the kernel space is spanned by gauge-free variables $\{w^I\}=\{\bQ_J,\bQ_A,\bQ_G\}$ (step 3).  Finally, we obtain the Lagrangian $L=\tilde{\bQ}_C\dot{\bPhi}-H$ with, now, the Hamiltonian function 
	\begin{align}
		\begin{aligned}
			H=&\,\frac{(\tilde{\bQ}_C-\tfrac{\msY}{2}\bPhi)^T\msC^{-1}(\tilde{\bQ}_C-\tfrac{\msY}{2}\bPhi)}{2}+\frac{\bPhi^T\msL^{-1}\bPhi}{2}\\
			&-\sum_i E_{Ji}\cos(2\pi\Phi_i/\Phi_Q),
		\end{aligned}
	\end{align}
	equivalent to Eq.~(46) in \cite{Rymarz:2021}. Canonical quantization follows in the standard way, promoting the conjugate variables to quantum operators (observe that there are no islands, and thus $\Phi_i\in \mathbb{R}$). We stress again that none of the ideal lumped-elements considered in Fig.~\ref{fig:Elements_noR}, including ideal capacitors or inductors, exist in reality, but, combined, can be used to represent large ranges of frequency response of (superconducting) electrical circuits.

	\subsection{Revision of the microscopically-inspired topological ansatz for two-terminal lumped elements}
	The spectra of macroscopic quantum flux and charge operators have been a subject of intense debate within the circuit QED community for some time now (for a summary of the debate, see, e.g., \cite{ParraRodriguez:2024b}). Aware of this debate, most practitioners have (implicitly) adopted the following approach (see e.g.  Ref.~\cite{Osborne_v2:2024}).
	
	Initially, the classical Lagrangian and Hamiltonian dynamics of the circuit are derived under the assumption that the original classical branch manifold is $\mathcal{M}_{2B} = \mathbbm{R}^{2B}$, and consequently, $\mathcal{M}_S = \mathbbm{R}^{2N}$ (with $N < B$). Discrete translational symmetries in the Hamiltonian function are then identified. Symplectic transformations are also performed (still under the assumption that all variables are extended) to identify the so-called \emph{periodic} variables (i.e., \emph{extended} coordinates along which the Hamiltonian function is periodic). These coordinates are then wrapped around to construct compact configuration submanifolds by the quotienting  the original $\mathbbm{R}$ by the discrete symmetry, resulting in pairs of $\mathbbm{R} \times \mathbbm{R}$ (for the non-periodic directions) and $\mathbbm{R} \times S^1$ (for the previously periodic directions). Finally, these pairs are promoted to quantum operators with real spectra (as in the quantum harmonic oscillator), as well as to pairs with discrete and continuous, but compact spectra (as in the quantum rotor, i.e., number and phase operators). 
	
	The identification of those periodicities is not completely trivial, however, and failing in carrying it out the results can be rather misleading. In particular, if the symplectic form (or its dual Poisson bracket) is not written according to all periodicities the resulting quantization will be incorrect. We have therefore proposed Ref.~\cite{ParraRodriguez:2024a}  a systematic procedure to first identify the number of compact directions and then obtain a geometric description that includes the topological structure.  
	\label{subsec:micro-topological}
	\begin{figure}[ht!]
		\centering  \includegraphics[width=.85\columnwidth ]{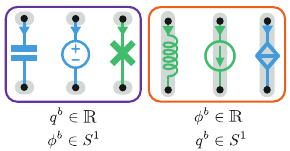}
		\caption{A microscopically-consistent topological ansatz for the branch manifold of lumped elements was put forward in Ref.~\cite{ParraRodriguez:2024a}. Here, }
		\label{fig:topological_ansatz}
	\end{figure}
	
	In fact, through the approach presented in Ref.~\cite{ParraRodriguez:2024a}, microscopically consistent classical Hamiltonian dynamics with configuration submanifolds of compact topology can be achieved by assuming specific nontrivial topologies in the initial branch manifold $\mcl{M}_{2B}$ for the various classes of lumped elements commonly encountered in superconducting circuits. We note that we do not include assignments for elementary ideal multiports (transformers and nonreciprocal constraints), for which additional subtleties will have to be taken into account. 
	
	A summary of these axiomatic rules is illustrated in Fig.~\ref{fig:topological_ansatz}, where capacitors, Josephson junctions, and voltage sources are assigned branch variables $q^b \in \mathbb{R}, \phi^b \in S^1$, while inductors, phase-slip junctions, and current sources are treated conversely. Once these assumptions are fixed, the geometrical method provides consistent Hamiltonians in which the aforementioned discrete symmetries will be directly implemented. This reduces the risk of errors and makes the approach potentially amenable to automation, see discussion below.
	
\begin{figure}[ht!]
		\centering  \includegraphics[width=.65\columnwidth ]{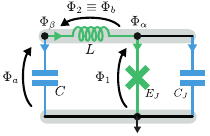}
		\caption{A circuit comprising two superconducting islands and no superconducting loop. The solution of the Kirchhoff constraints can provide different bases of network fluxes (in the figure) and charges (not shown). Imposing the topological ansatz for the branch variables in Fig.~\ref{fig:topological_ansatz}, only $\{\Phi_a,\Phi_b\}\in S^1\times\mathbbm{R}$ and $\{\Phi_1,\Phi_2\}\in S^1\times\mathbbm{R}$ are consistent parametrizations of the reduced flux submanifold. On the other hand, the set of node fluxes $\{\Phi_\alpha, \Phi_\beta\}\in\mathbbm{R}^2$ can only parametrize a (KVL) integral submanifold  if all  the branch variables are assumed to be extended.}
		\label{fig:topological_ansatz_RD}
	\end{figure}

	Let us illustrate the method with the circuit in Fig.~\ref{fig:topological_ansatz_RD}. Here, the KVL constraints are 
	\begin{align}
		\dd \phi_J+\dd \phi_L-\dd \phi_{C}=0,\\
		\dd \phi_J-\dd \phi_{C_J}=0
	\end{align}
	whereas the KCL ones are 
	\begin{align}
		\dd q_C-\dd q_L=0,\\
		\dd q_L-\dd q_J-\dd q_{C_J}=0.
	\end{align}
	Taking into account the topological branch manifold ansatz in Fig.~\ref{fig:topological_ansatz}, we can parametrize the constrained submanifold in multiple ways. For instance, a solution to  these equations is written as
	\begin{align}\label{eq:phi_sol_RD_1}
		\begin{aligned}
			\phi_J&=\phi_{C_J}=\Phi_1\in S^1 [\Phi_Q],\\
			\phi_L&=\Phi_2\in \mathbbm{R},\\
			\phi_C&=(\Phi_1+\Phi_2)\mod \Phi_Q, 
		\end{aligned}
	\end{align}
	and 
	\begin{align}
		\begin{aligned}
			q_J&=-Q_1 \in \mathbbm{R}\\
			q_C&=Q_2\in \mathbbm{R}\\
			q_L&=-Q_2 \mod (2e),\\
			q_{C_J}&=Q_1+Q_2.
	\end{aligned}\end{align} 
	As there are compact directions more than one coordinate patch would be required, of course. Yet, because of the simple $\mathbb{R}^3\times S^1$ topology of the integral manifold, both are directly inferred from the expressions above.
	Thus, the pull-back of the two-form is the canonical one (with canonical pairs of coordinates, and the standard shortcut of writing $\mathrm{d}\alpha$ for the not exact one form corresponding to compact $S^1$ directions, with $\alpha$ the angular variable in one patch)
	$\omega=\sum_{i=1,2}\dd Q_i\wedge \dd \Phi_i$
	and the Hamiltonian reads
	\begin{align}
		H=\frac{(Q_1+Q_2)^2}{2C_J}+\frac{Q_2^2}{2C}-E_J\cos(\varphi_1)+\frac{\Phi_2^2}{2L}.\label{eq:H_RD_1}
	\end{align}
	Here and henceforward we use $\varphi_i$ to denote the phase variable corresponding to the flux variable $\Phi_i$, namely $\varphi_i= 2\pi\Phi_i/\Phi_Q$.
	
	\begin{figure*}
		\centering\includegraphics[width=.85\textwidth ]{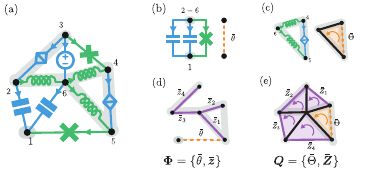}
		\caption{(a) Example of a circuit comprising two superconducting islands (SIs) and one superconducting loop (SL). The algorithm in the text to solve the Pfaff equations subsequently is equivalent to finding separated tree sets for the flux and the charge variables. For instance, to find the kernel of $\msF_{\text{loop}}$ in (\ref{eq:F_loop_PE_cEx}),  we first find a sub-tree for the compact branches bridging separated SIs (b), and then complete with a sub-tree for the extended branches (d). Dually, to find a network basis of charges, we first find a basis of compact charges in SLs (c), and then complete with a basis of extended charges for the rest of the network (e).}
		\label{fig:Topological_ansatz_solving_Pfaff}
	\end{figure*}
	
	A different parametrization of the geometrical solution to the KCLs and KVLs, describing a patch of the same manifold, is
	\begin{align}\label{eq:phi_sol_RD_2}
		\begin{aligned}
			\phi_C&=\Phi_a\in S^1[\Phi_Q],\\
			\phi_L&=\Phi_b\in \mathbbm{R},\\
			\phi_J&=\phi_{C_J}=(\Phi_a-\Phi_b) \mod \Phi_Q,
		\end{aligned}
	\end{align}
	and 
	\begin{align}
		\begin{aligned}
			q_J&=-Q_a\in \mathbbm{R}\\
			q_{C_J}&=-Q_b\in \mathbbm{R},\\
			q_C&=Q_a+Q_b,\\
			q_L&=-(Q_a+Q_b) \mod (2e).\\
	\end{aligned}\end{align}
	Again, this results in the following canonical, i.e., two-form $\omega=\sum_{\mu=a,b}\dd Q_\mu\wedge \dd \Phi_\mu$, and the Hamiltonian
	\begin{align}
		H=\frac{Q_b^2}{2C_J}+\frac{(Q_a+Q_b)^2}{2C}-E_J\cos(\varphi_a-\varphi_b)+\frac{\Phi_b^2}{2L}.\label{eq:H_RD_2}
	\end{align}
	Observe that, having solved Kirchhoff's constraints with the topological ansatz, we have derived one Hamiltonian function on a manifold, that is expressed in two different functional forms, Eqs.~(\ref{eq:H_RD_1}) and ~(\ref{eq:H_RD_2}), according to different parameterizations. Both account for the same dynamics with canonical coordinates, one of which in each case describes a compact direction ($\phi_1$ in Eq.~(\ref{eq:H_RD_1}) and $\phi_a$ in Eq.~(\ref{eq:H_RD_2})). These functional forms, if understood as functions on $\mathbb{R}^4$, are periodic in the corresponding variable.
	
	Now, let us comment on the possible pitfalls, following the standard node-flux~\cite{Devoret:1997}  analysis, if one assumes from the outset that all variables are real. By choosing a  ground node as specified in Fig.~\ref{fig:topological_ansatz_RD}, we  solve fluxes as
	\begin{align}\label{eq:phi_sol_RD_node}
		\begin{aligned}
			\phi_C&=\Phi_\alpha\in \mathbbm{R},\\
			\phi_J&=\phi_{C_J}=\Phi_\beta\in \mathbbm{R},\\
			\phi_L&=(\Phi_\beta-\Phi_\alpha),
		\end{aligned}
	\end{align}
	and charges as
	\begin{align}
		\begin{aligned}
			q_{C_J}&=Q_\alpha\in \mathbbm{R},\\
			q_C&=Q_\beta\in \mathbbm{R},\\
			q_J&=-(Q_\alpha+Q_\beta),\\
			q_L&=-Q_\beta, \\
	\end{aligned}\end{align}
	such that the two-form is already in canonical form $\omega=\sum_{\xi=\alpha,\beta}\dd Q_\xi\wedge\dd \Phi_\xi$ and the Hamiltonian is 
	\begin{align}
		H=\frac{Q_\alpha^2}{2C_J}+\frac{Q_\beta^2}{2C}+\frac{(\Phi_\beta-\Phi_\alpha)^2}{2L}-E_J\cos(\varphi_\alpha).
	\end{align}
	A first pitfall is that frequently this Hamiltonian has been described as lacking periodicity. In fact, this Hamiltonian possesses a discrete translational symmetry along the $(1,1)$  direction of the fluxes \cite{PRXComment}. A different possible pitfall is the following: as $Q_\alpha$ is conjugate to an apparently angular variable, it could be understood that it is to be quantized as a discrete charge number operator. This is wrong. The periodicity requires a translation also in the symplectic orthogonal spatial coordinate, thus it must involve also $Q_\beta$, and $Q_\alpha$ is not the generator of translations in the periodic direction.
	
	There are two ways to avoid these  problems. The first one is to quantize on a $\mathbb{R}^2$ configuration space, identify  the periodicity, and use Bloch's theorem to obtain an effective Hamiltonian. The second one is to carry out a symplectic transformation in $\mathbb{R}^4$ to take the Hamiltonian to either form~(\ref{eq:H_RD_1}) or form~(\ref{eq:H_RD_2}), and then quantize. In this way,  periodicity is immediately identified, and the periodicity direction coincides with one of the coordinates, thus precluding confusion. 
	
	By adopting the topological ansatz from the outset, on the other hand, we ensure that no discrete symmetries are overlooked during the variable reduction process, as they are inherently accounted for by definition. 
	
	We now show that there is a systematic process to achieve that goal. A general hamiltonian reduction process can lead to complicated topological structures, but the problem we address is simple enough, in that the only possible topologies that the KVL/KCL reduction leads to are flat space times tori. The first step is to identify the dimension of the tori, and the second one to write an adequate parameterization of the integral manifold.
	
	Systematically, we consider the KCL/KVL exterior differential system under the additional condition that the extended branch variables are fixed. We are left with a linear exterior differential system that only involves compact directions. In fact, we can understand it as an exterior system, as the coefficients are constant over the initial manifold, of the form $\mathsf{D}\mathrm{d}\boldsymbol{\theta}=0$ (with the by now standard abuse of notation for the canonical one-form in $\Lambda(S^1)$), and the number of compact directions in the integral manifolds of the KCL/KVL Pfaff system will be the dimensionality of the kernel of $\mathsf{D}$. 
	
	More explicitly, by row and column arithmetic operations we can  write the loop matrix $\msF_{\text{loop}}$, such that the KVL Pfaff system reads
	\begin{align}
		\msF_{\text{loop}}\dd{\bphi}= \msD_{\text{loop}} \dd\bsb{\theta}+\msE_{\text{loop}} \dd {\bz} =0\,,\label{eq:F_loop_PE_cEx}
	\end{align}
	and setting $\mathrm{d}\boldsymbol{z}$ to 0 provides us with the desired subsystem for compact fluxes. The number of compact flux directions will be the dimension of $\ker\msD_{\text{loop}}$. 
	
	The next step is to compute a parameterization of the integral manifolds, in as algorithmic a manner as possible. By reordering the loop matrix, it can be written in the form  
	\begin{align}
		\msF_{\text{loop}}=\begin{pmatrix}
			0&\tilde{\msD}_{\text{loop}}&\msE_{\text{loop}}
		\end{pmatrix}\,,
	\end{align}
	where the 0 pertains to the kernel of $\msD_{\text{loop}}$. In other words, the solution of the KVL Pfaff equations as an exterior system can be expressed as 
	\begin{align}
		{\bphi}=\bar{\theta}^i\begin{pmatrix}
			\bsb{k}_{i}\\0\\0
		\end{pmatrix}+\bar{z}^j\bsb{\
		}\begin{pmatrix}
			0\\\bsb{k}_{j}^\perp\\\bsb{b}_{j}
		\end{pmatrix},\end{align}
	where $\ker \msD_{\text{loop}} = \text{span}\left\{ \bsb{k}_i \right\}$ (after suitable padding with zeroes) and $(\ker \msD_{\text{loop}})^\perp = \text{span}\left\{ \bsb{k}_i^\perp \right\}$ (again after suitable padding with zeroes). In this way variables  $\bar{\theta}^i\in S^1$ whereas $\bar{z}^i\in \mathbbm{R}$. The same process is to be carried out for the cutset matrix $\msF_{\text{cut}}$.
	
	This algebraic procedure is equivalent to finding a reduced tree where inductive-type branches have been shunted, and then completing the tree with a basis of inductive-type branches, as depicted in Fig.~\ref{fig:Topological_ansatz_solving_Pfaff}. The KVL analysis is shown in Fig.~\ref{fig:Topological_ansatz_solving_Pfaff} (b) and (d). A graphically dual analysis  pertains to the cutset matrix $\msF_{\text{cut}}$, and  solves consistently the Pfaff equations with the maximum number of compact charges, as depicted in the example of Fig.~\ref{fig:Topological_ansatz_solving_Pfaff} in (c) and (e).

	\section{Geometrical description of transmission lines}
	\label{sec:TLs-geom-description}
	In the previous section we summarized the geometrical approach to
	construct Hamiltonian descriptions of \emph{lumped} element
	circuits introduced in~\cite{ParraRodriguez:2024a} and presented the necessary computations by way of example. Here we extend the systematics to transmission lines (TLs), revisiting alternative discrete models from which to obtain a universal continuum model. We describe as well how to connect TLs to one-port lumped-element networks. To provide further insight, we explicitly show how zero modes in the discrete model relate to those in the continuous model. 
	
	The use of voltage and current quantities for transmission lines dates back to the XIX. century \cite{Heaviside:1971} (ch. IV), and one straightfowardly passes on to local flux and charge fields. As is well known, the quasi-static approximation yields the telegrapher's equations for TEM propagation. We shall recover those from the following formalism.
	\begin{figure}[ht!]
		\centering
		\includegraphics[width=1\columnwidth ]{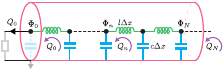}
		\caption{A discretization of a (semi-infinite) transmission line without differential capacitors at the end(s). With or without them, the chain tends to the same continuous model (cf. Sec.~\ref{sec:Alternative_disc}).}
		\label{fig:TL_QPhi}
	\end{figure}
	
	In order to properly identify some subtleties in the construction,
	most particularly the computation of the zero vectors (including gauge modes), we shall start with a discrete model of a semi-infinite transmission
	line. Let $c$ and $\ell$ be the capacitance and  inductance per unit
	length, and let $\Delta x$ be a small length that will be eventually
	taken to zero. Following the  recipe above
	~\cite{ParraRodriguez:2024a}, and using the node fluxes and loop charges depicted in
	Fig. \ref{fig:TL_QPhi} to express the branch variables ($q_n^c =Q_{n+1}-Q_{n}$ and $\phi_n^l=\Phi_{n+1}-\Phi_{n}$), we obtain  the two-form (\ref{eq:twoform_restricted})
	\begin{subequations}\label{eq:discreteometl}
		\begin{align}
			\omega_{\TL}=&\,\frac{1}{2}\sum_{n=0}^\infty\left(\dd Q_{n+1}-\dd Q_n\right)\wedge\dd \Phi_{n+1}\nonumber
			\\
			&+\frac{1}{2}\sum_{n=0}^\infty\left(\dd \Phi_{n+1}-\dd \Phi_n\right)\wedge\dd Q_{n}\label{eq:omegaTLdiscrete}\\
			=&\,\frac{1}{2}\dd Q_0\wedge\dd \Phi_0+\sum_{n=0}^\infty\left(\dd Q_{n+1}-\dd Q_n\right)\wedge\dd \Phi_{n+1}\,.\label{eq:cleandiscreteomTL}
		\end{align}
	\end{subequations}
	In the same discrete model the energy function for this transmission
	line reads
	\begin{align}
		\label{eq:hamtldiscrete}
		h_{\TL}=\sum_{n=0}^\infty\left[\frac{\left(Q_{n+1}-Q_n\right)^2}{2c\Delta x}+\frac{\left(\Phi_{n+1}-\Phi_n\right)^2}{2\ell\Delta x}\right].
	\end{align}
	Observe that here, due to absence of transformers and ideal NR elements, the basis of node fluxes and loop charges is a basis of the kernel of the Kirchhoff's constraints (KCL and KVL), which allows us to express all the branch variables in a first reduced set. Combining both symplectic and energy information, we construct the Lagrangian (\ref{eq:L_method})  $L_{\TL}=L_\omega -h_{\TL}$
	where \begin{align}
		L_\omega=\frac{1}{2}Q_0\dot{\Phi}_0+\sum_n \left(Q_{n+1}-Q_n\right)\dot{\Phi}_{n+1}.
	\end{align}
	
	Naturally, this Lagrangian dynamics is subjected to a standard continuous limit $\Delta x\rightarrow 0$,
	identifying
	$x_n= n \Delta x$ and $L=N\Delta x$ (which in the case of a semi-infinite line becomes $L\rightarrow\infty$), with $N\to\infty$, and having
	\begin{align}
		\label{eq:contlimit}
		\sum_{n=0}^\infty \Delta x f(n \Delta x)\to \int_0^L\dd x\,f(x)\,.
	\end{align}
	
	To avoid confusion between the integral measure $\dd x$ and the differential forms on the $Q(x,t)$, $\Phi(x,t)$ field-space, we shall denote the latter with $\delta$ inside the integrals. In this way, the two-form and the Lagrangian can be written in the continuous limit as
	\begin{align}
		\omega_{\TL}&= \frac{1}{2}\dd Q_0\wedge
		\dd \Phi_0+\int_{\mathbbm{R}^+}\dd x\,\delta
		Q'(x)\wedge\delta\Phi(x)\,,\label{eq:twoform_TL_cont}\\
		L_{\TL}&= L_{\omega,\TL}- h_{\TL}[Q(x),\Phi(x)],\label{eq:Lagrangian_TL_cont}
	\end{align}
	where 
	\begin{align}
		L_{\omega,\TL}&=\frac{1}{2}Q_0\dot{\Phi}_0+\int_{\mathbbm{R}^+} \dd x\, Q'(x)\dot{\Phi}(x)\\
		&\equiv \frac{1}{2}\int_{\mathbbm{R}^+} \dd x \left(Q'(x)\dot{\Phi}(x)+\Phi'(x)\dot{Q}(x)\right)+\partial_t(\cdot),\nonumber\\
		h_{\TL}&=\int_{\mathbbm{R}^+}\dd x\left(\frac{Q'(x)^2}{2c}+\frac{\Phi'(x)^2}{2l}\right).
	\end{align}
	
	Here, and in the rest of the article, we denote with $\Phi_0\equiv\Phi(0)$ (and $Q_0\equiv Q(0)$) the values of the fields at the end of the line, except where clarity demands otherwise. We use from here on the notation $h_{\TL}$, that we introduced for the Hamiltonian function  of the discretized version  of the transmission line, also for the Hamiltonian functional of the continuum. In general we shall not indicate explicitly its functional dependence unless clarity requires it. Observe the two distinct presentations of $L_{\omega,\TL}$ (for the semi-infinite TL). While both presentations will be used in this work, the first one proves to be more effective in analysing zero modes in the model, whereas the second one is preferred later on when we implement separation of variables to describe the TLs in terms of modes. We remind the reader that additional boundaries can systematically be taken into account without posing further problems.  See an example in Sec.~\ref{sec:quasi_lumped_circuit_examples}, which builds upon previous work~\cite{ParraRodriguez:2018}.
	\subsection{TLs connected to one-port networks}
	\label{sec:TLs-1port-networks}
	
\begin{figure}[ht!]
		\centering \includegraphics[width=1\columnwidth]{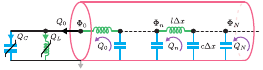}
		\caption{Transmission line coupled to a nonlinear LC oscillator, with the TL ended in a differential inductor.}
		\label{fig:NL-LC_TL_seminf}
	\end{figure}

	Now that we have the discrete and continuous Lagrangian descriptions of the transmission line in first order, let us extend this result to circuits consisting of TLs coupled to lumped elements (at points). To illustrate this, we will employ a (nonlinear) LC oscillator as a canonical example and connect it to the termination of a semi-infinite transmission line, as depicted in Fig.~\ref{fig:NL-LC_TL_seminf}. Using the natural variables to describe the TL, the node fluxes and the loop charges, and the branch charges of the additional external elements, we can write the total discrete (not yet canonical) two-form
	\begin{align}
		\omega &=
		\omega_{\TL}+\omega_{\mathrm{NL}}\nonumber\\
		&=\omega_{\TL}+\frac{1}{2}\dd \Phi_0\wedge\dd Q_L+\frac{1}{2}\dd Q_C\wedge\dd\Phi_0.
	\end{align}
	The additional contribution to the Hamiltonian is, again with branch variables, $H_{\mathrm{NL}}=h_c(Q_C)+h_l(\Phi_0)$.
	The use of branch charge variables allows us to illustrate that the familiar equivalences in circuit theory also apply here: swapping the outer inductance and capacitance branches yields the same two-form and Hamiltonian.
	
	Following our procedure, we must incorporate the additional KCL at the connection $\dd Q_C+\dd Q_L = \dd Q_0$ to reduce the set of variables while connecting inner and outer dynamics. Clearly,  solving $\dd Q_C$ in terms of $\dd Q_L$ and $\dd Q_0$ will be equivalent to solving $\dd Q_L$ in terms of $\dd Q_C$ and $\dd Q_0$, as we will see next. However, to simplify the subsequent steps of the procedure a particular choice might be more useful.
	
	Let us first consider the case where we solve the current $\dd Q_L$, such that the restricted two-form (\ref{eq:twoform_restricted}) and the Hamiltonian read
	\begin{subequations}
		\label{eq:NL-LC_TL_solving_dQL}
		\begin{align}
			\omega &=
			\dd Q_C\wedge\dd \Phi_0+\sum_{n=0}^\infty\left(\dd Q_{n+1}-\dd Q_n\right)\wedge\dd \Phi_{n+1},\\
			H&= h_{\TL}+h_c(Q_C)+h_l(\Phi_0)\,.\label{eq:H_NL-LC_TL_solving_dQL}
		\end{align}
	\end{subequations}
	Observe that, with this choice, the (loop) charge $Q_C$ is canonically conjugate to $\Phi_0$. As can be seen in the figure, there is one $Q_n$ for each $\Phi_n$, and we have the additional charge $Q_C$. It follows that the kernel of $\omega$ must be nontrivial. We are searching for vectors $\boldsymbol{W}$ such that
	$\omega(\boldsymbol{W},\cdot)=0$. The basis for the space of one forms is the set of flux and charge coordinate one-forms, which are independent. We see that the one forms for the fluxes appear independently, so we cannot have a zero vector with flux components. We search for zero vectors with only components in the charge directions, i.e.,
	\begin{align}
		\label{eq:WLCexample}
		\boldsymbol{W}=W^C\frac{\partial}{\partial Q_C}+\sum_{n=0}^\infty
		W^n\frac{\partial}{\partial Q_n}\,.
	\end{align}
	When we apply $\omega$ to this vector, it results in
	\begin{align}
		\label{eq:omeagaonW}
		\omega(\boldsymbol{W},\cdot)= W^C\dd \Phi_0+\sum_{n=0}^\infty\left(W^{n+1}-W^n\right)\dd \Phi_{n+1}\,.
	\end{align}
	As the one-forms $\dd \Phi_n$ are linearly independent, for this to be zero all the components must be zero, $W^C=0$ and, for all $n\geq0$, $W^{n+1}=W^n$. We have thus computed the kernel of $\omega$
	as the one  dimensional space generated by
	\begin{align}
		\label{eq:kernelomegaLC}
		\boldsymbol{W}=\sum_{n=0}^\infty\frac{\partial}{\partial Q_n}\,.
	\end{align}
	It now behoves us to compute the associated constraint given the
	Hamiltonian (\ref{eq:H_NL-LC_TL_solving_dQL}), and it is seen
	to be identically zero, $\boldsymbol{W}(H)\equiv0$, no matter what the (nonlinear) oscillator energy content is. We are thus in the presence of a \emph{gauge} zero mode.
	
	We could have identified the zero mode directly by noticing that a common shift $Q_n\to Q_n+q(t)$ leaves both the one-form and the Hamiltonian unchanged. Therefore we have a set of redundant descriptions for the same state of the system. Naturally, we can have a presentation in terms of \emph{fields} by taking the same continuous limit of the two-form and the Hamiltonian,
	\begin{subequations}
		\label{eq:contform}
		\begin{align}
			\omega &= \dd Q_C\wedge
			\dd \Phi_0+\int_{\mathbbm{R}^+}\dd x\,\delta
			Q'(x)\wedge\delta\Phi(x)\,,\label{eq:twoformcont}\\
			H &= h_c(Q_C)+h_l(\Phi_0)+h_{\TL}[Q(x),\Phi(x)]\,,\label{eq:hamcontinuous}
		\end{align}
	\end{subequations}
	where we have again used the notation $\Phi_0\equiv \Phi(0)$ and $Q_0\equiv Q(0)$ for the values of the fields at the ends. It is now natural to observe that the zero mode of the two-form in this language is
	\begin{align}
		\label{eq:functionalzeromode}
		\boldsymbol{W}=\int_{\mathbbm{R}^+}\dd \xi\,\frac{\delta}{\delta Q(\xi)}\,,
	\end{align}
	since the derivative with respect to coordinates becomes a functional derivative, see Appendix section \ref{sec_diff_forms_app} for details. Alternatively, observe that the charge field only enters the two-form and the Hamiltonian as $Q'$, and a global time dependant shift $Q(x,t)\to Q(x,t)+q(t)$ does not change the expression of the dynamics.
	
	Were we to solve the current $\dd Q_C$ (implementing KCL at the end) we would have the equivalent expressions 
	\begin{subequations}  \label{eq:NL-LC_TL_solving_dQC}
		\begin{align} %
			\begin{split}
				\omega=&\left(\dd Q_0-\dd Q_L\right)\wedge\dd \Phi_0\\
				&+\sum_{n=0}^\infty\left(\dd Q_{n+1}-\dd Q_n\right)\wedge\dd \Phi_{n+1}\,,\label{eq:omega_NL-LC_TL_solving_dQC}\\ 
			\end{split}\\
			\begin{split}
				\quad H&=h_{\TL}+h_c(Q_0-Q_L)+h_l(\Phi_0)\label{eq:H_NL-LC_TL_solving_dQC}
			\end{split}
		\end{align}
		The main difference is that the zero mode of the two-form takes now the form
		\begin{align}
			\label{eq:zeromodediscreteextended}
			\boldsymbol{W}=\frac{\partial}{\partial
				Q_L}+\sum_{n=0}\frac{\partial}{\partial Q_n}\,.
		\end{align}
	\end{subequations}
	It is again a gauge generator, $\boldsymbol{W}(H)\equiv0$ with the Hamiltonian~(\ref{eq:H_NL-LC_TL_solving_dQC}). The common shift that leaves the two-form and the Hamiltonian invariant now has to be carried out for $Q_L$ as well.
	
	Let us now concentrate on the gauge aspect of this example in either presentation. The goal of a non-redundant hamiltonian description of these systems can be achieved by \emph{gauge fixing}, i.e., by choosing one amongst all the equivalent shifts. For instance, in relation to the set \eqref{eq:NL-LC_TL_solving_dQL} we can choose the shift $q(t)=Q_C(t)-Q_0(t)$ to displace all $Q_n$. This choice corresponds with the identification of $Q_0$ with $Q_C$, so it is equivalent to simply eliminating $Q_C$ by that identification.  Had we decided to eliminate $Q_C$ to arrive at Eqs. \eqref{eq:NL-LC_TL_solving_dQC}, there exists a gauge choice that gives us exactly the same result, namely $q(t)= -Q_L(t)$, which amounts to setting $Q_L(t)=0$. That is, in both cases we have the gauge-fixed  Lagrangian 
	\begin{align}\label{eq:GaugeFixedContinuum}
		L= L_{\TL}- h_c(Q_0)-h_l(\Phi_0)\,, 
	\end{align}
	with $L_{\TL}$ as in Eq. \eqref{eq:Lagrangian_TL_cont}, which is regular.
	
	In this example it is fruitful to carry out this gauge reduction at this point, even if for no other reason than to prove the equivalence of both presentations above. 
	Nonetheless, it is sometimes convenient to keep a redundant description until a later stage of the analysis. In particular, with elimination of $Q_L$ and with no gauge fixing we would have 
	\begin{align}\label{eq:NoGaugeFixingContinuum}
		\begin{aligned}
			L =&\, Q_C\dot{\Phi}_0+ \int_{\mathbbm{R}^+}\dd x\, Q'(x)\dot{\Phi}(x) - h_c(Q_C)\\
			& -h_l(\Phi_0)-\int_{\mathbbm{R}^+}\dd x\left(\frac{Q'(x)^2}{2c}+\frac{\Phi'(x)^2}{2l}\right)   
		\end{aligned}
	\end{align}
	in the continuum limit. Indeed, this is the perspective we shall use for transmission lines in general: we shall present them in the continuum, without gauge fixing. Before we look at the multi-line case we examine two possible issues. First, whether this result hinges on the discretization we have chosen for TL. Second, we look at new zero modes that only appear in the continuum limit.
	
	\subsection{Alternative discretizations}
	\label{sec:Alternative_disc}
\begin{figure}[ht!]
		\centering \includegraphics[width=1\columnwidth]{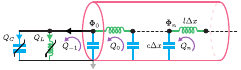}
		\caption{Alternative discretization of the circuit in Fig.~\ref{fig:NL-LC_TL_seminf} where the TL ends in a differential capacitance $c\Delta x$.}
		\label{fig:NL-LC_TL_seminf_ADiscrete}
	\end{figure}

	As we have pointed out above (Fig. \ref{fig:TL_QPhi}), alternative discretization models of transmission lines do exist. We examine here the impact of one such alternative, in the context of coupling to a (nonlinear) oscillator as in the previous subsection. Namely, consider that the shaded capacitor, also with capacity $c\Delta x$, is also present, see Fig.~\ref{fig:NL-LC_TL_seminf_ADiscrete}, such that there is an additional adjacent loop charge, $Q_{-1}$. We obtain a new two-form for the transmission line, $\tilde{\omega}_{\TL}={\omega}_{\TL}+\left(\dd Q_0-\dd Q_{-1}\right)\wedge\dd \Phi_0/2$, and there is an additional term for the line Hamiltonian as well, $\tilde{h}_{\TL}={h}_{\TL}+\left(Q_0-Q_{-1}\right)^2/2 c\Delta x$. 
	Let us now connect to the nonlinear oscillator. Before carrying out any computation, we stress that the additional differential capacitance  $c\Delta x$ is in parallel with the nonlinear capacitor with energy function $h_c(Q_C)$ (and nonlinear inductance $h_l(\Phi_0)$). Thus, the final description should be exactly as before, but with the external capacitor being substituted by an effective one, that is equivalent to the two parallel ones. In fact, we shall now compute that equivalence in this formalism, and show how it becomes the standard reduction if the external capacitor is linear, by following our method.
	
	The total two-form and Hamiltonian are in this case $\omega=\omega_{\mathrm{NL}}+\tilde{\omega}_{\TL}$ and  $H= H_{\mathrm{NL}}+\tilde{h}_{\TL}$ respectively, clearly different from those introduced in the previous section. However, we have yet to impose the Kirchhoff current law at the connecting node, which in this case reads $\dd Q_C+\dd Q_L=\dd Q_{-1}$. After doing so, the total two-form becomes
	\begin{align}
		\omega =& \left(\dd Q_C+\dd Q_0-\dd Q_{-1}\right)\wedge\dd\Phi_0\nonumber\\
		&\,+\sum_{n=0}^\infty \left(\dd Q_{n+1}-\dd Q_n\right)\wedge\dd\Phi_{n+1}\,.
	\end{align}
	The kernel of $\omega$ is in this case of dimension two. We choose as the basis of the kernel to be the gauge vector
	\begin{align}
		\boldsymbol{W}_g = \sum_{n=-1}^\infty\frac{\partial}{\partial Q_n}\,,
	\end{align}
	such that $\boldsymbol{W}_g(H)\equiv0$,
	and the non-gauge
	\begin{align}
		\boldsymbol{W}_n = \frac{\partial}{\partial Q_C}+ \frac{\partial}{\partial Q_{-1}}\,.
	\end{align}
	The condition $\boldsymbol{W}_n(H)=0$ becomes
	\begin{align}
		Q_{-1} = Q_0- c\Delta x h_c'(Q_C)\,.
	\end{align} 
	
	Thus the equivalent capacitor has the total energy $h_c(Q_c)+ c\Delta x \left[h'_c(Q_C)\right]^2/2$.
	The explicit expression for the reduced two-form and Hamiltonian are therefore
	
	\begin{align}
		\begin{split}
			\omega =&\,\left[1+c\Delta x h_c''(Q_C)\right]\dd Q_C\wedge\dd\Phi_0\\
			&+\sum_{n=0}^\infty\left(\dd Q_{n+1}-\dd Q_n\right)\wedge\dd\Phi_{n+1}, 
		\end{split}\label{eq:reducedlalternative_twoform}\\
		\begin{split}
			H &=\, h_c(Q_C)+h_l(\Phi_0)+ \frac{c\Delta x \left[h'_c(Q_C)\right]^2}{2}\\
			&+\sum_{n=0}^\infty\left[\frac{\left(Q_{n+1}-Q_n\right)^2}{2c\Delta x}+\frac{\left(\Phi_{n+1}-\Phi_n\right)^2}{2\ell\Delta x}\right].\label{eq:reducedlalternativeHamiltonian}
		\end{split}      
	\end{align}
	
	If the external capacitor were linear, with capacity $C$, it would be convenient to define a new variable $\tilde{Q}= (1+c\Delta x/C)Q_C$, and the two capacitive energy terms in the first line of Eq. \eqref{eq:reducedlalternativeHamiltonian} would be $\tilde{Q}^2/2(C+ c\Delta x)$, as corresponds to the equivalent capacitor. In fact, the idea of equivalent capacitive (or inductive) elements arising from dynamical constraints $\boldsymbol{W}(H)=0$ is more general than this limited example. Thus, if there is no value of $Q_C$ for which $1+ c\Delta x h''_c(Q_C)$ vanishes then $Q_C$ can be solved from the equation $\tilde{Q}= Q_C+ c\Delta x h_c'(Q_C)$, with a corresponding effective capacitive energy $\tilde{h}_c(\tilde{Q})= h_c(Q_C)+c\Delta x\left[h'_c(Q_C)\right]^2/2$. As the discrete network under exam is different from the previous one, it should not be surprising that the dynamics, at this discretized level, are different.
	
	However, in the continuum limit $\Delta x\to0$, we recover precisely the Lagrangian presented above, Eq. \eqref{eq:NoGaugeFixingContinuum}, without gauge fixing for $\boldsymbol{W}_g$. We have thus proved that both discretizations are equivalent for the study of the coupling of the transmission line to a network.
	
	\subsection{Zero modes in the continuum limit}
	\label{sec:GaugeContinuum}
	There is a subtlety in the Lagrangian in the continuum limit, Eq.~\eqref{eq:NoGaugeFixingContinuum}. The gauge zero mode that we have found in all discretizations, and whose continuum limit has been presented in Eq.~\eqref{eq:functionalzeromode}, could be understood as saying that the physical field is actually $q(x)=Q'(x)$,  as only $Q'(x)$ appears in the Lagrangian. Alternatively, in the discretized version we could have defined $\dd q_{n+1}= \left(\dd Q_{n+1}-\dd Q_n\right)/\Delta x$, and in terms of that variable we would have no zero mode.  Furthermore, had we used $Q_C$ and $q_n$, with $n\geq1$, these would be  canonically conjugate (up to a constant) with $\Phi_0$ and $\Phi_n$, respectively. 
	
	Now, in taking the continuum limit we will break this canonical structure, as now $Q_C$ and $q(0)$ would be conjugate to the same variable. And indeed, in the continuum limit expression of the two-form,
	\begin{align}
		\omega = \dd Q_C\wedge\dd\Phi_0+\int_{\mathbbm{R}^+}\dd x\, \delta q(x)\wedge\delta\Phi (x)\,,
	\end{align}
	there is an additional zero mode, 
	\begin{align}
		\boldsymbol{W}= \frac{\partial}{\partial Q_C}-\frac{\delta}{\delta q(0)}\,.
	\end{align}
	On applying this zero mode to the continuum Hamiltonian we obtain the constraint
	\begin{align}\label{eq:whContinuum}
		\boldsymbol{W}(H) &= h'_c(Q_C)- \frac{1}{c}q(0)\nonumber\\
		&= h'_c(Q_C)-\frac{1}{c}Q'(0)=0\,.
	\end{align}
	This constraint encapsulates the Kirchhoff voltage law at the connection, clearly. We had accounted for the KVL in setting out the discretized versions, but the continuum limit requires its restatement, in the form of a dynamical constraint. Thus, if we desire a strictly reduced dynamical description we would need to impose it in the Lagrangian.
	
	Had we started from the gauge fixed continuum Lagrangian in Eq. \eqref{eq:GaugeFixedContinuum} we would also run into this constraint, that is seen as independent from the gauge mode $\boldsymbol{W}_g$. In fact, we would recover this constraint from  \eqref{eq:GaugeFixedContinuum} by computing the variational equations of motion. We remind the reader of the origin of the dynamical constraints $\boldsymbol{W}(H)=0$: they are consistency conditions on the set of equations, $\omega_{\alpha\beta}\dot{z}^\beta=\partial_\alpha H$, see Eq.\eqref{eq:newcosntraints} and its context. The geometrical perspective has been added to investigate their consistency and integrability, as well as having a systematic way of identifying them. Thus, an equation such as \eqref{eq:whContinuum} is also to be found as part of the variational Euler--Lagrange equations.

	We shall make use of this fact at various points in what follows, in particular in subsection \ref{sec:quadr-hamilt-diag}, Eq. \eqref{eq:qcslaving}. Instead of looking at the zero modes that arise from demanding that $Q_0$ be the limit $\lim_{x\to0} Q(x)$, we will not gauge fix the resulting Lagrangians, and we shall find the additional zero modes by computing the variational equations and checking consistency.
	
	There is yet another consequence to be drawn from this example. Namely, that the external (linear or nonlinear) oscillator  \emph{directly} coupled to the TL cannot be separated from it, and dresses the excitations of the fields. In general, this will introduce nonlinearities. If there are linear coupling elements, however, a separation between a linear sector and the external oscillator can be found, such that linear modes couple to the external degrees of freedom. The linear sector includes \emph{dressed} modes of the line. After a summary of the formalism for TLs, we take this issue up in greater generality in Sec.~\ref{sec:TLs-multiport-networks}.
	
	\subsection{Summary of transmission lines}
	\label{subsec:summary_TLs}In conclusion, we can treat TLs as an element in parity with lumped ones in the construction of first order Lagrangians susceptible of symplectic reduction to reach a Hamiltonian description. Equally to the r\^ole that lumped capacitors and inductors play in the formalism of~\cite{ParraRodriguez:2024a}, we assign to a TL the two form (\ref{eq:twoform_TL_cont}), equivalent (by integration by parts) to 
	\begin{align}
		\omega_{\TL}&=\int_{\mcl{I}}\dd x\,\left(\delta Q'(x)\wedge\delta\Phi(x)+\delta Q(x)\wedge\delta\Phi'(x)\right),
	\end{align}
	with contributions to the Lagrangian (\ref{eq:L_method})
	\begin{align}
		L_{\omega,\TL}=\frac{1}{2}\int_{\mcl{I}}\left(\dd x\,Q'(x)\dot{\Phi}(x)+\Phi'(x)\dot{Q}(x)\right)\label{eq:Lw_TL_Interval}
	\end{align} 
	and an energy functional, 
	\begin{align}
		h_{\TL}= \int_{\mcl{I}}\dd x\left(\frac{Q'(x)^2}{2c}+\frac{\Phi'(x)^2}{2l}\right),\label{eq:h_TL_Interval}
	\end{align}
	\begin{figure}[ht!]
		\centering \includegraphics[width=1\linewidth]{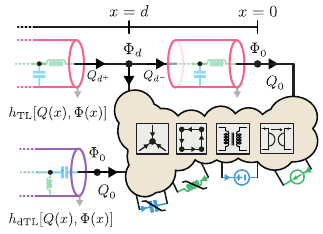}
		\caption{The quasi-lumped elements under consideration in this work extend naturally those in Ref.~\cite{ParraRodriguez:2024a}. In addition to the lumped elements in Fig.~\ref{fig:Elements_noR}, we have transmission lines ($\mcl{B_{\TL}}$) and dual-transmission lines (continuous limit of a left-handed metamaterial) ($\mcl{B_{\text{dTL}}}$). Continuous models of other 1D lumped-element lattices could be equally treated with the method.}
		\label{fig:Elements_TL_dTL_H_omega}
	\end{figure}where $\mcl{I}\subseteq \mathbbm{R}$. In this manuscript we mainly focus on scenarios where the TLs are connected to lumped elements at $x=0$. However, the two contributions to the Lagrangian (\ref{eq:Lw_TL_Interval}) and (\ref{eq:h_TL_Interval}) are written in a form useful for multiple connection points. In doing so, one needs to keep track of the charge field direction when the homogeneous TL is interrupted with a point connection, see Fig.~\ref{fig:Elements_TL_dTL_H_omega}. Furthermore, and following the same procedure explained in this section, it is straightforward to arrive at the analogous description for the dual transmission line (dTL), which is a continuous limit for the left-handed $LC$-staircase~\cite{Veselago:1968} (exchanging  capacitors for inductors, and vice-versa) with Lagrangian $L=L_{\omega,{\text{dTL}}}-h_{\text{dTL}}$, where
	\begin{align}
		L_{\omega,\text{d}\TL}=\frac{1}{2}\int_{\mcl{I}}\left(\dd x\,Q(x)\dot{\Phi}'(x)+\Phi(x)\dot{Q}'(x)\right),
	\end{align} 
	and the energy functional is 
	\begin{align}
		h_{\text{d}\TL}= \int_{\mcl{I}}\dd x\left(\frac{Q(x)^2}{2\bar{c}}+\frac{\Phi(x)^2}{2\bar{l}}\right).
	\end{align}
	
	Observe that $\bar{c}$ and $\bar{l}$ have different units from those of the standard TL, namely, [F m] and [H m], respectively. As is well known, such ideal continuous media do not exist in reality. However, they may be useful approximations for obtaining analytical results, for instance, when they are coupled to superconducting qubits~\cite{Egger:2013,Indrajeet:2020,Liberal:2021}. 
	
	\subsection{TL with topological ansatz}
	\label{subsec:TL-topological}
	The application of the topological ansatz of Ref.~\cite{ParraRodriguez:2024a} and Subsec.~\ref{subsec:micro-topological} has to be handled carefully in the continuous limit of infinite elements. In summary, the internal variables are found to be extended, and the character of the end-of-the-line variables is highly dependent on the topological character of the mode of termination.  
	
	To understand these statements, consider first a simple two-port that cannot be reduced to purely inductive or purely capacitive. If we construct a one port from closing one of the ports with a lumped element, the constraint equations can fully determine the topology assignment of the remaining port, and it will depend on that of the inserted lumped element (see App.~\ref{sec:pedagogical} for an explicit example). A finite length TL is a two port, and participates from this feature. Furthermore, in a discretized model of a TL we have KVL/KCL constraint equations that can be separated into bulk and boundary families. In our exposition above in this section we have solved the bulk constraints by using a flux node and loop charge solution, without explicit regard to their topological impact. The structure of the bulk KCL is 
	\begin{align}
		\mathrm{d}q_k^c+\mathrm{d}q_{k-1}^l=\mathrm{d}q_k^l\,,
	\end{align}
	using branch charges and an obvious notation. In this Pfaff equation, when considering the topological assignments, we are presented with two $S^1$ and one $\mathbb{R}$ charges. Were this equation isolated, the integral manifolds would have the topology $\mathbb{R}\times S^1$. As these are concatenated, we will have at most one $S^1$ charge, while for each node we introduce one extended charge to parameterize the manifold. Now, crucially, only if both boundary KCL conditions allow for the survival of an $S^1$ charge will there be one such in the final account. As stated, therefore, the  charge variables internal to the TL will be extended, and at most one $S^1$ charge might appear. As to the KVL, the bulk constraint is $\mathrm{d}\phi^c_{k-1} +\mathrm{d}\phi_{k-1}^l=\mathrm{d}\phi_k^c$, involving two compact and one extended flux variables, with exactly the same consequences as for charges. For a more explicit analysis, see App.~\ref{sec:pedagogical}.
	
	This is particularly important in the context of Josephson junctions connected to TLs: we cannot determine that there is a charge observable with discrete spectrum without analysis of the other termination of the TL. It is important to notice that this observation matches with the intuition that those discrete charge operators correspond to the existence of superconducting islands \cite{Devoret:2021}, the determination of which requires precisely a global study of the TL.
	
	\section{Transmission lines connected to multiport lumped-elements}
	\label{sec:TLs-multiport-networks}
	Quantizable hamiltonian descriptions of the circuit examined  in subsection \ref{sec:TLs-1port-networks}, if the external capacitor were linear,  could be achieved by  using just a second order Lagrangian written in terms of only node-flux degrees of freedom, while taking care of the possible divergent issues related to the necessity of having (ultra-violet) convergent energy properties in the lumped elements at the end~\cite{ParraRodriguezPhD:2021}. We will now turn to more complex circuits where the first-order presentation of the Lagrangian  introduced here is instrumental in obtaining exact Hamiltonian dynamics, more concretely circuits with TLs connected to (frequency-dependent) nonreciprocal elements, extending previous results in~\cite{ParraRodriguezPhD:2021,ParraRodriguez:2022}.
	
	\begin{figure}[h!]
		\centering	             \includegraphics[width=\columnwidth ]{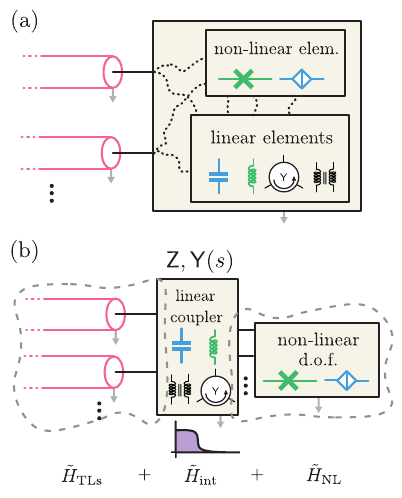}
		\caption{Two different paradigms of connectivity of lumped elements to transmission lines. (a) TLs arbitrarily connected to a general network of linear and nonlinear lumped elements. (b) TLs connected to nonlinear degrees of freedom through purely linear couplers, which allows for a blackbox characterization possibly followed by a subsequent adiabatic elimination in the spirit of Ref.~\cite{Solgun:2019,Labarca:2023}.  Because electrical coupling involves constraints, the linear coupler will generically behave as filter, as depicted symbolically, and the effective couplings after elimination will generically present regular behaviour for high energies. In this kind of configuration, the coupler itself may or may not have internal degrees of freedom, see examples in Sec.~\ref{sec:quasi_lumped_circuit_examples}. Observe that in either case, the total Hamiltonian description of the system is such that the transmission line modes are always dressed by the coupler structure, thus $\tilde{H}_{\text{TL}}$ will have a \emph{dressed}  mode eigenbasis involving the lumped elements they connect to.}
		\label{fig:TLs_to_LE_NR_networks}
	\end{figure}
	
	The core problem that this article   tackles is the determination of  Lagrangian and Hamiltonian descriptions of dynamics for circuits comprising multiple TLs interconnected by multiport lossless circuits, which can be created from lumped elements arranged in arbitrary configurations, i.e., keeping on a second plane the related issue of the level of phenomenology used to derive such a topological connection, see Fig.~\ref{fig:TLs_to_LE_NR_networks}(a).
	
	Within this general context, the construction of the Lagrangian, involving the elimination of redundancies, must adhere to a procedure similar to the one put forward in the previous section. Naturally, analogous singularities can potentially arise, whether they are linear and can be systematically addressed using the method outlined in Ref.~\cite{ParraRodriguez:2024a}, or nonlinear in nature, stemming from approximate circuit representations. An example of the latter occurs in small Josephson junctions with negligible parallel capacitors connected in series to a linear inductor. It is worth noting that connecting transmission lines (TLs) to such circuits will not resolve the latter singularity issue. This conclusion is not exclusive to the procedure we put forward. See for instance the  requirement of well-posedness \cite{Chua:1980} in modeling nonlinear devices, with explicit reference to the possibility of nonunique solutions, as the issue was well known in classical circuit theory \cite{Chua:1980a,Roska:1981}.

	Various solutions are currently being explored~\cite{Rymarz:2023,Miano:2023,ParraRodriguez:2024a}, involving controlled perturbation and adiabatic theory to eliminate higher harmonics that play a negligible role. These procedures, which solely involve the local degrees of freedom of the nonlinear network, can naturally be applied in a preliminary step preceding the work described here. Additional exploration in this direction will be deferred for future investigations. Within the present work, we will assume that the standard nonlinear elements in cQED, the Josephson (phase-slip) junctions do always come with  associated parallel capacitances (series inductances), or alternatively that  certain adequate conditions for invertibility in the nonlinear context are met.
	
	In this section, our primary focus centers on scenarios involving the connection of TLs to NL networks through passive, lossless linear systems (the blackbox) characterized by immittance matrices ($\msY(s)$ or $\msZ(s)$)~\cite{Newcomb:1966}, as illustrated in Fig.~\ref{fig:TLs_to_LE_NR_networks}(b). By employing the canonical fraction decompositions for the linear blackbox, one shows that the crux in deriving a canonical quantum Hamiltonian for this system boils down to analysing a circuit that comprises multiple interconnected lines via parallel or series capacitors, inductors, and ideal nonreciprocal elements \cite{ParraRodriguezPhD:2021,ParraRodriguez:2022}. Given that these represent well-known dual cases, we will refer the curious reader to Sec. 5 in~\cite{ParraRodriguezPhD:2021} for further details on the series configuration and concentrate solely on the parallel arrangement in this manuscript.
	
	\subsection{TLs connected through a frequency-dependent NR blackbox}
	\label{sec:quadr-hamilt-diag}
\begin{figure}[h]
		\centering    \includegraphics[width=\linewidth]{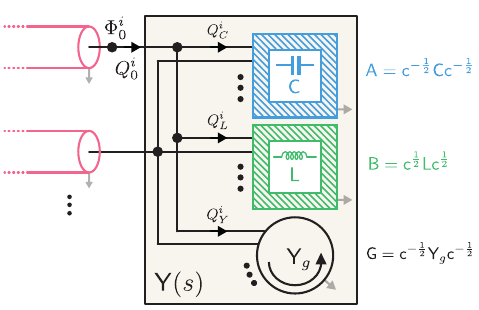}
		\caption{ Transmission lines connected through generic capacitors ($\msC$), inductors ($\msL$) and ideal nonreciprocal elements ($\msY_g$) in a parallel configuration, and their rescaled matrices $\msA$, $\msB$ and $\msG$ , respectively.}
		\label{fig:TLs_ABYp}
	\end{figure}

	As depicted in Fig.~\ref{fig:TLs_ABYp}, we consider a configuration of $N$ transmission lines interconnected through a passive lossless linear system characterized by a canonical admittance matrix (in Laplace space) of the form  $\msY(s)= \msL^{-1}/s+\msC s +\msY_g$, where $\msC$ and $\msL$ ($\msY_g$) are symmetric (anti-symmetric) $s$-independent matrices. This matrix presents just one zero and  and one infinity poles, yet  we can naturally introduce additional finite-frequency poles represented as $\sum_k\msY_k(s)=\sum_k\frac{\msD_k s+ \msE_k}{s^2+\Omega_k^2}$ without increasing the complexity of this particular system. See circuit examples below, for linear systems with such inner NR poles. 
	
	We shall now use the formalism introduced in~\cite{ParraRodriguez:2024a} and adapted here to TLs to arrive at a first order Lagrangian in terms purely of the charge and flux fields, including their boundary values. This Lagrangian will be quadratic in the fields, and by the application of our formalism will describe consistent reduced dynamics for the whole system. Furthermore, because of the positivity of the capacitance and inductance densities, together with the passivity of the linear and time independent blackbox, the dynamics it describes will be energetically stable. Taking these three properties together, it will be equivalent to a system of uncoupled harmonic oscillators, the normal modes of the full system. We extend the formalism presented in~\cite{ParraRodriguez:2018,ParraRodriguezPhD:2021} to the present case. A crucial element in the extension is that we look for diagonalization of the corresponding Hamiltonian in such a way that the modes are not coupled through the time derivative terms. In other words, that we have canonically conjugate variables in the diagonalization. 
	
	The first step is to construct the appropriate Lagrangian that describes the system. We work directly with the continuous description of the TLs and use differential form notation as above~\cite{ParraRodriguez:2024a}. As we have pointed out, it is frequently convenient to carry out the Kirchhoff law constraints in a stepwise fashion. In this case, by using branch current variables $\bQ_C$, $\bQ_L$ and $\bQ_Y$ as depicted in Fig.~\ref{fig:TLs_ABYp}, where $\bQ_C^T=\left(Q_{{C_1}},Q_{{C_2}},\cdots,Q_{{C_M}}
	\right)$, and similarly for the other two, we can express   the current and voltage equations at the boundary between the TLs and the discrete network as
	\begin{subequations}
		\label{eq:eq_bc_ABYp_T}
		\begin{align}
			\dQ_0&=\dQ_L+\dQ_C+\dQ_Y,\\
			\dPhi_C&=\dPhi_L=\dPhi_Y=\dPhi_0.
		\end{align}
	\end{subequations}
	Here, $\bPhi_0^T\equiv\bPhi^T(0)=\left(\Phi^1(0),\Phi^2(0),\cdots, \Phi^M(0))\right)$ and $\bQ_0^T\equiv\bQ^T(0)=\left(Q^1(0),Q^2(0),\cdots ,Q^M(0)\right)$ for clarity, and the flux vectors for the network elements are also introduced for convenience. The two-form (\ref{eq:twoform_restricted}) is now expressed as
	\begin{align}
		\omega=\omega_{\TL}+\omega_{L}+\omega_{C},\nonumber
	\end{align}
	where 
	\begin{align}
		\omega_{\TL}=\int_{\mathbbm{R}^+}(\delta{\bQ}')^T(x)\wedge\delta{\bPhi}(x)+\frac{1}{2}\dQ_0^T\wedge\dPhi_0,
	\end{align}
	the multi-line two-form, is the generalization of the  single-line two-form in Eq.~(\ref{eq:twoform_TL_cont}). The other two terms, $\omega_L$ and $\omega_C$, are the contribution from the capacitive and inductive boxes after the implementation of both the Kirchhoff's constraints at the boundary (\ref{eq:eq_bc_ABYp_T}), and the current-voltage relation of the ideal NR element, i.e.,  $\dQ_Y=\msY_g \dPhi_0$. Thus, they simplify to
	\begin{align}
		\omega_L&=\frac{1}{2}\dPhi_L^T\wedge\dQ_L\nonumber\\
		&=\frac{1}{2}\dPhi_0^T\wedge(\dQ_0-\dQ_C-\msY_g\dPhi_0),\nonumber\\
		\omega_C&=\frac{1}{2}\dQ_C^T\wedge\dPhi_C=\frac{1}{2}\dQ_C^T\wedge\dPhi_0,\nonumber
	\end{align}
	such that the overall we have
	\begin{align}
		\omega=&\,\omega_{\TL}+ \frac{1}{2}\dPhi_0^T\wedge\left(\dQ_0-\msY_g\dPhi_0-2\dQ_C\right).\label{eq:precan_two-form_CLY}
	\end{align}
	In other words, we are solving $\bQ_L$ and $\bQ_Y$ from the constraint relations, as they are in fact integrable. Substituting the expression for  the charges $\bQ_L$ and $\bQ_Y$  as well as the fluxes $\bPhi_L$ in the energy function, we obtain 
	\begin{align}    \label{eq:BoxesLagrangian}
		L=&\,\int_{\mathbbm{R}^+}\dd x\,(\bQ')^T(x)\dot{\bPhi}(x)-\bPhi^T_0\left( \dot{\bQ}_C+\frac{1}{2}\msY_g\dot{\bPhi}_0\right)\nonumber\\
		&-\left(h_{\TL}+\frac{1}{2}\bQ_C^T\msC^{-1}\bQ_C+\frac{1}{2}\bPhi_0^T\msL^{-1}\bPhi_0\right),
	\end{align}
	where  
	\begin{align}
		\begin{aligned}
			h_{\TL}=&\frac{1}{2}\int_{\mathbbm{R}^+}\dd x\left[(\bQ'(x))^T\msc^{-1}\bQ'(x)\right.\\
			&\qquad\qquad\left.+(\bPhi'(x))^T\msl^{-1}\bPhi'(x)\right].   
		\end{aligned}
	\end{align}
	On computing the equations of motion involving terms $\dot{\bPhi}_0$ (from variations of $\delta {\bQ_C}$ and $\delta {\bQ_0}$) that flow from this Lagrangian, we obtain for the boundary the voltage continuity equation in the TL equivalent to the voltages in the capacitance network (and by construction to the other two elements)
	\begin{align}\label{eq:Voltage_BC_L_ABG}
		\begin{aligned}
			\dot{\bPhi}_0&=\msC^{-1}\bQ_C,\\
			\dot{\bPhi}_0&=\msc^{-1}\bQ'_0.
		\end{aligned}
	\end{align}
	We observe that the combination of these two equations shows us that $\bQ_C$ is slaved to the derivative of the charge field at the edge of the TL, 
	\begin{align}
		\label{eq:qcslaving}
		\bQ_C=\msC\msc^{-1}\bQ'_0.
	\end{align}
	To be clear, our current approach involves the elimination of variables by a direct examination of the equations of motion, rather than relying on the zero vector method, as mentioned in Subsection \ref{sec:GaugeContinuum}. In order to ensure that substituting Eq. \eqref{eq:qcslaving} into the Lagrangian results in an equivalent system, we must confirm that the other equation at the boundary, which incorporates $\dot{\bQ}_C$, can be derived from the Lagrangian after the elimination process. In other words, that this is a consistent reduction.  This confirmation indeed holds, enabling us to proceed with the incorporation of the slaving condition Eq. \eqref{eq:qcslaving} into the Lagrangian.  We do not reduce further by gauge fixing as the duality symmetry between charge and flux will be helpful in the next part of the analysis.
	
	At this point we have achieved our first goal, the construction of a reduced first order Lagrangian involving only the fields. In other words, we observe that we can understand the system as incorporating the parallel connectors into boundary conditions, or alternatively that the degrees of freedom are those of the line, albeit dressed by the coupling elements. To make it easier to reach our second goal, a diagonalization with canonically conjugate variables, we will rewrite this Lagrangian in an equivalent way. First, we rescale both the flux and charge fields. To avoid introducing any additional symbols, we directly implement this rescaling through the following substitutions: $\msc^{-1/2}\bQ\rightarrow\bQ$ and $\bPhi\rightarrow\msc^{-1/2}\bPhi$. 
	The  Lagrangian reads then
	\begin{align}
		&L=\,\int_{\mathbbm{R}^+}\dd x\,(\bQ')^T(x)\dot{\bPhi}(x)+ \left(\msA\bQ'_0 +\frac{1}{2}\msG\bPhi_0\right)^T\dot{\bPhi}_0\nonumber\\
		&\qquad-\left(\tilde{h}_{\TL}+\frac{1}{2}(\bQ'_0)^T\msA\bQ'_0+\frac{1}{2}\bPhi_0^T\msB^{-1}\bPhi_0\right),\label{eq:L_ABG_integrated}
	\end{align}
	where 
	\begin{align}
		\tilde{h}_{\TL}=\int_{\mathbbm{R}^+}\dd x \left(\frac{(\bQ'(x))^2}{2}+\frac{(\bPhi'(x))^T\mDelta\bPhi'(x)}{2}\right).\nonumber
	\end{align}
	
	Here we have introduced the rescaled matrices $\msA=\msc^{-1/2}\msC \msc^{-1/2}$, $\msB=\msc^{1/2}\msL \msc^{1/2}$ and  $\msG=\msc^{-1/2}\msY_g \msc^{-1/2}$ as well as the squared-velocity matrix $\Delta= \msc^{-1/2}\msl^{-1}\msc^{-1/2}$. Integrating by parts, and up to a total time-derivative term, the Lagrangian can be written in the alternative equivalent form 
	\begin{align}
		L=&\,L_{\omega,{\msA}{\msB}{\msG}}-H_{{\msA}{\msB}{\msG}}\label{eq:Lfinal_ABG}\\
		=&\,\frac{1}{2}\int_{\mathbbm{R}^+}\dd x\,\left[(\bQ')^T(x)\dot{\bPhi}(x)+ (\bPhi')^T(x)\dot{\bQ}(x)\right]\nonumber\\
		&+\frac{1}{2}(\msA\bQ_0')^T\dot{\bPhi}_0+\frac{1}{2}\bPhi_0^T\left(\dot{\bQ}_0-\msA\dot{\bQ}_0'-\msG\dot{\bPhi}_0\right)\nonumber\\
		&+\int_{\mathbbm{R}^+}\dd x\,\left[\frac{\bQ^T(x)\bQ''(x)}{2}+\frac{\bPhi^T(x)\mDelta\bPhi''(x)}{2}\right]\nonumber\\
		&+\frac{1}{2}\bPhi_0^T\left(\mDelta\bPhi'_0-\msB^{-1}\bPhi_0-\msG\bQ'_0\right)\nonumber\\
		&+\frac{1}{2}\left(\bQ_0-\msA\bQ'_0-\msG\bPhi_0\right)^T\bQ'_0,\nonumber
	\end{align}
	that is more adequate for our purposes. To complete the last two lines we have also added and substracted 
	$\bPhi_0^T\msG\bQ'_0/2$, and made use of the anti-symmetry of matrix  $\msG$. The first two lines, adding up to $L_{\omega,{\msA}{\msB}{\msG}}$, are the $\omega_{\alpha\beta}z^\alpha\dot{z}^\beta/2$ part of a standard first-order FJ Lagrangian~(\ref{eq:L_method}), while the other three lines, $-H_{{\msA}{\msB}{\msG}}$, are the energy terms, $-H(z)$.  
	
	Our objective is the diagonalization of this structure. That is fundamentally the identification of the normal modes for the linear equations of motion, i.e. separation of variables by demanding $\exp(i\Omega t)$ behavior for the fields. After some manipulations, we see that the normal forms for the flux fields must be nontrivial solutions of  the Sturm--Liouville-like system
	\begin{align}\label{eq:SLlike}
		-\mDelta \mathbf{f}''&=\Omega^2\mathbf{f}\,,\\
		-\mDelta \mathbf{f}_0'&= \Omega^2 \mathbf{f}_0 -i \Omega \msG \mathbf{f}_0-\msB^{-1} \mathbf{f}_0\,.\nonumber
	\end{align}
	Crucially the would-be eigenvalue appears also in the boundary condition. No complete general theory, in parallel with standard Sturm--Liouville analysis, exists for this case. The important case of linear dependence on the eigenvalue ($\Omega^2$) in the boundary condition is amenable to a general description, as first expounded by Walter~\cite{Walter:1973} and Fulton~\cite{Fulton:1977}. This was used in \cite{ParraRodriguez:2018,ParraRodriguezPhD:2021} in the context of transmission lines. Notice however that \eqref{eq:SLlike} does not fall in that category, as $\Omega$ appears linearly, and not only  in the form of $\Omega^2$, in the boundary condition, so new tools are required. For our case we will profit from the fact that we are starting from a variational principle. Thus we will seek for a presentation of the energy functional $H_{{\msA}{\msB}{\msG}}$ as a quadratic form, building on ~\cite{ParraRodriguez:2018,ParraRodriguezPhD:2021}, and simultaneously we will look for an operator presentation of $L_{\omega,{{\msA}{\msB}{\msG}}}$.

	To that purpose, we consider the Hilbert space $\mcl{H}=\left[L^2(\mathbbm{R}^+)\otimes\mathbb{C}_{\mSigma}^{2N}\right]\oplus\mathbb{C}_{\msA^{-1}}^{N}\oplus\mathbb{C}_{\msB}^{N}$, with elements
	$\bmW=(\bW,\bw)\in\mcl{H}$ and inner product 
	\begin{align}\label{eq:inner_product_AB}
		\begin{aligned}
			\langle\bmW_{1},\bmW_2\rangle=&\int_{\mathbbm{R}^+}\dd x \bW_1^\dag (x)\mSigma\bW_2(x)\\
			&+\bw_{11}^\dag\msA^{-1}\bw_{21}+\bw_{12}^\dag\msB\bw_{22},    
		\end{aligned}
	\end{align}
	where $\mSigma=\mathrm{diag}(\mone,\mDelta^{-1})$ and $\bw_i^T=(\bw_{i1}^T,\bw_{i2}^T)$. We now observe that the operator $\mcL$ defined by its action and domain,
	\begin{align}
		\label{eq:DiffOp_ABG}
		\begin{aligned}
			\mcL\bmW&=\left(-\mDelta\bW'',\tilde{\bw}=\begin{pmatrix}
				-\mDelta\bU'+\msB^{-1}\bU+\msG\bV'\\
				-\msB^{-1}\bV'
			\end{pmatrix}_0\right),\\
			\mcl{D}(\mcL)&=\{\left(\bW\equiv\begin{pmatrix}
				\bU\\ \bV
			\end{pmatrix}(x),\bw\right), \bmW\in\mathcal{H}, \mathcal{L}\bmW\in\mathcal{H},\\
			&\qquad\bW,\bW'\in \text{AC}^1(\mathbbm{R}^+)\otimes\mathbb{C}_{\mSigma}^{2N},\\
			&\quad\bw=\begin{pmatrix}
				\msA \bU\\
				\bV-\msA\bV'-\msG\bU
			\end{pmatrix}_0
			\}\,
		\end{aligned}
	\end{align}
	respectively, where $\bU_0\equiv\bU(0)$ and $\bV_0\equiv\bV(0)$ and analogously for the column vectors, is a positive self-adjoint operator; refer to App.~\ref{sec:L_op_app} for the proof and additional details. Furthermore, its (generalized) eigenvectors form a (generalized) basis in $\mcl{H}$. From previous work~\cite{ParraRodriguezPhD:2021} we know that the eigenvalues are degenerate, with homogeneous degeneracy $2N$. Denote the eigenvalues as $\Omega^2$, the degeneracy index for an eigenspace as $\epsilon\in\left\{1,\ldots,2N\right\}$, and the eigenvectors as $\bmW_{\Omeps}=\left(\bW_{\Omeps},\bw_{\Omeps}\right)$. Thus, the wavefunction of  $\bmW(t)\in\mcl{H}$ is written in this basis as
	\begin{align}
		\label{eq:wexpansion}
		\bmW(x,t)=\sum_\epsilon\int_{\mathbbm{R}^+}\dd \Omega\,X^{\Omeps}(t)\bmW_{\Omeps}(x),
	\end{align}
	where, implicitly, we have collected the fields $\bW^T=(\bPhi^T,\bQ^T)(x,t)$. 
	We will occasionally use collective indices $\alpha$ or $\beta$, with  Einstein's convention, to denote $\Omega\epsilon$ double indices and the corresponding sum and integral.
	
	We see now that the energy functional in the  Lagrangian \eqref{eq:Lfinal_ABG}, $H\left[\bQ,\bPhi\right]$, can be written in the form
	\begin{align}
		\label{eq:enefuncasinner}
		H\left[\bQ,\bPhi\right]=\frac{1}{2}\langle\bmW,\mcL\bmW\rangle.
	\end{align}
	We expand the fields in the Lagrangian in terms of  an eigenbasis of $\mcL$, and we find
	\begin{align}
		L = \frac{1}{2}\omega_{\alpha\beta}X^\alpha\dot{X}^\beta-\frac{1}{2}\Omega^2 \left(X^{\Omeps}\right)^2.\label{eq:finallag}
	\end{align}
	
	Crucially, the matrix element
	\begin{align}
		\label{eq:omab}
		\begin{aligned}
			\omega_{\alpha\beta}=& \int_{\mathbbm{R}^+}\dd x\,\left[\left(\bV'_\alpha\right)^T\bU_\beta+\left(\bU'_\alpha\right)^T\bV_\beta\right]\\
			& +\left(\msA\bV'_\alpha(0)\right)^T \bU_\beta(0)\\
			&+\bU_\alpha^T(0)\left(\bV-\msA\bV'-\msG\bU\right)_\beta(0)\\
			\equiv&\,\,\langle i\mcT\bmW_\alpha,\bmW_\beta\rangle
		\end{aligned}
	\end{align}
	does define a symmetric operator $\mcT$. 
	The antisymmetry $\omega_{\beta\alpha}=-\omega_{\alpha\beta}$ is shown by integration by parts and making use of the definition of the domain (cf. Eq.~(\ref{eq:DiffOp_ABG}))
	together with the antisymmetry of $\msG$. By inspection, we see that the symmetric operator $\mcT$ in question  is defined by its action on  its domain $\mcl{D}(\mcT)$ (effectively including$\mcl{D}(\mcL)$) as
	\begin{align}
		\mcT \bmW= \left(\begin{pmatrix}
			-i\bV'\\-i\mDelta\bU'
		\end{pmatrix},\check{\bw}=\begin{pmatrix}
			-i\msA \bV'_0\\
			-i\msB^{-1}\bU_0
		\end{pmatrix}\right).
	\end{align}
	One can readily check that this operator acts on elements of $\mcl{D}(\mcL)$ as a square root $\mcT\sim\sqrt{\mcL}$, i.e., $\mcL \bmW_{\Omega \epsilon}=\mcT^2 \bmW_{\Omega \epsilon}=\Omega^2 \bmW_{\Omega \epsilon}$. The operator defined above is not self-adjoint as such, but is essentially self-adjoint, see proof in  App.~\ref{sec:T_op_app}.
	
	Borrowing nomenclature from other areas, we term $\mcL$ the single-excitation hamiltonian operator. As to $\mcT$, we proposed that it be called the duality operator~\cite{ParraRodriguezPhD:2021,ParraRodriguez:2022}. Observe that it maps $(\bPhi,\bQ)$ to, essentially, $(\bQ', \mDelta\bPhi')$. This is in fact the nonlocal realization of electromagnetic duality for TEM waves.
	
	As $\mcL$ and $\mcT$ effectively commute, we can study them blockwise, in each eigenspace of $\mcL$. In one such eigenspace $\omega_{\alpha\beta}= \omega_{(\Omega\epsilon_1)(\Omega\epsilon_2)}=:\omega^\Omega_{\epsilon_1\epsilon_2}$ is an antisymmetric $2N\times 2N$ matrix. We desire to find a basis in which this matrix is $\Omega \msJ$, with $\msJ$ the canonical symplectic matrix. As $\mcT$ is essentially selfadjoint, with real eigenvalues, and the square of those eigenvalues in the eigenspace is $\Omega^2$, it follows that the eigenvalues of $\mcT$ are $\pm \Omega$. They appear with equal degeneracy $N$. We use latin indices $\lambda$ to denote orthogonal eigenvectors in each eigenspace, ranging from 1 to $N$. Thus, we are considering the eigenbasis of the $\Omega^2$ eigenspace of $\mcL$ determined by the (orthonormal) vectors that satisfy
	\begin{align}
		\mcT\bmW^\pm_{\Omega\lambda}= \pm\Omega \bmW^\pm_{\Omega\lambda}\,.
	\end{align}
	Let us now introduce another orthonormal basis, with $N$ vectors of type $F$ and $N$ vectors of type $G$ in the  $\Omega^2$ eigenspace  of $\mcL$,
	\begin{align}
		\bmW^F_{\Omega\lambda}&= \frac{1}{\sqrt{2}}\left(\bmW^+_{\Omega\lambda}+ i \bmW^-_{\Omega\lambda}\right)\,,\\
		\bmW^G_{\Omega\lambda}&= \frac{-i}{\sqrt{2}}\left(\bmW^+_{\Omega\lambda}- i \bmW^-_{\Omega\lambda}\right),
	\end{align}
	such that $\mcT \bmW^{F,G}_{\Omega\lambda}=\pm i \Omega \bmW^{G,F}$. That is, we find the basis for which $\mathcal{T}$ is expressed as $\Omega \sigma^y$. We denote the expansion of a generic element of the Hilbert space in this basis as 
	\begin{align}
		\bmW = \sum_{\lambda=1}^N\int_{\mathbbm{R}^+}\dd\Omega\,\left(F^{\Omega\lambda}\bmW^F_{\Omega\lambda}+ G^{\Omega\lambda}\bmW^G_{\Omega\lambda}\right)\,.
	\end{align}
	Expanding the fields in the  Lagrangian in this form we obtain (by setting aside a total time derivative)
	\begin{align}\begin{aligned}
			L=&\sum_{\lambda=1}^{N}\int_{\mathbbm{R}^+}\dd \Omega\left(\Omega G^{\Omega \lambda}\dot{F}^{\Omega\lambda}\right.\\
			&\qquad\qquad\left.-\frac{\Omega^2 ((F^{\Omega\lambda})^2+(G^{\Omega\lambda})^2)}{2}\right). 
		\end{aligned}
	\end{align}
	Just to be explicit, we now carry out the nonsymplectic rescaling  $\tilde{F}^{\Omega\lambda}= F^{\Omega\lambda}$, and $\tilde{G}^{\Omega\lambda}=\Omega G^{\Omega\lambda}$, to obtain  canonical pairs of conjugated variables, with Poisson bracket  $\{\tilde{F}^{\Omega\lambda},\tilde{G}^{\Omega'\lambda'}\}=\delta(\Omega-\Omega')\delta^{\lambda\lambda'}$, and the Lagrangian is in its final form of a sum of canonical harmonic oscillator Lagrangians,
	\begin{align}
		\begin{aligned}
			L=&\sum_{\lambda=1}^{N}\int_{\mathbbm{R}^+}\dd\Omega \left( \tilde{G}^{\Omega \lambda}\dot{\tilde{F}}^{\Omega\lambda}-\frac{(\tilde{G}^{\Omega\lambda})^2+(\Omega\tilde{F}^{\Omega\lambda})^2}{2}\right).\label{eq:L_TLs_ABG_canonical} 
		\end{aligned}
	\end{align}
	Canonical quantization now follows, promoting the canonical pairs now to quantum operators in the standard way ($\{\cdot,\cdot\}\rightarrow \frac{-i}{\hbar}[\cdot,\cdot]$) and rewriting them in terms of annihilation and creation operators, i.e., $\tilde{F}^{\Omega\lambda}=i\sqrt{\frac{\hbar}{2\Omega}}(a^\dag_{\Omega\lambda}-a_{\Omega\lambda})$ and $\tilde{G}^{\Omega\lambda}=\sqrt{\frac{\hbar \Omega}{2}}(a_{\Omega\lambda}+a^\dag_{\Omega\lambda})$, such that the final normal ordered  Hamiltonian becomes 
	\begin{align}\label{eq:hamquantized}
		H&=\colon\sum_{\lambda=1}^{N}\int_{\mathbbm{R}^+}\dd \Omega\,\frac{(\tilde{G}^{\Omega\lambda})^2+(\Omega\tilde{F}^{\Omega\lambda})^2}{2}\colon\\
		&\stackrel{q.}{=}\sum_{\lambda=1}^{N}\int_{\mathbbm{R}^+}\dd \Omega\,\hbar \Omega\, a^\dag_{\Omega\lambda}a_{\Omega\lambda},
	\end{align}
	i.e., discarding the (infinite) zero-point fluctuations, and $\left[a_{\Omega\lambda},a^\dag_{\Omega'\lambda'}\right]=\delta(\Omega-\Omega')\delta_{\lambda \lambda'}$.
	
	These results extend the findings presented in Refs.~\cite{ParraRodriguezPhD:2021, ParraRodriguez:2022}, encompassing the broadest category of linear connections.  It is crucial to emphasise that our novel approach of expressing the circuit Lagrangian in first-order form played a pivotal role in effectively eliminating the non-dynamical component entirely, without necessitating any additional symplectic transformations. For a comprehensive discussion, please refer to Chap. 5 of Ref.~\cite{ParraRodriguezPhD:2021}. In that context, the operators $\mcL$ and $\mcT$ were indeed correctly recognized; however, they were employed in conjunction with a redundant formulation of Lagrangian (\ref{eq:Lfinal_ABG}) in second order. Extracting the exclusively dynamical sector from this version posed a formidable challenge, and in practice  remained unaccomplished. The complexity of the task stemmed from the necessity of harmonising finite and infinite-dimensional symplectic subspaces within a single symplectic transformation, which proved to be a daunting endeavour. We must also remark that to find the canonical eigenbasis of this problem, one must diagonalise the degenerate subspace of dimension $2N$, a task which could potentially require numerics, even when considering transmission lines of infinite length.
	
	Finally, it is worth saying that the analysis just presented contains the fundamental issues towards the most general linear connection. However, typically,  a flat nonreciprocal response in either $\msY$ or $\msZ$ matrices is nonphysical. Nonetheless, our results remain non-singular even within this limit. In later examples, we will discuss simpler and more physically-motivated scenarios.
	
	\subsection{TLs connected to nonlinear elements through a NR blackbox}\label{subsec:TLs_NR_BB_NL}
	
	We are now poised to delve into the central objective outlined at the outset of this section, which involves the formulation of Hamiltonian models for circuits that incorporate transmission lines coupled to external nonlinear degrees of freedom using a universal nonreciprocal coupler. Drawing on the previous subsection, obtaining such Hamiltonian dynamics should become straightforward because the connection is through  lossless linear systems, represented by their canonical expansions. Nevertheless, even in this simple scenario, linear-nonlinear constraints can potentially appear in the process~\cite{Miano:2023,Rymarz:2023}. 
	
	\begin{figure}[h]
		\centering    \includegraphics[width=
		\linewidth]{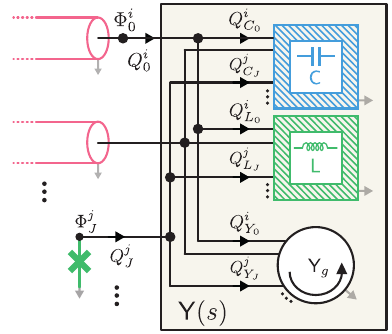}
		\caption{TLs directly connected through a frequency-dependent canonical admittance matrix to Josephson junctions. The standard parallel capacitor of the JJ ($C_J$) is embedded in the associated diagonal element of matrix $\msC$.}\label{fig:TLs_ABYp_JJs}
	\end{figure}
	For the sake of concreteness, and without loss of generality, let us consider the circuit of Fig.~\ref{fig:TLs_ABYp_JJs} where $N$ transmission lines are coupled through a nonreciprocal blackbox to a Josephson junction (JJ), while  the parallel capacitance $C_J$ to the JJ is considered as part of the box, i.e., it is part of the associated diagonal entry of the capacitance matrix $\msC$. Recall that here we are not considering internal resonances in the box, as their inclusion would not fundamentally complicate the task of deriving a quantum Hamiltonian description. 
	
	Following the recipe of the previous subsection, we will construct an enlarged Lagrangian with extra degrees of freedom which will couple through linear terms to one of the type of Eq. (\ref{eq:Lfinal_ABG}). 
	
	This parallel coupling means that the capacitive, inductive and circulator (NR element) boxes have the same number of ports. Denote the respective port charge differentials with $\dQ_C$, $\dQ_L$ and $\dQ_Y$ respectively. Similarly for the port flux differentials, $\dPhi_C$, $\dPhi_L$ and $\dPhi_Y$ respectively. The ports are connected either to the TLs or to the JJs, in two sets of nodes with node flux differentials $\dPhi_0$ at the line nodes and $\dPhi_J$ at the JJ nodes. The currents at those sets of nodes are given by the inflowing charge differentials $\dQ_0$ and $\dQ_J$ respectively. We organize the differentials for the connecting nodes in  $\dQ_P$ and $\dPhi_P$  sets, 
	\begin{align}
		\dQ_P=\begin{pmatrix}
			\dQ_0\\\dQ_J
		\end{pmatrix},\quad \dPhi_P=\begin{pmatrix}
			\dPhi_0\\\dPhi_J
		\end{pmatrix}\,.
	\end{align}
	Then the KCLs and KVLs are expressed in the relations
	\begin{align}
		\begin{aligned}
			\dQ_P&=\dQ_{L}+\dQ_{C}+\dQ_{Y},\\
			\dPhi_P&=\dPhi_C=\dPhi_L=\dPhi_Y.
		\end{aligned}
	\end{align} 
	Before enforcing these Kirchhoff constraints we shall construct a pre-canonical two-form with these variables, namely 
	\begin{align}\label{eq:two-form_TLs_ABG_JJs}
		\omega&=\omega_{\TL}+\omega_{C}+\omega_{L}+\omega_{J}
	\end{align}
	where $\omega_C=\frac{1}{2}\dQ_{C}^T\wedge\dPhi_P$, $\omega_L=\frac{1}{2}\dPhi_P^T\wedge\dQ_L$ and $\omega_J=\frac{1}{2}\dQ_J^T\wedge\dPhi_J$. With $M\in\{C,L,Y\}$ we organize the vectors of charge differentials as 
	\begin{align}
		\dQ_M=\begin{pmatrix}
			\dQ_{M_0}\\\dQ_{M_J}
		\end{pmatrix},
	\end{align}
	and making use of the Kirchhoff constraints we arrive at
	\begin{align}
		\omega_C&=
		\frac{1}{2}\left(\dQ_{C_0}^T\wedge\dPhi_0+\dQ_{C_J}^T\wedge\dPhi_J\right),\nonumber\\
		\omega_L&=
		\frac{1}{2}\dPhi_0^T\wedge(\dQ_0-\dQ_{C_0}-\msY_0 \dPhi_0+\msY_{0J}\dPhi_J)\nonumber\\
		&+\frac{1}{2}\dPhi_J^T\wedge(\dQ_J-\dQ_{C_J}-(\msY_J\dPhi_J+\msY_{0J}^T\dPhi_0)),\nonumber\\
		\omega_J&=\frac{1}{2}\dQ_J^T\wedge\dPhi_J.\nonumber
	\end{align}
	We have   also used the decomposition of the direct nonreciprocal matrix 
	\begin{align}
		\msY_g=\begin{pmatrix}
			\msY_0&-\msY_{0J}\\(\msY_{0J})^T&\msY_J
		\end{pmatrix},
	\end{align}
	connecting $\dQ_Y=\msY_g\dPhi_Y$. Regrouping all the elements, the total two-form can be written as $\omega=\omega_{{\msA_{0}}{\msB_{0}}{\msG}_{0}}+\omega_{J\text{int}}$, where $\omega_{{\msA_{0}}{\msB_{0}}{\msG}_{0}}$ is the pre-canonical two-form of previous section, Eq.~(\ref{eq:precan_two-form_CLY}) except for the substitution of $\dQ_{C_0}$ for $\dQ_C$, which arises from the coupling of the linear box and the TLs, while 
	\begin{align}
		\omega_{J\text{int}}= \left(\dQ_{C_J}+\frac{1}{2}\msY_J\dPhi_0+\msY_{0J}^T\dPhi_0\right)\wedge\dPhi_J
	\end{align}
	includes couplings to the Josephson fluxes as well as local terms of the latter.
	
	In other words, we compute the total Lagrangian $L=L_\omega-H$, where the term associated with the two-form is
	\begin{align}
		\begin{aligned}
			L_\omega=&  L_{\omega,\TL}+\frac{1}{2}\bPhi_0^T(\dot{\bQ}_0-\msY_0\dot{\bPhi}_0-2\dot{\bQ}_{C_0})\\
			&\,+\bQ_{C_J}^T\dot{\bPhi}_J-\frac{1}{2}\bPhi_J^T\msY_J\dot{\bPhi}_J+\bPhi_0^T\msY_{0J}\dot{\bPhi}_J
		\end{aligned}
	\end{align}
	and the energy term is
	\begin{align}
		H&=h_{\TL}+\frac{1}{2}\bQ_C^T\msC^{-1}\bQ_C+\frac{1}{2}\bPhi_P^T\msL^{-1}\bPhi_P+h_J(\bPhi_J).
	\end{align}
	Here $h_J(\bPhi_J)=-\sum_i E_{Ji}\cos(\varphi_{Ji})$ are the inductive energy terms of the Josephson junctions. We shall use the following notation for block decomposition of the inverse (symmetric) capacitance and inductance matrices,
	\begin{align}
		\msC^{-1}=\begin{pmatrix}
			\msC_0^{-1}&\msC_{0J}^{-1}\\
			\msC_{J0}^{-1}&\msC_{J}^{-1}
		\end{pmatrix},\quad \msL^{-1}=\begin{pmatrix}
			\msL_0^{-1}&\msL_{0J}^{-1}\\
			\msL_{J0}^{-1}&\msL_{J}^{-1}
		\end{pmatrix}.
	\end{align}
	We now proceed to systematic reduction. First, observe that $\bQ_J$ appears neither in the two-form nor in the energy function. Thus they are purely gauge variables that are directly discarded. As before, we do not make any gauge choice at this point for the gauge constraint arising from the charge field appearing only through the spatial derivative.
	Next, and along the same line as in the previous section, i.e., by looking at the equations of motion to identify zero vectors and impose their dynamical constraints in one go, we eliminate the redundant $\bQ_{C_0}$ charges using the voltage boundary condition
	\begin{align}\label{eq:voltage_BC_TLs_ABG_JJs}
		\msc^{-1}\bQ_0'=\dot{\bPhi}_0=\msC_0^{-1}\bQ_{C_0}+\msC_{0J}^{-1}\bQ_{C_J}.
	\end{align}
	Assuming that the submatrix $\msC_{0}^{-1}$ is invertible (stray capacitances at the connections will always provide such condition), $\bQ_{C_0}=\msc^{\frac{1}{2}}\msA_0(\msc^{-\frac{1}{2}}\bQ_0'-\tilde{\msA}_{0J}^{-1}\bQ_{C_J})$ where $\msA_0=\msc^{-\frac{1}{2}}\msC_0\msc^{-\frac{1}{2}}$ and $\tilde{\msA}_{0J}^{-1}=\msc^{\frac{1}{2}}\msC_{0J}^{-1}$. Making use of this expression, rescaling the fields $\bQ=\msc^{\frac{1}{2}}\tilde{\bQ}$ and $\bPhi=\msc^{-\frac{1}{2}}\tilde{\bPhi}$ (and removing the tildes for simplicity) in $L_\omega$,  we obtain
	\begin{align}\label{eq:Lw_TLs_ABG_JJs_subs_QC0}
		L_\omega&=L_{\omega,\TL}+\frac{1}{2}(\msA_0(\bQ_0'-\tilde{\msA}_{0J}^{-1}\bQ_{C_J}))^T\dot{\bPhi}_0\\
		&+\frac{1}{2}\bPhi_0^T(\dot{\bQ}_0-\msG_0\dot{\bPhi}_0-\msA_0(\dot{\bQ}_0'-\tilde{\msA}_{0J}^{-1}\dot{\bQ}_{C_J}))\nonumber\\
		&+\bQ_{C_J}\dot{\bPhi}_J-\frac{1}{2}\bPhi_J^T\msY_J\dot{\bPhi}_J+\bPhi_0^T\tilde{\msG}_{0J}\dot{\bPhi}_J,\nonumber\\
		&=L_{\omega,{\msA_0}{\msB_0}{\msG_0}}+L_{\omega,J\text{int}}.
	\end{align}
	where $\msG_{0}=\msc^{-\frac{1}{2}}\msY_0\msc^{-\frac{1}{2}}$ and $\tilde{\msG}_{0J}=\msc^{-\frac{1}{2}}\msY_{0J}$, and equivalently in $H$ 
	\begin{align}
		H=&\,h_{\TL}+(\bQ_0'-\tilde{\msA}_{0J}^{-1}\bQ_{C_J})^T\msA_0(\tilde{\msA}_{0J}^{-1}\bQ_{C_J})\\
		&+\frac{1}{2}(\bQ_0'-\tilde{\msA}_{0J}^{-1}\bQ_{C_J})^T\msA_0 (\bQ_0'-\tilde{\msA}_{0J}^{-1}\bQ_{C_J})\nonumber\\
		&+\frac{1}{2}\bQ_{C_J}^T\msC_{J}^{-1}\bQ_{C_J}+\frac{1}{2}\bPhi_P^T\msL^{-1}\bPhi_P+h_J(\bPhi_J).\nonumber\\
		=&\,h_{\TL}+\frac{1}{2}(\bQ_0')^T\msA_0\bQ_0'+\frac{1}{2}\bPhi_0^T\msB_0^{-1}\bPhi_0\\
		&+\frac{1}{2}\left(\bPhi_0^T\tilde{\msB}_{0J}^{-1}\bPhi_J+\bPhi_J^T\tilde{\msB}_{J0}^{-1}\bPhi_0\right)+\frac{1}{2}\bPhi_J^T\msL_J^{-1}\bPhi_J\nonumber\\
		&+\frac{1}{2}\bQ_{C_J}^T\tilde{\msC}_J^{-1}\bQ_{C_J}+h_J(\bPhi_J)\nonumber\\
		=&\,H_{{\msA_0}{\msB_0}{\msG_0}}+H_{J\text{int}}\label{eq:H_TLs_ABG_JJs_subs_QC0}
	\end{align}
	where we have also defined the matrices $\msB_{0}=\msc^{\frac{1}{2}}\msL_{0}\msc^{\frac{1}{2}}$, $\tilde{\msB}_{0J}^{-1}=\msc^{-\frac{1}{2}}\msL_{0J}^{-1}$ and $\tilde{\msC}_J=\msC_J^{-1}-\msC_{J0}^{-1}\msC_0\msC_{0J}^{-1}$.
	
	We have introduced the notation $L_{\omega,{\msA_0}{\msB_0}{\msG_0}}$ (respectively $H_{{\msA_0}{\msB_0}{\msG_0}}$) to stress the identification with the corresponding part of the Lagrangian (respectively with the Hamiltonian) in Eq. \eqref{eq:Lfinal_ABG}. Indeed, we can now use the differential operator $\mcL$ from the previous section (with ${\msA_0}{\msB_0}{\msG_0}$ boundary), Eq. \eqref{eq:DiffOp_ABG}, to expand fields
	as
	\begin{align}\label{eq:PhiQExpansion}
		\begin{pmatrix}
			\bPhi\\\bQ
		\end{pmatrix}(x,t)=\sum_\lambda\int \dd \Omega\,\left(F^{\Omega\lambda}\bW_{\Omega\lambda}^F+G^{\Omega\lambda}\bW_{\Omega\lambda}^G\right)
	\end{align}
	so that 
	\begin{align}
		\begin{aligned}
			L_{\omega}&=\int \dd \Omega\,\Omega G^{\Omega\lambda}\dot{F}^{\Omega \lambda}+L_{\omega,J}\\
			&+\bPhi_0^T\left[\mGamma_{0J}^Q\dot{\bQ}_{C_J}+\tilde{\msG}_{0J}\dot{\bPhi}_J\right]
		\end{aligned}
	\end{align}
	where $\mGamma_{0J}^Q=\msA_0\tilde{\msA}_{0J}^{-1}$, and analogously with the Hamiltonian, to be examined in detail later. We have not yet expanded $\bPhi_0$ explicitly, for clarity. We have also introduced 
	\begin{align}
		L_{\omega,J}= \bQ_{C_J}^T\dot\bPhi_J-\frac{1}{2}\bPhi_J^T\msY_J\dot{\bPhi}_J.
	\end{align}
	Our task now is to look for an identification of canonically conjugate variables such that the expression of the Hamiltonian in those variables is not jumbled. 
	
	To that purpose, $L_{\omega,J}$ is brought to a canonical form  with the transformation
	\begin{align}\label{eq:dQ_CJ_dPhi_J_transform_YJ}
		\begin{pmatrix}
			\dQ_{C_J}\\\dPhi_J
		\end{pmatrix}=\begin{pmatrix}
			\mone&\msY_J^T/2\\0&\mone    
		\end{pmatrix}\begin{pmatrix}
			\dd\bar{\bQ}_{C_J}\\\dd\bar{\bPhi}_J
		\end{pmatrix}, 
	\end{align}
	so that we have 
	\begin{align}
		\begin{aligned}
			L_{\omega}=&\,\int \dd \Omega\,\Omega G^{\Omega\lambda}\dot{F}^{\Omega \lambda}+\bar{\bQ}_{C_J}^T\dot{\bar{{\bPhi}}}_J\\
			&+\bPhi_0^T\left(\mGamma_{0J}^Q\dot{\bar{{\bQ}}}_{C_J}+\mGamma_{0J}^\Phi\dot{\bar{{\bPhi}}}_{J}\right)
		\end{aligned}
	\end{align}
	where $\mGamma_{0J}^\Phi=(\tilde{\msG}_{0J}+\mGamma_{0J}^Q\msY_J)$. 
	
	Working in the $\Omega$ subspace, one can check that changing coordinates according to 
	\begin{align}\label{eq:FG_transf_TLs_ABG_JJs}
		\begin{aligned}
			F^{\Omega\lambda}&=\tilde{F}^{\Omega\lambda}-\frac{1}{\Omega}(\bU_{\Omega\lambda0}^G)^T\left(\mGamma_{0J}^{Q}\bar{\bQ}_{C_J}+\mGamma_{0J}^{\Phi}\bar{\bPhi}_{J}\right),\\
			G^{\Omega\lambda}&=\frac{1}{\Omega}\tilde{G}^{\Omega\lambda}+\frac{1}{\Omega}(\bU_{\Omega\lambda0}^F)^T\left(\mGamma_{0J}^{Q}\bar{\bQ}_{C_J}+\mGamma_{0J}^{\Phi}\bar{\bPhi}_{J}\right),
		\end{aligned}
	\end{align}
	removes the coupling between the two subsectors, and gives a canonical term $\tilde{G}\dot{\tilde{F}}$ for the first one. Here, $\bU_{\Omega\lambda0}^{F,G}\equiv\bU_{\Omega\lambda}^{F,G}(0)$. The finite sector is more involved, however. Here we have used the notation $\bU^{F/G}$ for the top component of $\bW^{F/G}$, as (in obvious compact form) $\bPhi= F\cdot\bU^F+ G\cdot\bU^G$, in line with Eq. \eqref{eq:PhiQExpansion}.
	
	Even without using any special properties of the basis of the operator $\mcL$ here, namely closed analytical integral formulas~\cite{ParraRodriguez:2018,ParraRodriguezPhD:2021} (see App.~\ref{sec:L_op_app}), we have arrived at a result susceptible of numerical solution. This comes about because the infinite dimensional part of the two-form has been decoupled from the finite part, which is now amenable to numerical analysis. Explicitly, the two-form term has transformed into 
	
	\begin{align}
		L_\omega=\int \dd \Omega \tilde{G}^{\Omega\lambda}\dot{\tilde{F}}^{\Omega\lambda}+\tilde{L}_{\omega,J},
	\end{align}
	where 
	\begin{align}
		&\tilde{L}_{\omega,J}=\,\bar{\bQ}_{C_J}^T\dot{\bar{{\bPhi}}}_J\\
		&\quad-\left(\mGamma_{0J}^Q{\bar{\bQ}}_{C_J}+\mGamma_{0J}^\Phi{\bar{\bPhi}}_{J}\right)\msU\left(\mGamma_{0J}^Q\dot{\bar{{\bQ}}}_{C_J}+\mGamma_{0J}^\Phi\dot{\bar{{\bPhi}}}_{J}\right)\nonumber
	\end{align}
	where we have defined the (integrated) anti-symmetric matrix
	\begin{align}
		\msU=\sum_\lambda\int \dd \Omega \frac{\bU_{\Omega\lambda0}^G(\bU_{\Omega\lambda0}^F)^T-\bU_{\Omega\lambda0}^F(\bU_{\Omega\lambda0}^G)^T}{\Omega}.
	\end{align}
	
	On the other hand, upon use of the expansion of fields  (\ref{eq:PhiQExpansion}), $H_{{\msA_0}{\msB_0}{\msG_0}}=\int \dd \Omega \frac{\Omega^2}{2}(F_{\Omega\lambda}^2+G_{\Omega\lambda}^2)$ as in previous section, and the full Hamiltonian (\ref{eq:H_TLs_ABG_JJs_subs_QC0}) transforms under (\ref{eq:dQ_CJ_dPhi_J_transform_YJ}) and (\ref{eq:FG_transf_TLs_ABG_JJs}) to yield $H=\tilde{H}_{{\msA_0}{\msB_0}{\msG_0}}+\tilde{H}_{\text{int}}+\tilde{H}_{J}$, with the dressed modes' term
	\begin{align}\label{eq:HA0B0G0}
		\tilde{H}_{{\msA_0}{\msB_0}{\msG_0}}=\int \dd \Omega \frac{(\tilde{G}^{\Omega\lambda})^2+\Omega^2(\tilde{F}^{\Omega\lambda})^2}{2},
	\end{align}
	the interaction 
	\begin{align}\label{eq:hintfinal}
		\tilde{H}_{\text{int}}=&\int \dd \Omega(\tilde{G}^{\Omega\lambda}\bU_{\Omega\lambda0}^F-\tilde{F}^{\Omega\lambda}\Omega\bU_{\Omega\lambda0}^G)^T\times\dots\nonumber\\
		&\times(\mGamma_{0J}^Q\bQ_{C_J}+\mGamma_{0J}^\Phi\bPhi_{J})\nonumber\\
		+&\int \dd \Omega(\tilde{F}^{\Omega\lambda}\bU_{\Omega\lambda0}^F+\Omega^{-1}\tilde{G}^{\Omega\lambda}\bU_{\Omega\lambda0}^G)^T\times\dots\nonumber\\
		&\times\tilde{\msB}_{0J}^{-1}\bar{\bPhi}_J,
	\end{align}
	and the dressed JJs energies
	\begin{align}
		&\tilde{H}_{J}=\frac{1}{2}\bar{\bQ}_{C_J}^T\msC_{J}^{-1}\bar{\bQ}_{C_J}+\frac{1}{2}\bar{\bPhi}_J^T\tilde{\msL}_J^{-1}\bar{\bPhi}_J+h_J(\bar{\bPhi}_J)\nonumber\\&+\bar{\bQ}_{C_J}^T\left((\mGamma_{0J}^Q)^T(\msA_0^{-1}\mGamma_{0J}^\Phi-\msU\tilde{\msB}_{0J}^{-1})-\frac{\tilde{\msC}_{J}^{-1}\msY_J}{2}\right)\bar{\bPhi}_J.
	\end{align}
	Here we defined a dressed inductive matrix 
	\begin{align}
		\tilde{\msL}_J^{-1}=&\,\msL_J^{-1}+(\mGamma_{0J}^\Phi)^T\msA_0^{-1}\mGamma_{0J}^\Phi+\frac{\msY_J\tilde{\msC}_J^{-1}\msY_J^T}{4}\nonumber\\&
		-(\mGamma_{0J}^\Phi)^T\msU\tilde{\msB}_{0J}^{-1}-(\tilde{\msB}_{0J}^{-1})^T\msU^T\mGamma_{0J}^\Phi.
	\end{align}
	The direct NR coupling between TLs translates in neither $\bU_{\Omega\lambda}^F$ nor $\bU_{\Omega\lambda}^G$ being zero, i.e., $\msU\neq 0$, and therefore, in induced coupling terms in the JJs subsector, both in $\tilde{L}_{\omega,J}$ and in $\tilde{H}_{J}$. Thus, in a final step, one needs to find numerically a canonical set of Darboux coordinates for the subsector, and transform both $\tilde{H}_{J}$ and $\tilde{H}_{\text{int}}$. Naturally, there will be parameter manifolds where such interaction could be negligible and thus, approximately discarded, e.g., outside of the ultra-strong coupling regime~\cite{FornDiaz:2017,Kockum:2019}.  In either case, canonical quantization will follow in the standard way.
	
	\subsection{Divergence-free properties of the quasi-lumped models}
	\label{subsec:Divergence-free_properties}
	We note that the coupling parameters derived through this method, as presented in Eq.~\eqref{eq:hintfinal}, are inherently free from divergence issues, rendering the need for renormalization superfluous~\cite{ParraRodriguezPhD:2021}. We give now three arguments/interpretations for this fact, one analytical, one based on the properties of admittance matrices, and finally a physical one. In the next section an explicit example will be computed and examined with respect to this issue.
	
	We start with the analytic argument.
	To be precise in this regard, one important condition we have imposed  is that the capacitive and inductive matrices coupling to the TLs be full rank. Were that not the case, some of the nonlinear degrees of freedom could merge with boundaries of the TLs, such that the conception of coupling constants would be nullified, as shown in  the analysis of an example in Sec.~\ref{sec:GaugeContinuum}. Under the assumption, the properties of operator $\mathcal{L}$ and its eigenfunctions will ensure the convergence of quantities such as the Lamb shift of a system transition $\omega_{jl}$, formally $\sum g_\Omega^2/(\Omega-\omega_{jl})$ (where $\sum$ may be a sum for finite-length TLs or a principal value integral for the infinite-length case). To see this fact, notice that the operator $\mathcal{L}$ involves the boundaries, and that the inner products required for it to be self-adjoint have a discrete part controlled by the capacitive matrix $\mathsf{A}$, as presented in Eq.~\eqref{eq:inner_product_AB}. The combination of these facts with the demand for orthonormal bases provides us with analytical sum rules, which can be compactly written as $(U_0^\alpha)^2=\mathsf{A}^{-1}$. It follows that $\bU_{\Omega\epsilon}(0)$ must tend to zero faster than $\Omega^{-1/2}$.  The couplings in Eq.~\eqref{eq:hintfinal} involve $\Omega \bU_{\Omega\lambda0}$, which could be divergent. However, that term goes together with $\tilde{F}^{\Omega\lambda}$. In quantization, as can be surmised from Eq.~\eqref{eq:HA0B0G0}, $\tilde{F}\sim\Omega^{-1/2}(a+a\dag)$. Therefore, the coupling arising from this term gives us $g_\Omega$ tending to zero, with some exponent $\Omega^{-\delta}$ with $\delta>0$. It follows that $g_\Omega^2/\Omega\sim\Omega^{-1-2\delta}$, and by the integral test the sum is convergent.  The terms with the structure $\tilde{G}U_0$ have an asymptotic behaviour $\tilde{G}U_0\sim \Omega^{1/2}\Omega^{-1/2-\delta}$ with positive delta, and we have the same situation. Notice that the coupling  through $\bar{\mathsf{B}}_{0J}^{-1}$ is even more regular, given this sum rule.
	
	Passing now to the admittance argument, given the full rank of $\mathsf{A}$ and $\mathsf{B}$ the admittance matrices will be filtering for high frequencies. Therefore, the high frequency modes entering from the TL will not excite the external nonlinear system. As a consequence, the couplings will decrease with frequency.
	
	The physical explanation for this phenomenon lies in the fact that the admittance box has to be understood as included in the system that couples to the nonlinear elements. The individual modes of that complete system are the dressed modes we have studied in Sec.~\ref{sec:quadr-hamilt-diag}. Now, even without a detailed study of those modes, we observe that two natural length-scales arise. The capacitive connector will present, because the capacitive matrix $\mathsf{C}$ is full rank (recall that $\msA=\msc^{-1/2}\msC \msc^{-1/2}$), some characteristic capacity scale, such as $C_\gamma^{-1}=\|\mathsf{C}^{-1}\|$. Analogously, a characteristic inductance $L_\gamma$ will also exist. Then two natural length scales appear, the first given by the characteristic connector capacity $C_\gamma$ divided by the characteristic scale of capacity per unit length of the TLs, $c_\gamma$, namely $\alpha=C_\gamma/c_\gamma$. The second one will pertain to the comparison of characteristic inductances, $\beta=L_\gamma/\ell_\gamma$.  These characteristic lengths reflect that in any dressed mode there is some energy in the boundary. Furthermore, they provide us with natural cutoffs in wavenumber, $k_\gamma^c=1/\alpha$ and  $k_\gamma^l=1/\beta$, or, in frequency terms, $\Omega_\gamma^c= 1/\alpha \sqrt{c_\gamma\ell_\gamma}$ and $\Omega_\gamma^l= 1/\beta \sqrt{c_\gamma\ell_\gamma}$. Using the characteristic impedance of the transmission lines, $Z_\gamma=\sqrt{\ell_\gamma/c_\gamma}$, we have $\Omega_\gamma^c=1/Z_\gamma C_\gamma$ and $\Omega_\gamma^l=Z_\gamma/L_\gamma$.

	\section{Quasi-lumped circuit examples}
	\label{sec:quasi_lumped_circuit_examples}
	Having worked out the fundamentally most cumbersome linear interaction between transmission lines and JJs, let us apply the methodology to a pair of two simpler and illustrative circuit examples. 
	\subsection{TL linearly coupled to a Josephson junction}
\begin{figure}[h]
		\centering   \includegraphics[width=0.9
		\linewidth]{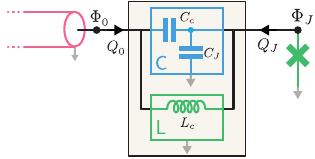}
		\caption{TL capacitively and inductively coupled to a (capacitively shunted) Josephson junction: a reciprocal particular case of that in  Fig.~\ref{fig:TLs_ABYp_JJs}. A full analysis of this circuit following the node-flux method was previously done in~\cite{ParraRodriguez:2018}.}\label{fig:L_ABp_JJ_example}
	\end{figure}

	In this first one, we recover  results obtained in Ref.~\cite{ParraRodriguez:2018} of a transmission line, capacitively and inductively coupled to a Josephson junction with the new formalism in this new alternative and faster way that circumvents the need to invert the infinite-dimensional kinetic term, see Fig.~\ref{fig:L_ABp_JJ_example}.
	
	Using the notation above, the total capacitance and inductive matrices read 
	\begin{align}
		\msC^{-1}=\begin{pmatrix}
			\frac{1}{C_s}&\frac{1}{C_J}\\[3pt]
			\frac{1}{C_J}&\frac{1}{C_J}
		\end{pmatrix},\quad \msL^{-1}=\frac{1}{L_c}(\mone-\sigma_x)
	\end{align}
	with the ``series" capacitance $C_s=C_c C_J/(C_c+C_J) 
	$.
	As one can observe, this falls under the category of circuits discussed in the previous Subsec.~\ref{subsec:TLs_NR_BB_NL}. In this particular case where we have one TL ($N=1$), $\dd Q_0=\dd Q_{C_0}+\dd Q_{L_0}$ and $\dd Q_J=\dd Q_{C_J}+\dd Q_{L_J}$, and the Lagrangian $L=L_\omega-H$ is decomposed in the following terms
	\begin{align}
		L_{\omega}&=L_{\omega,\TL}+\frac{1}{2}\Phi_0(\dot{Q}_0-2\dot{Q}_{C_0})+Q_{C_J}\dot{\Phi}_J\label{eq:Lw_ABp_JJ_example}\\
		H&=h_{\TL}+\frac{1}{2}\bQ_P^T\msC^{-1}\bQ_P+h_l(\bPhi_P)\label{eq:H_ABp_JJ_example}\\
		&=h_{\TL}+\frac{Q_{C_0}^2}{2C_s}+\frac{(2Q_{C_0}+Q_{C_J})Q_{C_J}}{2 C_J}+h_l(\bPhi_P)\nonumber
	\end{align}
	with $h_l(\bPhi_P)=\frac{\bPhi_P^T\msL^{-1}\bPhi_P}{2}-E_J\cos(\varphi_J)$, with $\bQ_P^T=(Q_0, Q_J)$, $\bPhi_P=(\Phi_0,\Phi_J)$ and $\varphi_J=2\pi\Phi_J/\Phi_Q$ ($\Phi_Q$ the flux quantum). Now, to bring it to the shape of the Lagrangian (\ref{eq:Lw_TLs_ABG_JJs_subs_QC0}) and (\ref{eq:H_TLs_ABG_JJs_subs_QC0}), we need to express $Q_{C_0}$ in terms of $Q_0'$. That we can do by solving the voltage boundary-condition equations (\ref{eq:voltage_BC_TLs_ABG_JJs}), which in this case  is
	\begin{align}
		\frac{Q_{C_0}}{C_c}+\frac{Q_{C_0}+Q_{C_J}}{C_J}=\dot{\Phi}_0=\frac{Q_0'}{c},
	\end{align}
	so $Q_{C_0}=a_0 (Q_0' + \sqrt{c}\tilde{a}_{0J}^{-1} Q_{C_J})$ where $a_0=C_s/c$, and  $\tilde{a}_{0J}=C_J/\sqrt{c}$. Introducing this last expression in the first-order term (\ref{eq:Lw_ABp_JJ_example}), and rescaling the fields through the substitutions $Q/\sqrt{c}\rightarrow Q$ and $\sqrt{c}\Phi\rightarrow \Phi$ we obtain 
	\begin{align}\label{eq:Lw_ABp_JJ_example_noQc0}
		\begin{aligned}
			L_\omega=&\,L_{\omega,\TL}+\frac{1}{2}\Phi_0\left(\dot{Q}_0-a_0(\dot{Q}_0'-\tilde{a}_{0J}^{-1} \dot{Q}_{C_J})\right)\\
			&+\frac{1}{2}\left(a_0(Q_0'-\tilde{a}_{0J}^{-1} Q_{C_J})\right)\dot{\Phi}_0+Q_{C_J}\dot{\Phi}_J.
		\end{aligned}
	\end{align}
	Including the solution of $Q_{C_0}$ and rescaling the fields as well in $H$, we obtain
	\begin{align}\label{eq:H_ABp_JJ_example_noQc0}
		\begin{aligned}
			H=&\,h_{\TL}+\frac{a_0(Q_0'-\tilde{a}_{0J}^{-1}Q_{C_J})^2}{2}+\frac{\Phi_0^2}{2b_0}\\
			&+\frac{a_0}{\tilde{a}_{0J}}(Q_0'-\tilde{a}_{0J}^{-1} Q_{C_J}) Q_{C_J}-\frac{\Phi_0\Phi_J}{\tilde{b}_{0J}}\\
			&+\frac{Q_{C_J}^2}{2 C_J}+\frac{\Phi_J^2}{2L_c}+h_J(\Phi_J)\\
			=&\,h_{\TL}+\frac{a_0 (Q_0')^2}{2}+\frac{(\Phi_0)^2}{2 b_0}-\frac{\Phi_0\Phi_J}{\tilde{b}_{0J}}\\
			&+\frac{Q_{C_J}^2}{2}\left(\frac{1}{C_J}-\frac{a_0}{\tilde{a}_{0J}^2}\right)+\frac{\Phi_J^2}{2L_c}+h_J(\Phi_J),
		\end{aligned}
	\end{align}
	{where $b_0= c L_c$ and $\tilde{b}_{0J}=\sqrt{c}L_c$}. Observe that it seems that the capacitive coupling has disappeared from the energy term in the line before the last. Indeed it has not,  as it is just encoded in the non-canonical term derived from the two-form (\ref{eq:Lw_ABp_JJ_example_noQc0}).
	
	Now, we can make the field expansion described generically in the previous section. For that we need the particular operator with action
	\begin{align}
		\mcL\bmW=\left(-\Delta \bW'',\begin{pmatrix}
			-\Delta U'+b_0^{-1}U\\-b_0^{-1}V'
		\end{pmatrix}_0\right)
	\end{align}
	and domain 
	\begin{align}
		\mcl{D}(\bmW)=\left\{\left(\bW,\begin{pmatrix}
			a_0 U\\V-a_0 V'
		\end{pmatrix}_0\right)\right\},
	\end{align}
	plus the obvious functional analytical requirements. 
	In this particular case, the doubled-space basis decomposes into two uncoupled components as $\mcL=\mcL_{a_0}\oplus\mcL_{b_0}$, and thus 
	\begin{align}
		\bmW_\Omega^F&=\left(\begin{pmatrix}
			U_{\Omega}\\0
		\end{pmatrix},\begin{pmatrix}
			a_0 U_{\Omega}(0)\\0
		\end{pmatrix}\right),\label{eq:bmW_OmegaF_ex1}\\
		\bmW_ \Omega^G&=\left(\begin{pmatrix}
			0\\V_\Omega
		\end{pmatrix},\begin{pmatrix}
			0\\V_{\Omega}(0)-a_0 V_{\Omega}'(0)
		\end{pmatrix}\right),
	\end{align}
	such that \begin{align}
		\begin{pmatrix}
			\Phi\\Q
		\end{pmatrix}(x,t)=\int \dd \Omega\,\left(F^{\Omega}\bW_\Omega^F+G^{\Omega}\bW_\Omega^G\right).
	\end{align}
	Observe that there is no index $\lambda$ because $N=1$. After the dust settles, we reach the canonical form of $L_\omega$,
	\begin{align}
		L_{\omega}=&\int \dd \Omega\, \left(\Omega G^\Omega-\frac{a_0}{\tilde{a}_{0J}} U_{\Omega}(0){Q}_{C_J}\right)\dot{F}^\Omega +Q_{C_J}\dot{\Phi}_J\nonumber\\
		=&\int \dd \Omega\, \tilde{G}^\Omega\dot{\tilde{F}}^\Omega+Q_{C_J}\dot{\Phi}_J
	\end{align}
	where we have used transformation (\ref{eq:FG_transf_TLs_ABG_JJs}), which simplifies to $\Omega G^\Omega=\tilde{G}^\Omega+\frac{a_0}{\tilde{a}_{0J}} U_{\Omega}(0)Q_{C_J}$ and $F^\Omega=\tilde{F}^\Omega$. Observe that we are already in canonical form. We can introduce the same transformation in the energy part such that it becomes now the useful Hamiltonian 
	\begin{align}
		H=&\int \dd \Omega\,\frac{\Omega^2}{2}\left[(F^{\Omega})^2+(G^{\Omega})^2\right]- \int \dd \Omega\, \frac{F^{\Omega}U_{\Omega}(0)\Phi_J}{\tilde{b}_{0J}}\nonumber\\
		&+\frac{Q_{C_J}^2}{2}\left(\frac{1}{C_J}-\frac{a_0}{\tilde{a}_{0J}^2}\right)+\frac{\Phi_J^2}{2L_c}+h_J(\Phi_J)\nonumber\\
		=&\int \dd \Omega\,\frac{1}{2}\left[(\tilde{G}^{\Omega})^2+\Omega^2(\tilde{F}^{\Omega})^2\right]\label{eq:H_TL_Cc_Lc_JJ_final}\\
		&+\int \dd \Omega\,\left(\frac{a_0 \tilde{G}^{\Omega}U_{\Omega}(0)Q_{C_J}}{\tilde{a}_{0J}}-\frac{\tilde{F}^{\Omega}U_{\Omega}(0)\Phi_J}{\tilde{b}_{0J}}\right)\nonumber\\
		&+\frac{Q_{C_J}^2}{2C_J}+\frac{\Phi_J^2}{2L_c}+h_J(\Phi_J),\nonumber
	\end{align}
	where we have used the (one-dimensional) integral rule $\int_{\mathbbm{R}^+} \dd \Omega\, U_{\Omega}^2(0)=(1/a_0)$, previously introduced in Refs.~\cite{Walter:1973,ParraRodriguez:2018,ParraRodriguezPhD:2021}, to simplify the coefficient accompanying $Q_{C_J}^2$. This is the final result of the classical computation, readying the problem  for standard canonical quantization; see for comparison Eq.~(3.65) of \cite{ParraRodriguezPhD:2021} with the analysis performed using the node-flux method.
	\begin{figure}[h]
		\centering   \includegraphics[width=1
		\linewidth]{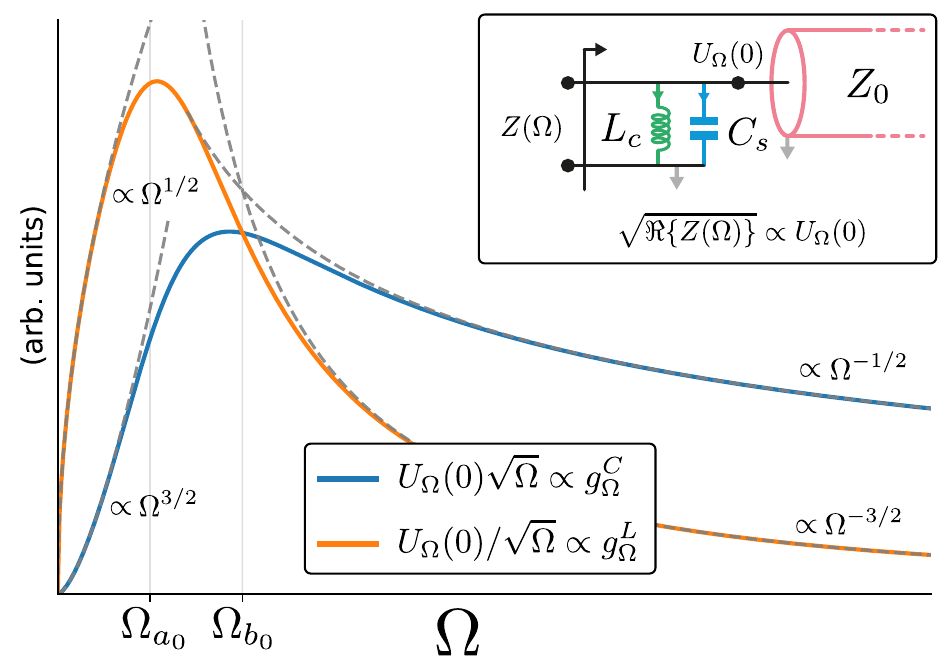}
		\caption{Coupling parameters for the circuit in Fig.~\ref{fig:L_ABp_JJ_example} when written in the form of (\ref{eq:H_quantized_ex_CcLc}). Saturation of both coupling parameters appears around $\Omega_{a_0}$ and $\Omega_{b_0}$, which are related to the quasi-lumped \emph{electric} and \emph{magnetic lengths} $C_s/c$ and $L_c/l$, respectively. These coupling parameters are directly related to the impedance of the inset circuit where the capacitance at the end is the total capacitance to ground seen by the TL in Fig.~\ref{fig:L_ABp_JJ_example}.}\label{fig:g_Omega_example}
	\end{figure}
	It is important to highlight that with our novel approach, the inversion of infinite dimensional kinetic terms is no longer necessary~\cite{ParraRodriguez:2018} (we are not performing a Legendre transform), allowing us to reach the final result faster. This improvement  is primarily due to an enhanced comprehension of the doubled eigenvalue problem, among other factors. In this particular case, the solution of the eigenvalue sub-problem $\mcL_{a_0} \bsb{\mcl{U}}_{\Omega}=\Omega^2 \bsb{\mcl{U}}_{\Omega}$  is enough to obtain the coupling parameters, which are controlled by the value of the generalized eigenvector at the coupling point~\cite{ParraRodriguez:2018}
	\begin{align}
		U_\Omega(0)&=\Delta^{-\frac{1}{4}}\sqrt{\frac{2\Omega^2}{\pi(\Omega^2+\Delta^{-1}\left(a_0 \Omega^2-1/b_0\right)^2)}}\label{eq:U_Omega0_example_Cc_Lc}\\
		&\propto\frac{1}{\sqrt{1+\left({\Omega}/{\Omega_{a_0}}-\Omega_{b_0}/\Omega\right)^2)}},\nonumber
	\end{align}
	where we have defined the lumped cutoff frequencies $\Omega_{a_0}=a_0^{-1} \Delta^{\frac{1}{2}}=1/(C_s Z_0)$ and $\Omega_{b_0}=b_0^{-1}\Delta^{-\frac{1}{2}}=Z_0/L_c$, with $Z_0\equiv\sqrt{l/c}$ the characteristic impedance of the TL. Naturally, $\bsb{\mcl{U}}_{\Omega}$ is equivalent to the restriction of $\bmW_{\Omega}^F$ in Eq.~(\ref{eq:bmW_OmegaF_ex1}) to its nontrivial sector. We recall that, in rewriting the quantized version of the Hamiltonian (\ref{eq:H_TL_Cc_Lc_JJ_final}) in terms of annihilation and creation operators $\tilde{G}^{\Omega}=\sqrt{\frac{\hbar \Omega}{2}}(a_{\Omega}+a_{\Omega}^\dag)$ and $\tilde{F}^{\Omega}=i\sqrt{\frac{\hbar }{2\Omega}}(a_{\Omega}-a_{\Omega}^\dag)$, 
	\begin{align}
		H/\hbar=&\,\tilde{H}_J/\hbar+\int \dd \Omega \, \Omega a_\Omega^\dag a_\Omega\label{eq:H_quantized_ex_CcLc}\\&+\int \dd \Omega \,[g_\Omega^C(a_\Omega+a_\Omega^\dag)Q_{C_J}+i g_\Omega^L(a_\Omega-a_\Omega^\dag)\Phi_J], \nonumber
	\end{align}
	and we obtain well-behaved capacitive $g^C_{\Omega}\propto U_{\Omega}(0)\sqrt{\Omega}$ and inductive $g^L_{\Omega}\propto U_{\Omega}(0)/\sqrt{\Omega}$ coupling parameters, i.e., with scalings $g^C_{\Omega}\rightarrow \Omega^{-1/2}$ and $g^L_{\Omega}\rightarrow \Omega^{-3/2}$ as $\Omega\rightarrow \infty$~\cite{ParraRodriguez:2018}, see Fig.~\ref{fig:g_Omega_example}. It is interesting to note that both coupling parameters (defined from $U_\Omega(0)$) are proportional to the $\Re\{Z(\Omega)\}$ of the circuit shown in the inset of Fig.~\ref{fig:g_Omega_example}. This circuit is constructed by viewing from the transmission line (TL) the lumped network depicted in Fig.~\ref{fig:L_ABp_JJ_example}, and independently determining the total capacitance to ground $C_s$ and the total inductance to open $L_c$. Alternatively, other equivalent Hamiltonian descriptions can be developed by considering the total environment observed by the pure Josephson element as a unique dissipative environment. The corresponding models could employ a single coupling parameter, either capacitive or inductive, for all frequencies~\cite{Caldeira:1983, Devoret:1997}, or alternatively, differentiate between capacitive and inductive responses across two distinct frequency sets~\cite{Mehta:2022} (exotic gauge choice). In these scenarios, the coupling parameter(s) would be a square root function of an immittance response seen from the JJ. 
	
	As a final message, it is worth realizing that the exact model found for this example can be greatly simplified by considering simpler divergent models, obtained by solving Dirichlet-type eigenvalue problems (without $a_0$ and $b_0$) and applying hard cutoffs at the computed frequencies $\Omega_{a_0}$ and $\Omega_{b_0}$. Note that these frequencies could potentially be higher or lower than those of the superconducting gap, and thus should be compared with them. Additionally, the quasi-lumped approximation suggests that the associated lengths $a_0$ and $\tilde{b}_0 = L_c/l$ should be smaller than the couplers' lengths for the model to remain valid. Overall, this analysis clearly indicates that renormalization techniques are unnecessary in this family of Hamiltonian models.

	\subsection{TL resonators coupled via a frequency-dependent circulator and to a JJ}
	\begin{figure*}
		\centering   \includegraphics[width=0.7
		\textwidth]{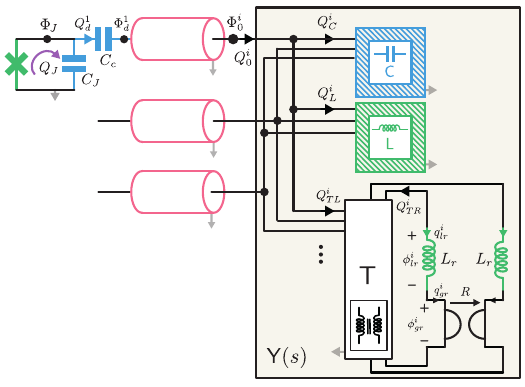}
		\caption{Josephson junction capacitively coupled to a transmission line, which in turn is coupled to two others via a frequency-dependent circulator. This circuit represents an extension to one analyzed in Ref.~\cite{ParraRodriguez:2022}.}\label{fig:JJ_Cc_TL_ABGk_TLs}
	\end{figure*}
	In Section~\ref{sec:TLs-multiport-networks} we have used our technique to analyze the most cumbersome problem of linear coupling between TLs and JJs. This involved taking into consideration ideal nonreciprocal coupling between all the different components. Although pedagogical, for practical purposes, a linear system will typically not require the use of direct nonreciprocal coupling. This is because stray capacitive and inductive effects at the terminals will filter the response. Consequently, a physically-motivated nonreciprocal admittance response will look as
	\begin{align}
		\msY(s)=\frac{\msL^{-1}}{s}+\msC s +\sum_k\frac{\msD_k s+ \msE_k}{s^2+\Omega_k^2},
	\end{align}
	where $\msD_k$ and $\msE_k$ are symmetric and anti-symmetric matrices, respectively. Let us then analyze a standard scenario where one has a Josephson junction coupled to a transmission line resonator, and this one in turn coupled to other two lines through a frequency-dependent circulator, see Fig.~\ref{fig:JJ_Cc_TL_ABGk_TLs}, with one nonreciprocal pole $k\in\{1\}$. Following our method, we can use  the full set of constraints (Kirchhoff's, transformer, and ideal gyrator), 
	\begin{align}
		\begin{aligned}
			\dQ_0&=\dQ_C+\dQ_L+\dQ_{TL},\\
			\dPhi_{TR}&=\dPhi_{lr}+\dPhi_{gr},\\
			\dQ_{TL}&=-\msT\dQ_{TR}= \msT\dQ_{gr},\\
			\dPhi_{TR}&=\msT^T\dPhi_{TL}=\msT^T\dPhi_{0},\\
			\dQ_{gr}&=\msY_{g}\dPhi_{gr},
		\end{aligned}
	\end{align}
	to implement the restriction of the two-form (\ref{eq:twoform_restricted}), obtaining 
	
	\begin{align}
		\omega=&\,\omega_{\TL}-\frac{1}{2}\dd Q_d^1\wedge \dd \Phi_d^1+\frac{1}{2}\dPhi_0^T\wedge(\dQ_0-2\dQ_C)\nonumber\\
		&\,-\dPhi_{gr}^T\wedge\msY_{g}\dPhi_{gr}+\dd Q_J\wedge \dd \Phi_J\\
		=&\,\omega_{{\msA_0}{\msB_0}a_d}+\omega_{gr}+\omega_{J},\nonumber
	\end{align}
	where $Q_d^1\equiv Q^1(d,t)$, and, as before, $\bQ_0\equiv\bQ(0,t)$. The gyrator matrix is $\msY_g=R^{-1}i\sigma_y$ while $\msT$ denotes a rectangular transformer matrix~\cite{Newcomb:1966}. Thus, the Lagrangian $L=L_{\omega}-H$ now consists of the associated first-order terms $L_\omega$, and the energy term 
	\begin{align}
		H=&\,H_{\TL}+\frac{1}{2}\bQ_C^T\msC^{-1}\bQ_C^T+ \frac{1}{2}\bPhi_0^T\msL^{-1}\bPhi_0\\
		&\,+\frac{1}{2}(\msT^T\bPhi_0-\bPhi_{gr})^T\msL_{r}^{-1}(\msT^T\bPhi_0-\bPhi_{gr})\nonumber\\
		&\,+\frac{1}{2 C_c}(Q_d^1)^2+\frac{1}{2 C_J}(Q_J-Q_d^1)^2+h_J(\Phi_J),\nonumber
	\end{align}
	where $\msL_r=\mone L_r$. 
	Introducing the solution of the EOMs for the voltages at the end $x=0$ ($\bQ_C=\msC\msc^{-1}\bQ_0'$) as in Eq.~(\ref{eq:qcslaving}), rescaling the fields in the same way as above, and rescaling the internal variables $\tilde{\bPhi}_{gr}=R^{-1/2}\bPhi_{gr}$, we obtain the first-order term 
	\begin{align}
		L_{\omega}=&L_{\omega,{\msA_0}{\msB_0}a_d}+\tilde{\Phi}_{gr}^1\dot{\tilde{\Phi}}_{gr}^2+Q_J\dot{\Phi}_J,
	\end{align}
	where $\msA_0=\msc^{-\frac{1}{2}}\msC\msc^{-\frac{1}{2}}$,  $\msB_0=\msc^{\frac{1}{2}}\msL\msc^{\frac{1}{2}}$ and $a_d=C_s/c^1$ ($C_s=C_J C_c/(C_c+C_J)$), which under the use of an eigenbasis of the associated $\mcL$ and $\mcT$ operators (see App.~\ref{sec:Details_circuit_ex2}) can be simplified with 
	\begin{align}
		L_{\omega,{\msA_0}{\msB_0}a_d}   &=\langle i \mcT \bmW_,\dot{\bmW}\rangle=\sum_{n,\lambda} \Omega_n G^{\Omega_n,\lambda}\dot{F}^{\Omega_n,\lambda}.
	\end{align}
	On the other hand, the energy term reads 
	\begin{align}
		H=&\,H_{{\msA_0}{\msB_0}a_d}+\bPhi_0^T\mGamma_{0,gr}\tilde{\bPhi}_{gr}+\frac{1}{2}\tilde{\bPhi}_{gr}^T\msOmega_{r}\tilde{\bPhi}_{gr}\nonumber\\
		&+\frac{\sqrt{c_1}}{C_J}Q_d^1 Q_J+\frac{Q_J^2}{2C_J}+h_J(\Phi_J),
	\end{align}
	where $\mGamma_{0,gr}=\msT\msL_r^{-1}R^{1/2}$ and $\msOmega_r=R/L_r\mone$, and the term of the dressed transmission lines contribute as 
	\begin{align}
		\begin{aligned}
			H_{{\msA_0}{\msB_0}a_d}=&\,H_{\TL}+\frac{1}{2}(\bQ_0')^T\msA_0\bQ_0'\\
			&+\frac{1}{2}\bPhi_0^T\msB_0\bPhi_0+\frac{a_d (Q_d^1)^2}{2}\\
			=&\langle\mcL\bmW,\bmW\rangle\\
			=&\sum_{n,\lambda}\frac{\Omega_n^2}{2}\left((F^{\Omega_n\lambda})^2+(G^{\Omega_n\lambda})^2\right).
		\end{aligned}
	\end{align}
	We can perform the analogous (simplified) transformation of the dressed TL coordinates (\ref{eq:FG_transf_TLs_ABG_JJs}) $\tilde{F}^{\Omega_n,\lambda}=F^{\Omega_n,\lambda}$ and $\tilde{G}^{\Omega_n,\lambda}=\Omega_n G^{\Omega_n,\lambda}$ and finally obtain a description in terms of canonical pairs of conjugate variables 
	\begin{align}
		L=&\sum_{n,\lambda} \tilde{G}^{\Omega_n,\lambda}\dot{\tilde{F}}^{\Omega_n,\lambda}+\Phi_{gr}^1\dot{\Phi}_{gr}^2+Q_J\dot{\Phi}_J-H,
	\end{align}
	and the Hamiltonian function
	\begin{align}
		H=&\,\sum\frac{1}{2}\left((\tilde{G}^{\Omega_n\lambda})^2+(\Omega_n\tilde{F}^{\Omega_n\lambda})^2\right)\nonumber\\
		&+\sum_{n,\lambda}\tilde{F}^{\Omega_n \lambda}\bU_{n\lambda}^T(0)\mGamma_{0,gr}\tilde{\bPhi}_{gr}\\
		&+\frac{\Omega_G}{2}\left((\tilde{\Phi}_{gr}^1)^ 2+(\tilde{\Phi}_{gr}^2)^ 2\right)\nonumber\\
		&+\sum_{n,\lambda}\frac{\sqrt{c_1}}{C_J}\tilde{G}^{\Omega_n\lambda}U_{n\lambda}^1(d) Q_J+\frac{Q_J^2}{2C_J}+h_J(\Phi_J)\nonumber
	\end{align}
	where we have used the boundary relation $V_{n\lambda}^1(d)=\Omega_n U_{n\lambda}^1(d)$. This Hamiltonian, with an infinite-dimensional dressed mode basis, is ready for canonical quantization. Explicit values for the coupling parameters can be obtained by solving the relevant eigenvalue problem (with ${\msA_0}{\msB_0}{a_d}$ boundary). As previously advanced, this Hamiltonian, which is a more involved version of those discussed in Fig.~\ref{fig:TLs_to_LE_NR_networks}(b), has by construction nondivergent properties, something that can be simply seen from the  argument that the Josephson junction \emph{sees} the infinite dimensional system through a band-pass filter from the shunting ($C_J$) and coupling ($C_c$) capacitances, see also Subsec.~\ref{subsec:Divergence-free_properties}.
	
	\section{(Non)reciprocal dissipative systems within the Caldeira-Leggett framework}
	\label{sec:C-L_models}
	This article mainly focuses on the application of the geometrical approach and the Faddeev-Jackiw method to quasi-lumped circuits. However, this technique transcends transmission line theory and can equally be applied to more general infinite-dimensional systems, for instance, 
	\begin{figure*}[t]
		\centering
		\includegraphics[width=1\textwidth]{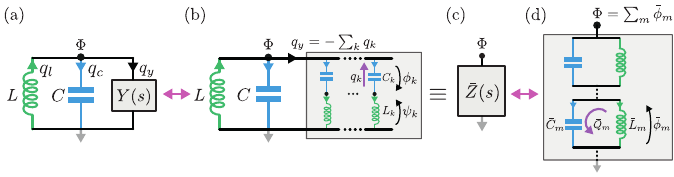}\caption{(a) An $LC$ oscillator coupled to a generic one-port admittance, and (b) this generic admittance decomposed in an infinite sum of harmonic oscillators. (c) The equivalent impedance of the total admittance $\bar{Z}^{-1}(s)=\bar{Y}(s)=s C+ 1/(sL)+Y(s)$, and (d) a continuous infinite expansion of such impedance in a normal mode basis.}\label{fig:CaldeiraLeggett}
	\end{figure*}
	lumped element circuits coupled to generic dissipative linear systems, modeled with a continuum of harmonic degrees of freedom~\cite{Feynman:1963,Caldeira:1983,YurkeDenker:1984}. Thus, in this section we will obtain canonical Hamiltonians of dissipative one-port (reciprocal) and two-port (nonreciprocal) environments whose continuum limit is well-behaved.  
	
	\subsection{One-port dissipative system}
	
	In this subsection, we will briefly derive the quantum Hamiltonian for a lumped network, here a parallel $LC$ oscillator, when it is coupled to a dissipative one-port environment. This environment is described by its causal admittance response $y(t)\propto \Theta(t)$, with a Laplace transform $Y(s)$, or its (causal) Fourier counterpart, obtained through the substitution $s=-i\omega+0^+$, see Fig.~\ref{fig:CaldeiraLeggett}(a). 
	
	Within the Caldeira-Leggett paradigm~\cite{Caldeira:1983}, one  replaces the environment response function with a dense infinite sequence of LC harmonic oscillators with response function $Y_k(s)=s/ L_k/(s^2+\Omega_k^2)$, i.e.,
	\begin{align}
		\tilde{Y}(\omega)=Y(-i\omega+0^+)=\lim_{\Delta \Omega\rightarrow0}\sum_k \,\tilde{Y}_k(\omega),
	\end{align}
	in such a way that the continuous real part of the response function $\tilde{Y}(\omega)$ is obtained from a dense sequence of delta distributions
	\begin{align}
		\Re \{\tilde{Y}_k(\omega)\}=\frac{\pi y_k\Omega_k}{2}(\delta(\omega-\Omega_k)+\delta(\omega+\Omega_k)),
	\end{align}
	where $y_k=1/(\Omega_k L_k)$. Here, the imaginary part follows straightforwardly from the Sokhotski–Plemelj theorem, see further details in~\cite{Vool:2017} and in Appendix~\ref{sec:details_dissip_multiport_NR_circuit} for a more general nonreciprocal construction. Once we have an explicit decomposition of the immittance, we can apply our method. 
	
	Following the notation in Fig.~\ref{fig:CaldeiraLeggett}(b), we compute the canonical two-form (\ref{eq:twoform_restricted})
	\begin{align}
		\begin{aligned}
			\omega&=\frac{1}{2}\left(\dd q_c\wedge\dd \Phi-\dd \Phi\wedge\dd q_l\right)\\
			&\quad+\frac{1}{2}\left(\sum_k \dd q_k\wedge\dd\phi_k-\dd\psi_k\wedge \dd q_k\right)\\
			&=\dd q_l\wedge\dd \Phi+\sum_k \dd q_k\wedge\dd \psi_k.
		\end{aligned}
	\end{align}
	where we have solved $\dd q_c=\dd q_l-\dd q_y$ and $\dd \phi_k=\dd \psi_k -\dd\Phi$, and the total current in the admittance is $\dd q_y=-\sum_k \dd q_k$. Thus, we construct the Lagrangian
	\begin{subequations}
		\begin{align}
			&\qquad\qquad\qquad L=q_l\dot{\Phi}+\sum_k q_k\dot{\psi}_k-H,\\
			&H=\sum_k\frac{q_k^2}{2 C_k}+\frac{\psi_k^2}{2 L_k}+\frac{(q_l+\sum_n q_n)^2}{2 C}+\frac{\Phi^2}{2 L},\label{eq:H_RLC_ql_phi_k}
		\end{align}
	\end{subequations}
	and the pairs of conjugated variables are read from the two-form term $\{\Phi,q_l\}=1$, and $\{\psi_k, q_{k'}\}=\delta_{k k'}$. This is   Hamiltonian dynamics equivalent to that obtained in Eq.~(3.31) of Ref.~\cite{Devoret:1997}, under a (symplectic) transformation, i.e., writing it in terms of $q_c$ and $\phi_k$. Then the first-order Lagrangian is
	\begin{align}
		L= q_c\dot{\Phi} +\sum_k  q_k \dot{\phi}_k-H,
	\end{align}
	where now the Hamiltonian becomes
	\begin{align}
		\label{eq:H_RLC_qc_psi_k}H=\sum_k\frac{q_k^2}{2 C_k}+\frac{(\Phi+\phi_k)^2}{2 L_k}+\frac{q_c^2}{2 C}+\frac{\Phi^2}{2 L}.
	\end{align}
	Observe that in both of the above Hamiltonian versions, Eqs.~(\ref{eq:H_RLC_qc_psi_k}) and (\ref{eq:H_RLC_ql_phi_k}), finding the eigenbasis of the compound system for a generic dissipative admittance $\tilde{Y}(s)$ is a cumbersome task. However, for pure resistors characterized by $Y(s)=1/R$ and modeled as a semi-infinite transmission line, optimal diagonalizations are more straightforward, as detailed in Ref.~\cite{ParraRodriguez:2018}. It is worth mentioning that significant analytical advancements have been made in both the weak and ultra-strong coupling regimes, e.g., in Refs.~\cite{Cattaneo:2021,Ashida:2022}.
	
	For this particular closed system, and if one is mainly interested in computing properties of the external variables, such as correlators of the node flux $\Phi$, Ref.~\cite{Devoret:1997} noticed that it is possible to use an equivalent response function, here the total impedance  $\tilde{Z}(s)=(s C+ 1/(sL)+Y(s))^{-1}$, see Figs.~\ref{fig:CaldeiraLeggett}(c,d). Now, we reach the dissipative response by a different diagonalising sequence 
	\begin{align}
		\bar{Z}(s)=\sum_m \bar{Z}_m(s)=\sum_m (s/\bar{C}_m)/(s^2+\bar{\Omega}_m^2).
	\end{align}
	For this representation, the total Lagrangian would naturally be $L=\sum_m \bar{Q}_m \dot{\bar{\phi}}_m-H$, with a canonical Hamiltonian
	\begin{align}
		H=\sum_m \frac{\bar{Q}_m^2}{2 \bar{C}_m}+\frac{\bar{\phi}^2_m}{2 \bar{L}_m},
	\end{align}
	upon which the standard continuous limit can be taken, see further details in Ref.~\cite{Vool:2017}.
	
	\subsection{Nonreciprocal multiport dissipative circuit}
	Building upon the previous analysis, we extend our focus to multiport dissipative environments through the use of canonical multiport fraction expansions of causal matrices~\cite{Newcomb:1966,Solgun:2015,ParraRodriguez:2018,ParraRodriguezPhD:2021}, instead of functions, as in~\cite{Caldeira:1983,YurkeDenker:1984,Vool:2017}. Here we will just concentrate in obtaining a Hamiltonian for the environment.
	
\begin{figure}[h]
		\centering
		\includegraphics[width=0.85\linewidth]{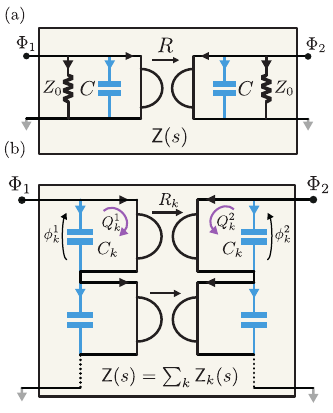}\caption{(a) A dissipative nonreciprocal oscillator coupled consisting of a gyrator (characterized by parameter $R$), capacitors $C$ and resistances $Z_0$. (b) A canonical impedance representation in terms of a continuum of nondissipative nonreciprocal harmonic oscillators.}\label{fig:NR_2port_CaldeiraLeggett}
	\end{figure}

	For the sake of completeness, we will work out in full detail the particular circuit in Fig.~\ref{fig:NR_2port_CaldeiraLeggett}(a) which may serve the reader as a guide for more complex NR dissipative environments. There, we have depicted a two-port linear nonreciprocal dissipative circuit characterized by the 2-by-2 admittance $\msY(s)=(s C+Z_0)\mone + R^{-1}i\sigma_y$, or impedance $\msZ(s)=\msY^{-1}(s)$ matrices in Laplace space, where $\sigma_y$ is the second Pauli matrix. Crucially, and within the Caldeira-Leggett framework~\cite{Caldeira:1983}, we can represent the quantum dissipative dynamics exerted by this system onto others, in the weak coupling regime, by replacing this circuit with a continuum of nondissipative nonreciprocal harmonic degrees of freedom. In practice, and working again in Fourier space ($s=-i\omega+0^+$) this process is tantamount to finding a canonical sequence of impedances $\tilde{\msZ}_k(\omega)=\msZ_k(-i\omega+ 0^+)$ which in the continuum limit must yield 
	\begin{align}
		\tilde{\msZ}(\omega)&=\lim_{\Delta\Omega \rightarrow 0}\sum_k\tilde{\msZ}_k(\omega),\\&=\frac{(-i\omega C+G_0)\mone-G\msJ}{(-i\omega C+G_0)^2+G^2},\label{eq:NRD_Z2p_omega}
	\end{align} 
	with $Y_0=Z_0^{-1}$, $G=R^{-1}$, and $\msJ=i\sigma_y$. 
	
	Working with the canonical multiport Cauer representations of nonreciprocal two-port systems~\cite{Newcomb:1966}, $\msZ_k(s)$ must take the form 
	\begin{align}
		\msZ_k(s)&= \frac{a_k s\mone+b_k \msJ}{s^2+\Omega_k^2}\label{eq:DNR_Zk_s_2port}
	\end{align}
	where $a_k=z_k \Omega_k$, $b_k=z_k \Omega_k^2$, $z_k=R_k$, and $\Omega_k=1/(R_k C_k^2)$. In circuital terms, such parameters describe a collection of nonreciprocal harmonic oscillators consisting of pairs of capacitors ($C_k$) coupled by a gyrator ($R_k$), see Fig.~\ref{fig:NR_2port_CaldeiraLeggett}(b). The causal Fourier transform of each element of the sequence yields
	\begin{align}
		\tilde{\msZ}_k(\omega)=&\frac{\pi a_k\mone}{2}\left(\delta(\omega-\Omega_k)+\delta(\omega+\Omega_k)\right)+\mcl{P}\frac{b_k\msJ}{\Omega_k^2-\omega^2}\nonumber\\
		+&\frac{i\pi b_k\msJ}{2\Omega_k}\left(\delta(\omega+\Omega_k)-\delta(\omega-\Omega_k)\right)-\mcl{P}\frac{i a_k\omega\mone}{\Omega_k^2-\omega^2}.\nonumber
	\end{align}
	
	To go to the continuum limit, we must understand the sums as Riemann integrals in the standard way, i.e., $\lim_{\Delta\Omega\rightarrow 0}\sum_k \Delta \Omega\rightarrow \int \dd \Omega$, while the sequences will be associated with functions $a_k=k\Delta\Omega a(k\Delta \Omega)$ and $b_k=k\Delta\Omega b(k\Delta \Omega)$. As in the previous one-port example, the frequencies of the harmonics are discretized as  $\Omega_k=k\Delta\Omega$. Finally, adding all contributions and taking the limit $\Delta \Omega\rightarrow 0$ we obtain the matrix
	\begin{align}
		\tilde{\msZ}(\omega)=&\lim_{\Delta \Omega\rightarrow 0}\sum_k \tilde{\msZ}_k(\omega)=\int \dd \Omega \tilde{\msZ}_\Omega(\omega)\\
		=&\,\frac{\pi\mone}{2}[a(\omega)+a(-\omega)]+\mcl{P}\int \dd \Omega \frac{b(\Omega)\msJ}{\Omega^2-\omega^2}\nonumber\\
		-&\frac{i\pi\msJ}{2\omega}[b(\omega)+b(-\omega)]-i\mcl{P}\int \dd \Omega \frac{a(\Omega)\mone}{\Omega^2-\omega^2},\nonumber
	\end{align}
	that matches the original Eq.~(\ref{eq:NRD_Z2p_omega}) upon the use of the appropriate $a(\omega)$ and $b(\omega)$ functions. See Appendix~\ref{sec:details_dissip_multiport_NR_circuit} for further details on the construction.
	
	Having obtained the relevant  discrete sequences that permit the decomposition in Fig.~\ref{fig:NR_2port_CaldeiraLeggett}(b), we can write the two-form (\ref{eq:twoform_restricted}) in terms of the inner branch fluxes and loop currents, as there is no further loop to the outside world here, 
	\begin{align}
		\omega=\sum_k\frac{1}{2}\dQ_k^T\wedge \dPhi_k,
	\end{align}
	where $\bQ_k=(Q_k^1,Q_k^2)$ and $\bPhi_k=(\Phi_k^1,\Phi_k^2)$. Using the current-voltage constraint imposed by the gyrators $\dPhi_k=\msZ_k\dQ_k$, where $\msZ_k=R_k\msJ$, we can simplify the two-form and write the Lagrangian 
	\begin{align}
		L=\sum_k \frac{1}{2}\bQ_k^T\msZ_k\dot{\bQ}_k-\frac{(Q_k^1)^2+(Q_k^2)^2}{2C_k},
	\end{align}
	which is almost in a canonical form. Indeed, by applying a rescaling  transformation only  on the flux variables, i.e.,  
	$q_k^{(a,b)}=Q_k^{(1,2)}$ and $\psi_k^{(a,b)}=\pm R_k Q_k^{(2,1)}$, we reach the canonical Lagrangian $L=\sum_k q^a_k\dot{\psi}^a_k-H$ with Hamiltonian
	\begin{align}
		\begin{aligned}
			H=&\sum_k\left(\frac{(q_k^a)^2}{2C_k}+\frac{(\psi_k^a)^2}{2C_k R_k^2}\right)\\
			\stackrel{\text{q.}}{=}&\sum_k \hbar \Omega_k a_k^\dag a_k.
		\end{aligned}
	\end{align}
	Here the canonical pairs of variables ($\{\Phi_k^a,q_{k'}^a\}=\delta_{kk'}$) are promoted to canonically conjugated quantum operators following the standard procedure. Continuous limits of this Hamiltonian can be implemented naturally. 
	
	\section{Conclusions \& Outlook}
	In summary, this article builds upon the foundational work presented in our prior publication \cite{ParraRodriguez:2024a} to develop an exact nonreciprocal quasi-lumped element circuit theory. Here, we have extended our geometrical first-order construction methodology to address quantization in the presence of infinite dimensional subsystems, enabling the formulation of Lagrangian and Hamiltonian descriptions for a broader spectrum of circuits. Our approach seamlessly incorporates transmission lines and a diverse array of lumped element components, ranging from linear to nonlinear capacitors, inductors, sources, transformers, and gyrators. This culminates in establishing a comprehensive framework for the modeling of  nonreciprocal quasi-lumped element circuits.
	
	One of the key contributions of our approach is the addition of generic linear blackbox devices with nonreciprocal behavior, serving as couplers between input/output waveguides and nonlinear degrees of freedom. These developments are particularly relevant in the context of microwave engineering and open up new avenues for designing and analysing circuits with unique functionalities. Furthermore, we have demonstrated that our models do not require renormalization and inherently possess high-energy cutoffs, thereby addressing important challenges in the field. Importantly, this approach represents a simplification and streamlining of previous work that involved doubled variables \cite{ParraRodriguez:2022,Egusquiza:2022}. We have also presented three different arguments to understand these high-energy cutoffs, including the classical electrical engineering one whereby the linear couplers act as low-pass/band-pass filters.
	
	The methodology presented in this article provides a widely applicable and direct avenue for the Hamiltonian modeling and understanding the behavior of nonreciprocal, infinite-dimensional electrical circuits based on discrete lumped element models. Special emphasis has been placed on nonreciprocal quasi-lumped superconducting circuits comprising transmission lines and Josephson junctions, explicitly showing how to implement more general divergence-free mode expansions of mixed flux and charge fields. This advance brings promise of applications across various fields, from quantum electronics to classical microwave engineering, and represents a valuable addition to the toolkit of physicists and engineers working in these domains. In particular, this work should have extensive applications in refining models for waveguide QED systems used in distributed quantum computation and simulation \cite{Paulisch:2016,Mirhosseini:2019,Kannan:2020,Sheremet:2023}, as well as in characterizing networks within high-impedance environments~\cite{Pechenezhskiy:2020,Crescini:2023,Ardati:2024,Kuzmin:2024}, especially given that the frequency saturation point decreases with increasing characteristic impedances. More importantly, this work should be foundational in analyzing, characterizing and designing very general nonreciprocal superconducting networks.
	
	Further open work not covered here, which will benefit from our findings, includes, for instance, the perturbative elimination of blackbox internal degrees of freedom when coupling TLs to external nonlinear systems (such as JJ-based qubits), particularly when adiabatic approximations are pertinent, to derive general input-output relations in terms of immittance parameters~\cite{Solgun:2019,Labarca:2024}. Finally, additional investigations will be necessary to devise a systematic procedure for resolving all nonlinear singularities in quasi-lumped element networks, addressing the intriguing question posed from different perspectives by Refs.~\cite{Rymarz:2023,Miano:2023,PRXComment}.
	
	\begin{acknowledgments}
		The authors thank Rémi Robin for pointing out an inaccuracy in the definition of the differential operator. A. P.-R. is funded by the Juan de la Cierva fellowship FJC2021-047227-I. I. L. E. acknowledges support by the Basque Government through Grant No. IT1470-22, and project PCI2022-132984 financed by MICIN/AEI/10.13039/501100011033 and the European Union NextGenerationEU/PRTR. 
	\end{acknowledgments}

	\appendix

	\section{Mathematical details}
	\label{sec:math_details_app}
	To ensure comprehensiveness, this mathematical appendix offers a concise overview to assist those readers who may be less experienced with computations involving differential forms. Additionally, we include previously established results from \cite{ParraRodriguezPhD:2021} regarding the differential operators for the multi-line problem, thus making the manuscript self-contained. We finish the appendix by presenting new mathematical results concerning the duality operator $\mathcal{T}$, required for the complete proof of the expression of first order Lagrangians of linear systems in the abstract form
	\begin{align}
		L=\frac{1}{2}\left\langle i\mathcal{T}\bmW,\dot{\bmW}\right\rangle -\frac{1}{2}\left\langle \bmW,\mathcal{L}\bmW\right\rangle\,.
	\end{align}

	\subsection{Differential forms}\label{sec_diff_forms_app}
	As the external differential
	forms formalism used in this article,
	standard as it is, might not be familiar to readers from all physics
	disciplines, let us be explicit as to the computation of the zero
	modes discussed, for instance, in Secs. \ref{sec:TLs-geom-description} and \ref{sec:TLs-multiport-networks}. For a proper introduction to differential forms, see for
	instance \cite{Flanders:1963}.
	
	We remind the reader that given a coordinate system the associated
	local bases of one-forms and vectors are related by duality,
	\begin{align}
		\label{eq:dualbasis}
		\dd z^\alpha\left(\frac{\partial}{\partial z^\beta}\right)=\delta^\alpha_\beta\,.
	\end{align}
	This extends to two-forms and tensors by looking at
	\begin{align}
		\label{eq:twoformtwotensor}
		\left[\dd z^\alpha\wedge\dd z^\beta\right]\left(\frac{\partial}{\partial
			z^\lambda}\otimes\frac{\partial}{\partial
			z^\mu}\right)=\delta^\alpha_\lambda\delta^\beta_\mu-\delta^\alpha_\mu \delta^\beta_\lambda\,,
	\end{align}
	and using linearity. This provides us with the contraction of a
	two-form with a vector, by extending linearly
	\begin{align}
		\label{eq:vectorcontraction}
		\left[\theta_1\wedge\theta_2\right]\left(\boldsymbol{V},\cdot\right)=\theta_1(\boldsymbol{V})\theta_2-\theta_2(\boldsymbol{V})\theta_1\,,
	\end{align}
	with $\theta_1$ and $\theta_2$ one-forms and $\boldsymbol{V}$ a
	vector. The wedge product is, for the contangent space at a point, the exterior product, fully antisymmetric. For differential forms it is extended from its local definition.
	
	In the main text we used the fact that the dynamical constraints of interest for the Lagrangian in Eq. \eqref{eq:BoxesLagrangian} can be obtained as part of the Euler--Lagrange equation to slave $\bQ_C$ to the variables for the TLs. In fact, we could have argued in terms of the zero modes of the two-form  in Eq. \eqref{eq:precan_two-form_CLY}, as we now show. We need to extend the duality relation Eq. \eqref{eq:dualbasis} to the functional context, 
	\begin{align}
		\label{eq:functionalduality}
		\delta z(\xi)\left[\frac{\partial}{\partial z(\zeta)}\right]= \delta(\xi-\zeta)\,,
	\end{align}
	using Dirac's delta. Remember that we are using the notation $\delta z(\xi)$ instead of $\dd z(\xi)$ to avoid confusion with the integration measure. Let us now understand the action of the integral part of a TL two-form,
	\begin{align}
		\omega_c= \int_{\mathbbm{R}^+}\dd x\,\delta Q'(x)\wedge\delta\Phi(x)\,,
	\end{align}
	on vectors of shape
	\begin{align}
		\boldsymbol{W}= \int_{\mathbbm{R}^+}\dd \xi\,W(\xi)\frac{\delta}{\delta Q(\xi)}\,.
	\end{align}
	The simplest way to do so is by using integration by parts to write
	\begin{align}
		\omega_c = -\delta Q(0)\wedge\delta\Phi(0)-\int_{\mathbbm{R}^+}\dd \,\delta Q(x)\wedge\delta\Phi'(x)\,,
	\end{align}
	and then apply it to $\boldsymbol{W}$ using linear extension as in Eq. \eqref{eq:vectorcontraction} to obtain
	\begin{align}
		\omega_c\left(\boldsymbol{W},\cdot\right) &= -W(0)\delta\Phi(0)-\int_{\mathbbm{R}^+}\dd x\,W(x)\delta\Phi'(x)\nonumber\\
		&=\int_{\mathbbm{R}^+}\dd x\,W'(x)\delta\Phi(x)\,,
	\end{align}
	where the last step is achieved by  again integrating by parts. Looking now at the two-form in Eq. \eqref{eq:twoformcont} we see directly that indeed \eqref{eq:functionalzeromode} is the zero mode.
	
	\subsection{Topological characterization for a family of 2-port networks}
	\label{subsec:top-TL-app}
	\begin{figure}
		\centering   \includegraphics[width=0.9
		\linewidth]{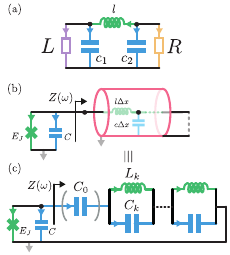}
		\caption{(a) The discretization of the transmission line involves a telescopic circuit composed of shunt capacitors and series inductors, which can be connected to external elements on the left and right sides. (b) A TL terminated on one side by a Josephson junction (with its parallel parasitic capacitance) and on the other side by either a short (solid) circuit or an open (dashed) circuit. (c) A canonical Foster fraction expansion of the circuit in (b) includes a pole at zero frequency and an infinite series of poles at harmonics, $Z(\omega) = \frac{i}{\omega C_0} + \dots $, represented by a series capacitance $C_0$ for the open-ended case (and $\lambda/2$ modes), or the absence of it for the short-circuited case ($\lambda/4$). The topology of the reduced dynamical manifold, upon imposing the topological ansatz, differs depending on whether the circuit is ended in short or open circuit, no matter whether it is modeled using (b) the field theory or (c) a mode-discretized version of it.
		}\label{fig:Topological_ansatz_terminated_TLs_Foster}
	\end{figure}
	\subsubsection{A pedagogical example}\label{sec:pedagogical}
	Consider a simple two-port element connected to two one-port elements as depicted in Fig.~\ref{fig:Topological_ansatz_terminated_TLs_Foster}(a). We want to understand the consequences of the topological ansatz for this example. The KCL/KVL system is readily computed to be
	\begin{subequations}
		\begin{align}
			\mathrm{d}\phi_{2}^c&=\mathrm{d}\phi_1^c+\mathrm{d}\phi^l\,,\label{eq:pedag_KVL_int}\\
			\mathrm{d}\phi_1^c&=\mathrm{d}\phi_L\,,\\
			\mathrm{d}\phi_2^c&=\mathrm{d}\phi_R\,,\\
			\mathrm{d}q^l&=\mathrm{d}q_1^c+\mathrm{d}q_L=-\mathrm{d}q_2^c-\mathrm{d}q^l\,.
		\end{align}
	\end{subequations}
	It is important to notice that, no matter the topological assignments to the left and right dipoles, there are restrictions on the topology of the integrated solution.
	Observe that Eq.~\eqref{eq:pedag_KVL_int} constraints together 2 compact and one 
	extended fluxes. The integral manifold of this constraint has the topology $S^1\times\mathbb{R}$.  One can solve in terms of $\phi_1^c$ ($S^1$) and $\phi^l$ ($\mathbb{R}$). Notice however that next we have to impose the other two KVL Pfaff equations. If both $\phi_L$ and $\phi_R$ were $S^1$, then we have indeed that the topology can be ensured to be $S^1\times\mathbb{R}$ at this point. If both are $\mathbb{R}$, then the integral manifold  is $\mathbb{R}\times\mathbb{R}$. The issue arises if one is compact and the other one extended. Without loss of generality, assume that the extended one is $\phi_L$. Then the integral manifold for the complete external Pfaff system is  $\mathbb{R}\times\mathbb{R}$. A similar analysis can be carried out for the KCL. 
	
	The conclusion we draw from this simple exercise is that we cannot have an unequivocal topological assignment for the port fluxes and charges for this simple two-port, that is neither purely capacitive nor purely inductive, and furthermore that coupling one of the ports to a one-port with inductive topology forces a completely extended flux set, independently of how the other port is connected: there are topological implications for one end arising from couplings at the other end.
	
	This conclusion is immediately applicable to a finite length transmission line, as depicted in Fig.~\ref{fig:Topological_ansatz_terminated_TLs_Foster}(b). After all, it is a two-port element with both inductive and capacitive properties. If it is not connected at both ends with compact flux dipoles the description will be in terms of extended fluxes.  In the following subsection we shall analyze two cases of one-port elements that result from the two-port transmission line, by imposing conditions on one of the ports.
	\subsubsection{Finite length Transmission Lines and topological characterization}
	Let us now consider finite discretized TLs, coupled on one end to a nonlinear
	oscillator involving a Josephson junction, and on the other presenting one of
	two possible conditions. Either it is terminated in a capacitor or in an
	inductor.
	
	Let us tackle first the capacitor ending case. We take $N$ capacitors with
	$c\Delta x$ capacity and $N$ inductors with $\ell\Delta x$ inductance. The
	line is terminated in the other end with a Josephson junction, with both
	inductive and capacitive branches. Thus the total number of branches is
	$2N+2$. We have to solve the KCL/KVL constraint system, having regard for the
	topological assignments. The leftmost loop informs us that
	$\mathrm{d}\phi_C=\mathrm{d}\phi_J$, such that the torus is reduced to one
	$S^1$. We next have to consider
	$\mathrm{d}\phi_J+\mathrm{d}\phi_0^l=\mathrm{d}\phi_1^c$. Here two variables,
	$\phi_J$ and $\phi_1^c$,
	are $S^1$, while $\phi_0^l$ is extended. In the reduction we have one $S^1$
	and one $\mathbb{R}$ variables. We can choose to describe the solution in
	terms of $\phi_J$ (with two patches, as it is $S^1$) and $\phi_0^l$, which is extended. This telescopes out in the next loops, with
	Pfaff equations
	$\mathrm{d}\phi_{k-1}^c+\mathrm{d}\phi_{k-1}^l=\mathrm{d}\phi_{k}^c$ for
	$N\geq k>1$, from which we observe that we have to add an additional
	$\phi_{k-1}^l$ (extended) for each step. All in all, we see that the integral
	manifold for fluxes is $S^1\times\mathbb{R}^N$, with possible coordinates
	$\phi_J$ and the inductors' branch fluxes.
	
	As to the KCL, 
	the capping equations are $\mathrm{d}q_J+\mathrm{d}q_C=\mathrm{d}q_0^l$ and
	$\mathrm{d}q_N^c=-\mathrm{d}q_{N-1}^l$, while in the bulk the conservation of
	charge reads $\mathrm{d}q_k^c+\mathrm{d}q_{k-1}^l=\mathrm{d}q_k^l$. Starting
	from the rightmost end, we see that we need an extended variable. The next one
	involves two $S^1$ and one extended, a priori, but one of the compact ones,
	$q_{N-1}^l$ in fact, must be expressed through an extended variable, so we
	have to introduce a new extended parameter. This telescopes out until the
	leftmost node, for which the constraint reads
	$\mathrm{d}q_0^l=\mathrm{d}q_J+\mathrm{d}q_C$. The two outer branch charges
	are extended, and the one for the inductor has to be expressed in terms of
	extended variables because of the previous analysis. Therefore, we are left
	with $N+1$ extended charge variables. These can be the loop charges and $q_J$. The exterior system is correspondingly solved by 
	\begin{subequations}
		\begin{align}
			\mathrm{d}q_k^c&=\mathrm{d}Q_k-\mathrm{d}Q_{k-1}\,,\qquad N> k\geq 1\,,\label{eq:qcdiffloops}\\
			\mathrm{d}q_N^c&=-\mathrm{d}Q_{N-1}\,,\label{eq:qcN}\\
			\mathrm{d}q_k^l&=\mathrm{d}Q_k\,,\qquad n\geq k\geq 1\,,\label{eq:qlequalqloop}\\
			\mathrm{d}q_C&=\mathrm{d}Q_0-\mathrm{d}q_J\,.
		\end{align}
	\end{subequations}
	Observe that in spite of the identity of differentials in Eq.~\eqref{eq:qlequalqloop}, the loop charges are extended variables, as is $q_J$. On the other hand, in what regards fluxes, we have the solution of the exterior system in the form
	\begin{subequations}
		\begin{align}
			\mathrm{d}\phi_C&=\mathrm{d}\phi_J\,,\\
			\mathrm{d}\phi_k^c&=\mathrm{d}\phi_J+\sum_{n=0}^{k-1}\mathrm{d}\phi_n^l\,,\qquad N\geq k\geq1\,.
		\end{align}
	\end{subequations}
	Here $\phi_J$ is compact and the inductance branch fluxes are extended. Making use of the solution of the exterior systems, and integrating Eqs.~\eqref{eq:qcdiffloops} and \eqref{eq:qcN} uncontroversially, we obtain the pair of symplectic form and Hamiltonian
	\begin{align}
		\omega&=\mathrm{d}\phi_J\wedge\mathrm{d}Q_J+\sum_{k=0}^{N-1}\mathrm{d}\phi_k^l\wedge\mathrm{d}Q_k\,,\\
		H&= -E_J\cos\left(\frac{2\pi\phi_J}{\Phi_Q}\right)+\frac{\left(q_J-Q_0\right)^2}{2C}\nonumber\\
		&\quad +\sum_{k=0}^{N-1}\frac{\left(\phi_k^l\right)^2}{2\ell\Delta x}+\sum_{k=1}^N\frac{\left(Q_k-Q_{k-1}\right)^2}{2c\Delta x}\nonumber\\
		&\qquad +\frac{Q_{N-1}^2}{2c\Delta x}\,.
	\end{align}
	It is immediate to observe that the symplectic form is canonical and nondegenerate. The coupling between the external nonlinar oscillator and the discretized line is capacitive. 
	
	So far we have only studied the discretized model. Let us now take the limit $N\to\infty$ and $\Delta x\to0$ while keeping $L=N\Delta x$ constant. To do so we introduce interpolating functions $Q(x)$ and $\Phi'(x)$, such that 
	\begin{align}
		Q_k &= Q(k\Delta x)\,,\\
		\phi_k^l&= \Delta x\, \Phi'(k\Delta x)\,.
	\end{align}
	The last term in the Hamiltonian will diverge unless $Q(L)=0$. Going back to the discretized model, this introduces a zero mode in the two-form, namely $\partial/\partial\phi_{N-1}^l$, whence a dynamical constraint appears, $\phi_{N-1}^l=0$, which in turn corresponds to $\Phi'(L)=0$ in the continuum limit. Thus, together with these boundary conditions, we obtain in the continuum limit
	\begin{align}
		\omega&= \mathrm{d}\phi_J\wedge\mathrm{d}q_J+\int_0^L\mathrm{d}x\,\delta\Phi'(x)\wedge\delta Q(x)\,,\\
		H&= -E_J\cos\left(\frac{2\pi\phi_J}{\Phi_Q}\right)+\frac{\left[q_J-Q(0)\right]^2}{2C}\nonumber\\
		&\qquad+\int_0^L\mathrm{d}x\,\left[\frac{1}{2\ell}\left(\Phi'\right)^2+\frac{1}{2c}\left(Q'\right)^2\right]\,.
	\end{align}
	Importantly, $\phi_J$ is a compact variable in these expressions. In order to use the formalism presented later in the main text some manipulations are warranted. Notice, for example, the existence of a gauge symmetry, given by the global shift of $\Phi(x)$ by a constant. Additionally, it might be convenient to rewrite the two form in a symmetric form. We leave these tasks to the initiative of the reader, as the thrust of this analysis is the compatibility of the topological assignments with 
	the symplectic description of transmission lines.
	
	We now address a discretization terminated with an inductor. We will have $N$ line capacitors, $N+1$ line inductors, and two branches for the Josephson junction. Altogether we are presented with $2N+3$ branches, and odd number, and after solving the KCL/KVL constraints we will be ineluctably led to a degenerate two-form. Let us analyze the topology of the integral flux manifold first. The leftmost loop provides us with the constraint $\mathrm{d}\phi_N^c=-\mathrm{d}\phi_N^l$. As this relates one compact and one extended direction, the solution will be parameterized with an extended variable. As one moves to the right, with consecutive constraints $\mathrm{d}\phi_k^c+\mathrm{d}\phi_k^l=\mathrm{d}\phi_{k+1}^c$, we see that the description of fluxes must be in terms of real variables. The final loop gives us $\mathrm{d}\phi_J=\mathrm{d}\phi_1^c-\mathrm{d}\phi_0^l$, so there is no compact direction in the integral flux manifold. In particular, we can parameterize the integral manifold with $N+1$ node fluxes, all of those extended. We shall use the notation $\Phi_J$ for the Josephson junction node flux, and $\Phi_k$ for the others. As to charges, we shall again use loop charges, which are extended variables, with analogous notation. Putting the solution of the exterior system in terms of these variables and integrating simply, we arrive at the two-form and Hamiltonian of the form
	\begin{align}
		\omega&= \mathrm{d}\Phi_J\wedge\left(\mathrm{d}Q_J-\mathrm{d}Q_0\right)+\sum_{k=1}^N\mathrm{d}\Phi_k\wedge\left(\mathrm{d}Q_{k-1}-\mathrm{d}Q_k\right)\,,\\
		H&= -E_J\cos\left(\frac{2\pi\phi_J}{\Phi_Q}\right)+\frac{\left(Q_J-Q_0\right)^2}{2C}\nonumber\\
		&\quad +\sum_{k=0}^{N-1}\frac{\left(\Phi_{k+1}-\Phi_k\right)^2}{2\ell\Delta x}+\sum_{k=1}^N\frac{\left(Q_k-Q_{k-1}\right)^2}{2c\Delta x}\nonumber\\
		&\qquad +\frac{\left(\Phi_1-\Phi_J\right)^2}{2\ell\Delta x}\,.
	\end{align}
	As mentioned earlier, $\omega$ is degenerate. A zero-mode is straightforwardly computed to be 
	\begin{align}
		\boldsymbol{W}_Q=\frac{\partial }{\partial Q_J}+\sum_{k=0}^N\frac{\partial }{\partial Q_k}\,.
	\end{align}
	This is actually a gauge mode, as $\boldsymbol{W}_Q(H)=0$.
	
	Regarding the continuum limit, the last term of the Hamiltonian must be kept in check as $\Delta x\to0$, which imposes the boundary condition $\Phi(0)=\Phi_J$.  On the other hand, we can use the gauge freedom in charges to shift all by $Q_J$, resulting in the pair
	\begin{align}
		\omega&= \mathrm{d}\Phi(0)\wedge\mathrm{d}Q(0)+\int_0^L\mathrm{d}x\,\delta\Phi(x)\wedge\delta Q'(x)\,,\\
		H&=-E_J\cos\left(\frac{2\pi\Phi(0)}{\Phi_Q}\right)+\frac{Q(0)^2}{2C}\nonumber\\
		&\qquad +\int_0^L\mathrm{d}x\,\left[\frac{1}{2\ell}\left(\Phi'\right)^2+\frac{1}{2c}\left(Q'\right)^2\right]\,.
	\end{align}
	Again, we could further treat this example by bringing the symplectic form to a more symmetric presentation and thus amenable to the second formalism we present here.
	
	In conclusion to these two examples, where we have examined the impact of the topology assignment on the symplectic formalism for TLs, we observe that the topological character of the final Hamiltonian description of the system is the result of the complete circuit, including in our example the open or closed boundary conditions of the finite length line. The topological character of the flux entering the inductive part of the Josephson junction is different in those two cases. It is to be noted that this conclusion is not predicated on a particular choice of coordinates to describe the system, but is rather a geometrical assertion. Clearly it is not a property simply inherent to the Josephson junction on its own, as its determination also requires knowledge about the circuit in regions possibly very far away, physically, from it. The composition of systems by tensoring or Cartesian multiplying the kinematics and adding Hamiltonians plus an interaction hamiltonian term is not the correct perspective because the connection is carried out through constraints in this circuital context. This is clearly to be seen in the topological aspects we have emphasized, of course, but also in the rather distinct Hamiltonians we have obtained.
	\subsubsection{First canonical Foster form}
	As the result above might not be intuitive for some readers, let us look at it from a different perspective. Namely, that of canonical Foster forms \cite{Foster:1924}. The finite length line with boundary conditions on one end is a one port , with impedance function $Z(s)$. The two cases have different $Z(s)$, of course. We can now look at the canonical Foster form of these impedances, whereby a lumped element circuit of canonical form provides us with the same impedance function. This is a sequence of oscillators, possibly prefaced by an inductor and a capacitor, as depicted in Fig.~\ref{fig:Topological_ansatz_terminated_TLs_Foster}(c). Crucially for our purposes, the first case considered above has a canonical form with a leading capacitor and a sequence of LC oscillators, while the second one has no leading capacitor. Let us now consider a lumped element circuit consisting of the parallel nonlinear oscillator given by the Josephson junction connected in series with a capacitor $C_0$ and a sequence of parallel LC oscillators, numbered with an index $k$. The KCL/KVL Pfaff constraints are immediate,
	\begin{align}
		\mathrm{d}\phi_k^c&=\mathrm{d}\phi_k^l\,,\\
		\mathrm{d}\phi_C&=\mathrm{d}\phi_J\,,\\
		\mathrm{d}\phi_{C0}&=-\mathrm{d}\phi_J-\sum_k\mathrm{d}\phi_k^l\,,\\
		\mathrm{d}q_k^l&=\mathrm{d}q_{C0}-\mathrm{d}q_k^c\,,\\
		\mathrm{d}q_C&=\mathrm{d}q_{C0}-\mathrm{d}q_J\,.
	\end{align}
	It is immediate to see that, with the topological assignments presented above, the integral manifold will be described by one compact variable ($\phi_J$, for example) while all other directions will be extended. If the number of LC oscillators were finite, $N$, say, we will have $N+2$ charge type variables and $N+1$ which are flux type. We see that the two-form will necessarily be degenerate. Using as variables $\phi_J$,  $q_J$, $Q_k$ such that $\mathrm{d}Q_k=\mathrm{d}q_k^l$, and $\phi_k^l$, together with $q_{C0}$, we have that the two form is 
	\begin{align}
		\omega&=\mathrm{d}\phi_J\wedge\mathrm{d}q_J+\sum_k\mathrm{d}\phi_k\wedge\mathrm{d}Q_k\,.
	\end{align}
	We readily identify the zero mode vector $\boldsymbol{W}_Q=\partial/\partial q_{C0}$. The dynamical constraint is not trivial, 
	\begin{align}
		\left[\frac{1}{C}+\frac{1}{C_0}+\sum_k\frac{1}{C_k}\right]q_{C0}=\frac{q_j}{C}+\sum_k \frac{Q_k}{C_k}\,,
	\end{align}
	and introduces a coupling after reduction. We note that techniques exist to achieve further simplification for important cases such as the finite-length transmission lines under consideration here (\cite{ParraRodriguezPhD:2021}, Sec. 3.3). A possible algebraic alternative is to compute the impedance of the one port given by the conductive part of the Josephson junction in parallel with the original $Z(s)$, and then computing its Foster first canonical form. This will simplify the Hamiltonian of the full system, while providing us with a canonical presentation of the symplectic form.
	
	On the other hand, if the leading capacitor were not present then the integrable manifold would have no compact direction. We see that the geometric analysis of the Foster forms of finite length transmission lines coupled to Josephson junctions is compatible with their discretization along the lines of the telegrapher's equation, also in what concerns the topological assignments.
	\subsection{Differential operator $\mcL$ for the semi-infinite line}
	\label{sec:L_op_app}
	Here we provide a summary of the results obtained in the App. D in~\cite{ParraRodriguezPhD:2021} on the differential operator $\mcL$ for the boundary problem involving capacitors, inductors and ideal NR elements (${\msA}{\msB}{\msG}$) on one end ($x=0$ in all TLs), used for finding a complete basis on which to expand the charge and flux fields. Here we will assume that $\msA$ and $\msB$ are full-rank matrices. We refer the interested reader to that Appendix to see extra details and proofs for cases where such an assumption is not fulfilled.
	\subsubsection{Self-adjointness and positivity}
	Most notably, the $\mcL$ operators used in this manuscript are positive and self-adjoint. We restrict ourselves to the semi-infinite line case, with one boundary. Natural extensions with several of these boundaries have already been explored ~\cite{ParraRodriguez:2018,ParraRodriguezPhD:2021}. 
	
	{\bf Assertion 1:}
	\emph{Let $\msA$, $\msB$, $\mDelta$ be real positive-definite (full-rank) $N\times N$ symmetric matrices, and $\msG$ an $N\times N$ real skew-symmetric matrix. The differential operator $\mcL$ defined on its domain,  
		\begin{align}
			\mcl{D}(\mcL)=&\left\{\left(\bW\equiv\begin{pmatrix}\bU\\\bV\end{pmatrix},\bw\right), \bmW\in\mathcal{H}, \mathcal{L}\bmW\in\mathcal{H}\right.\nonumber\\ 
			&\qquad \bW,\bW'\in AC^1(\mathbbm{R}^+)\otimes\mathbbm{C}^{2N},\\
			&\quad\left.\bw=\begin{pmatrix}
				\msA \bU\\\bV-\msA \bV' - \msG \bU
			\end{pmatrix}_0\in\mathbbm{C}^{2N}\right\},\nonumber
		\end{align}
		and action 
		\begin{align}
			\mcL \bmW &=\left(-\mDelta\bW'',\tilde{\bw}=\begin{pmatrix}
				\msB^{-1}\bU+\msG \bV'-\mDelta\bU'\\-\msB^{-1}\bV'
			\end{pmatrix}_0\right),\label{eq:L_op_appendix_full}
		\end{align}
		is self-adjoint with respect to the inner product 
		\begin{align}
			\langle \bmW_1,\bmW_2\rangle&=\int_{\mathbbm{R}^+}dx \,\bW_1^\dag \mSigma \bW_2 +\bw_1^\dag\mGamma\bw_2,
	\end{align}}
	with $\mSigma=\text{diag}(\mone,\mDelta^{-1})$, $\mGamma=\text{diag}(\msA^{-1},\msB)$\textit{, and }$\{\bU_0,\bV_0\}\equiv \{\bU(0),\bV(0)\}$.
	
	{\bf Proof:}
	For $\bmW_1$ to be in $\mcl{D}(\mcL^\dag)$, there must exist a $\bmW_3$ such that \begin{align}
		r=&\langle \bmW_1,\mcL\bmW_2\rangle-\langle\bmW_3,\bmW_2\rangle=0.
	\end{align}
	holds $\forall\bmW_2\in\mcl{D}(\mcL)$, in which case one defines $\bmW_3=\mcL^\dag\bmW_1$. By integration by parts  
	\begin{align}
		\begin{aligned}
			r=&\left(\bW_1^\dag \mSigma\mDelta\bW_2' -(\bW_1^\dag)' \mDelta\mSigma\bW_2\right)_0&\\
			&-\bw_1^\dag\mGamma \begin{pmatrix}
				\mDelta\bU_2'-\msB^{-1}\bU_2-\msG \bV_2'\\\msB^{-1}\bV_2'
			\end{pmatrix}_0&\\
			&-\bw_3^\dag \mGamma\begin{pmatrix}
				\msA \bU_2\\\bV_2-\msA \bV_2' - \msG \bU_2
			\end{pmatrix}_0,&
		\end{aligned}
	\end{align}
	where $\bw_1^\dag=\begin{pmatrix}
		\bsb{a}^\dag&\bsb{b}^\dag
	\end{pmatrix}$, and $\bw_3^\dag=\begin{pmatrix}
		\bsb{c}^\dag&\bsb{d}^\dag 
	\end{pmatrix}$. Explicitly, the four equations for the action and domain of the adjoint operator are
	\begin{align}
		\begin{aligned}
			\bsb{a}^\dag&=\bU_1^\dag(0)\msA ,\\
			\bsb{d}^\dag&=-(\bV_1^\dag(0))'\msB^{-1},\\
			\bsb{b}^\dag&=(\bV_1^\dag(0) +\bsb{a}^\dag\msA^{-1}\msG+\bsb{d}^\dag\msB\msA)\\
			&=\bV_1^\dag(0)+\bU_1^\dag(0)\msG-(\bV_1^\dag)'\msA,\\
			\bsb{c}^\dag&=\bsb{a}^\dag\msA^{-1}\msB^{-1}-(\bU_1^\dag)'(0)\mDelta+\bsb{d}^\dag\msB\msG\\
			&=(\bU_1^\dag)(0)\msB^{-1}-(\bU_1^\dag)'(0)\mDelta-(\bV_1^\dag)'(0)\msG,
		\end{aligned}
	\end{align}
	whose unique solution corresponds to $\bw_1\in\mcl{D}(\mcL)$ and $\bw_3=\tilde{\bw}_1$, that is to say, the same domain and action of the original operator.
	
	Additionally, by again employing integration by parts, it can be verified that $\mcL$ is a monotone (accretive) operator,
	\begin{align}
		\begin{aligned}
			\langle \bmW, \mcL\bmW\rangle=&\int_{\mathbbm{R}^+} dx\,(\bW^\dag)'\mSigma\mDelta\bW'\nonumber\\&+\bU_0^\dag\msB^{-1}\bU_0+(\bV_0^\dag)'\msA \bV_0'\geq0    
		\end{aligned}
	\end{align}
	$\forall \bmW\in \mcl{D}(\mcL)$ as $\mSigma\mDelta$, $\msB^{-1}$ and $\msA$ are real positive symmetric matrices.

	\subsubsection{Eigenbasis}
	A complete orthogonal basis for the flux and charge fields can be constructed from the spectral decomposition of the previous self-adjoint operator, whose domain is dense in the Hilbert space $\mcl{H}=\left[L^2(\mathbbm{R}^+)\otimes\mathbb{C}_{\mSigma}^{2N}\right]\oplus\mathbb{C}_{\msA^{-1}}^{N}\oplus\mathbb{C}_{\msB}^{N}$. We have denoted $\Omega$ as the continuous frequency parameter selecting a degenerate subspace and $\lambda$ the inner index. Using the boundary equations for the eigenvalue problem $\mcL\bmW_{\Omega\lambda}=\Omega^2\bmW_{\Omega\lambda}$, the eigenvectors forming the basis for the later conjugate pairs of coordinates are $\bmW_{\Omega\lambda}^{F,G}=(\bW^{F,G},\bw^{F,G})_{\Omega\lambda}$, where  
	\begin{widetext}
		\begin{align}
			\label{eq:basis_L_ABG}
			\begin{aligned}
				\bW_{\Omega\lambda}^F(x)=&\sqrt{\frac{2}{\pi m_\lambda}}\left[\cos\left(\Omega x\mDelta^{-\frac{1}{2}}\right)\begin{pmatrix}
					\mDelta^{-\frac{1}{2}}\be\\ \mDelta^{\frac{1}{2}}(\tilde{\msG}\be-(\Omega\tilde{\msE})^{-1}\br)
				\end{pmatrix} + \sin\left(\Omega x\mDelta^{-\frac{1}{2}}\right)\begin{pmatrix}
					\tilde{\msG}\br+(\Omega\tilde{\msE})^{-1}\be\\ \br
				\end{pmatrix}\right]_{\Omega\lambda},\\
				\bW_{\Omega\lambda}^G(x)=&\sqrt{\frac{2}{\pi m_\lambda}}\left[\cos\left(\Omega x\mDelta^{-\frac{1}{2}}\right)\begin{pmatrix}
					-\mDelta^{-\frac{1}{2}}\br\\ -\mDelta^{\frac{1}{2}}(\tilde{\msG}\br+(\Omega\tilde{\msE})^{-1}\be)
				\end{pmatrix}+ \sin\left(\Omega x\mDelta^{-\frac{1}{2}}\right)\begin{pmatrix}
					\tilde{\msG}\be-(\Omega\tilde{\msE})^{-1}\br\\ \be
				\end{pmatrix}\right]_{\Omega\lambda},
			\end{aligned}
		\end{align}
	\end{widetext}
	and $\bw^{F,G}$ are constructed from the corresponding values of $\bW(x)$ at the end(s) ($x=0$), following the structure in $\mcl{D}(\mcL)$, see further examples in~\cite{ParraRodriguezPhD:2021}. Here, $\tilde{\msE}^{-1}=\tilde{\msB}^{-1}-\Omega^2\tilde{\msA}$,  where $\tilde{\msA}=\mDelta^{-\frac{1}{2}}\msA\mDelta^{-\frac{1}{2}}$, $\tilde{\msB}^{-1}=\mDelta^{-\frac{1}{2}}\msB^{-1}\mDelta^{-\frac{1}{2}}$, and $\tilde{\msG}=\mDelta^{-\frac{1}{2}}\msG\mDelta^{-\frac{1}{2}}$. $m_\lambda$ are the double degenerate eigenvalues (with eigenvectors $(\be^T,\br^T)_{\Omega\lambda}$ and $(-\br^T,\be^T)_{\Omega\lambda}$) of the matrix
	\begin{align}
		\msM_{\msN}&=\begin{pmatrix}
			\msM&\msN\\-\msN&\msM
		\end{pmatrix}=\msM_{\msN}^T,
	\end{align}
	with 
	\begin{align}
		\msM&= \mDelta^{-\frac{1}{2}}+ \frac{1}{\Omega^2}\tilde{\msE}^{-1}\mDelta^{\frac{1}{2}}\tilde{\msE}^{-1}+\tilde{\msG}^T\mDelta^{\frac{1}{2}}\tilde{\msG},\nonumber\\
		\msN&=(\Omega\tilde{\msE})^{-1}\mDelta^{\frac{1}{2}}\tilde{\msG}-\tilde{\msG}^T \mDelta^{\frac{1}{2}}(\Omega\tilde{\msE})^{-1} ,\nonumber
	\end{align}
	such that the orthonormal basis has relations
	\begin{align}
		\begin{aligned}
			\langle \bmW_{\Omega\lambda}^F,\bmW_{\Omega'\lambda'}^F\rangle&=\frac{\delta_{\Omega \Omega'}}{\sqrt{m_{\lambda}m_{\lambda'}}}\begin{pmatrix}
				\be\\ \br
			\end{pmatrix}_{\Omega\lambda}^T\mcM\begin{pmatrix}
				\be\\ \br
			\end{pmatrix}_{\Omega\lambda'}\\
			&=\delta_{\Omega \Omega'}\delta_{\lambda \lambda'},\\
			\langle \bmW_{\Omega\lambda}^G,\bmW_{\Omega'\lambda'}^G\rangle&=\frac{\delta_{\Omega \Omega'}}{\sqrt{m_{\lambda}m_{\lambda'}}}\begin{pmatrix}
				-\br\\ \be
			\end{pmatrix}_{\Omega\lambda}^T\mcM\begin{pmatrix}
				-\br\\ \be
			\end{pmatrix}_{\Omega\lambda'}\\
			&=\delta_{\Omega \Omega'}\delta_{\lambda \lambda'}.\nonumber
		\end{aligned}
	\end{align}
	
	It can be easily shown that this basis for the fields $\bmW^{F,G}_{\Omega\lambda}$ satisfies the properties $\mcT \bmW^{F,G}_{\Omega\lambda}=\pm i \Omega \bmW^{G,F}$.
	
	\subsubsection{Integral identities}\label{sec:integral_identities}
	In the simplification of the expressions of Secs.~\ref{subsec:TLs_NR_BB_NL} and ~\ref{sec:quasi_lumped_circuit_examples} we are making use of the following exact integral identities
	\begin{align}\label{eq:A_sum_rule}
		\msA^{-1}&=\int_{\mathbbm{R}^+}\dd \Omega \bU_{\Omega\epsilon}(0)\bU_{\Omega\epsilon}^T(0)\\
		&=\int_{\mathbbm{R}^+}\dd \Omega \left(\bU_{\Omega\lambda0}^F(\bU_{\Omega\lambda0}^F)^T+\bU_{\Omega\lambda0}^G(\bU_{\Omega\lambda0}^G)^T\right), \nonumber
	\end{align}
	and 
	\begin{align}
		&\msB=\int_{\mathbbm{R}^+}\dd \Omega \frac{\bV_{\Omega\epsilon}'(0)(\bV_{\Omega\epsilon}')^T(0)}{\Omega^4}\\
		&=\int_{\mathbbm{R}^+}\dd \Omega \frac{(\bV_{\Omega\lambda0}^F)'((\bV_{\Omega\lambda0}^F)')^T+(\bV_{\Omega\lambda0}^G)'((\bV_{\Omega\lambda0}^G)')^T}{\Omega^4},\nonumber
	\end{align}
	which can be obtained by expanding elements of the Hilbert space with null value on the open interval $\mathbbm{R}^+$, readily, 
	\begin{align}
		\bmW^{(0,1_i)}=\left(\begin{pmatrix}
			\bU\\\bV
		\end{pmatrix}(x)=\begin{pmatrix}
			0\\0
		\end{pmatrix},\begin{pmatrix}
			0\\\vdots\\1_i\\\vdots\\0
		\end{pmatrix}\right),
	\end{align}
	on a complete basis, and use identities from the eigenvalue problem~\cite{ParraRodriguezPhD:2021}. We note that the above integral identities generalize Theorem 2 in Ref.~\cite{Walter:1973}.
	\subsection{Duality operator $\mcT$ for the semi-infinite line}
	\label{sec:T_op_app}
	In this subsection we offer additional insights into the \emph{duality} operator $\mcT$ presented in Subsec.~\ref{sec:quadr-hamilt-diag}, which is a modified version to that previously defined in~\cite{ParraRodriguezPhD:2021}. 
	
	{\bf Assertion 2:}
	\textit{Let $\msA$, $\msB$, $\mDelta$ be real positive-definite (full-rank) $N\times N$ symmetric matrices, and $\msG$ an $N\times N$ real skew-symmetric matrix. The differential operator $\mcT$ defined by its domain and action
		\begin{align}
			\label{eq:topdef}
			\mcl{D}(\mcT)&=\left\{\left(\bW\equiv\begin{pmatrix}
				\bU\\ \bV
			\end{pmatrix},\bw\right)\in \mcl{H}, \mathcal{T}\bmW\in\mathcal{H}, \right.\nonumber\\
			&\left.\bW\in AC^1(I),\bw=\begin{pmatrix}
				\msA \bU\\
				\bV-\msA\bV'-\msG\bU
			\end{pmatrix}_0
			\right\},\nonumber\\
			\mcT \bmW&= \left(\begin{pmatrix}
				-i\bV'\\-i\mDelta\bU'
			\end{pmatrix},\begin{pmatrix}
				-i\msA \bV'_0\\
				-i\msB^{-1}\bU_0
			\end{pmatrix}\right),
		\end{align}
		is essentially self-adjoint.}
	
	{\bf Proof:}
	The fastest way to argue that this is not a self-adjoint operator is to observe that by integration by parts only $\bU_0$
	and $\bV_0$ will appear, as it is a first order operator, and in
	the domain and the action only those two and $\bV_0'$. Therefore, there are only three vectorial equations, and we need four to fully fix domain and action of $\mcT^\dag$ without ambiguity. 
	
	On the other hand, it is easy to show that is a symmetric (hermitian) operator, and we can study its deficiency indices, i.e., the dimension of the eigenspaces of
	$\mcT^\dag$ with eigenvalues in the upper and lower complex
	half-spaces. When one uses adimensionalized operators, as is usual in
	mathematics, one looks at the dimension of the eigenspace of
	eigenvalue $i$ (respectively $-i$), since it is the same
	dimensionality for an eigenvalue $z$ with $\mathrm{Im}(z)>0$
	(resp. $\mathrm{Im}(z)<0$). 
	
	In our case, it will prove convenient to use a dimensional reference, since
	$\left[\mcT\right]=\left[\mDelta\right]^{1/2}/L$, with $L$
	length. As $\mDelta$ is a squared velocity, $\mcT$ has
	dimensions of inverse time, and we shall use $\Omega$ to denote the
	reference frequency. Thus, we want to compute the dimension of the
	eigenspaces $\mcT^\dag\bmW=\pm i \Omega \bmW$.
	
	An obvious first step is to write explicitly $\mcT^\dag$. This is achieved by examining the equations
	\begin{align}
		\label{eq:adjoint}
		\left\langle\bmW_a,\mcT\bmW\right\rangle-\left\langle\bmW_b,\bmW\right\rangle=0\,,
	\end{align}
	since $\bmW_a\in\mcl{D}\left(\mcT^\dag\right)$ if and only if there
	exists $\bmW_b\in\mcl{H}$ such that for all
	$\bmW\in\mcl{D}(\mcT)$ Eq. \eqref{eq:adjoint}
	holds. We then say that
	$\bmW_b=\mcT^\dag\bmW_a$. Carrying out this
	computation for the case at hand provides us with the domain and
	action of $\mcT^\dag$, as follows
	\begin{align}
		\mcl{D}\left(\mcT^\dag\right) &=\left\{\left(\bW\equiv\begin{pmatrix}
			\bU\\ \bV
		\end{pmatrix},\bw\right)\in \mcl{H}, \bW\in AC^1(I),\right.\nonumber\\
		&\left.\qquad\bw=\begin{pmatrix}
			\msA \bU_0\\
			\bw_2
		\end{pmatrix}
		\right\},\label{eq:tdag}\\
		\mcT^\dag\bmW&= \left(
		\begin{pmatrix}
			-i\bV'\\-i\mDelta\bU'
		\end{pmatrix},
		\begin{pmatrix}
			i\bw_2- i \bV_0+ i \msG \bU_0\\
			-i\msB^{-1}\bU_0
		\end{pmatrix}\right).\nonumber
	\end{align}
	Observe that $\bw_2$ is not determined by $\bW(0)$. Further observe
	that  $\mcT^\dag$, restricted to
	$\mcl{D}(\mcT)$, does coincide with $\mcT$, again
	proving that it is a symmetric (hermitian) operator.
	
	We now have to study the eigenvector equation
	\begin{align}
		\label{eq:plusindex}
		\mcT^\dag\bmW= i\Omega \bmW\,,
	\end{align}
	for positive frequencies $\Omega$.
	The system of differential equations, together with normalizability,
	provides us with
	\begin{align}
		\label{eq:soldiffeqs}
		\bU(x) &= e^{-x\Omega\mDelta^{-1/2}}\bU(0)\,,\\
		\bV(x) &= e^{-x\Omega\mDelta^{-1/2}}\mDelta^{1/2}\bU(0)\,,
	\end{align}
	whence $\bV(0)= \mDelta^{1/2}\bU(0)$. Using the discrete equations
	now, we have
	\begin{align}
		\label{eq:w2infirst}
		\bw_2= -\frac{1}{\Omega}\msB^{-1}\bU(0)\,.
	\end{align}
	The remaining linear equation reads
	\begin{align}
		\label{eq:consraint}
		\left(\Omega^2 \msA +\Omega \mDelta^{1/2} -\Omega \msG + \msB^{-1}\right)\bU(0)=0\,.
	\end{align}
	Now, we are assuming $\msA$, $\msB$ and $\mDelta$ symmetric and positive definite, thus full rank. Take $\Omega$ large, so that $\Omega
	a_{\mathrm{min}}$ is larger than the spectral radii of $\mDelta^{1/2}$
	and $\msG$, and additionally $\Omega^2
	a_{\mathrm{min}}>1/b_{\mathrm{min}}$, where $a_{\mathrm{min}}$ and
	$b_{\mathrm{min}}$ are the smallest (positive) eigenvalues of $\msA$
	and $\msB$ respectively. Then the matrix inside the brackets in
	Eq. \eqref{eq:consraint} is full rank. Thus the corresponding index of
	$\mcT$ is zero, making use of the independence of the dimension
	on $\Omega$. The other index is also zero following the same reasoning.
	
	In summary, $\mcT$ defined in Eqs. \eqref{eq:topdef} is
	symmetric, and its deficiency indices are $(0,0)$. In other words, it
	is \emph{essentially self-adjoint} (see App. C2 in
	\cite{Galindo:2012}, for instance). As a consequence, the closure of
	$\mcT$ is self-adjoint.
	
	Let us now turn to the study of the matrix elements of $\mcT$ (and of its closure!) in the eigenbasis of $\mcL$ following the subsequent steps:
	\begin{enumerate}
		\item Use the collective index $\alpha$ to indicate
		$\Omega\epsilon$. We will actually use $\Omega_\alpha$ to indicate
		that part of the multicomponent index. Thus, $\bmW_\alpha$,
		and it holds that
		$\mcL\bmW_\alpha=\Omega^2_\alpha\bmW_\alpha$.
		\item By construction,
		$\bmW_\alpha\in\mcl{D}(\mcT)$. We want to show
		that $\mcT\bmW_\alpha\in \mcl{D}(\mcT)$ as
		well. As to the vector function part there is no issue, since
		$\bW_\alpha$ is actually smooth and normalizable. Thus we are left
		with checking on the boundaries that
		\begin{align}
			\label{eq:chek1}
			&\begin{pmatrix}
				-i\msA \bV'_\alpha(0)\\ -i \mathsf{B}^{-1}\bU_\alpha(0)
			\end{pmatrix}=\\
			&\qquad\qquad\begin{pmatrix}
				-i\msA\left(\bV'_\alpha\right)_0\\
				-i\left(\mDelta\bU_\alpha'-\msA\left(\mDelta\bU_\alpha''\right)-\msG\left(\bV'_\alpha\right)\right)_0
			\end{pmatrix}\,.\nonumber
		\end{align}
		Clearly, the only potential issue appears in the second line. 
		\item First, observe that $-\mDelta
		\bU_\alpha''=\Omega_\alpha^2\bU_\alpha$, and since $\bU_\alpha\in
		AC^1$, this also holds on the boundary.
		\item Next, observe that one of the elements of the eigenequation for
		$\mcL$ reads
		\begin{align}
			\label{eq:eigcomponent}
			\Omega^2_\alpha\bU_\alpha(0)= -\mDelta \bU'_\alpha(0)+\msB^{-1}\bU_\alpha(0)+\msG\bV'_\alpha(0)\,.
		\end{align}
		After substitution of steps 3 and 4 in the second component of the
		RHS of Eq. \eqref{eq:chek1}, we see that indeed
		$\mcT\bmW_\alpha\in \mcl{D}(\mcT)$.
		\item That means that we can compute
		$\mcT^2\bmW_\alpha$, and again making use of $\mDelta
		\bU_\alpha''(0)=\Omega_\alpha^2\bU_\alpha(0)$ we see that
		\begin{align}
			\label{eq:tsquared}
			\mcT^2\bmW_\alpha= \Omega_\alpha^2\bmW_\alpha=\mcL\bmW_\alpha\,.
		\end{align}
	\end{enumerate}
	In summary, we see that, acting on the eigenbasis of $\mcL$,
	$\mcT$ commutes with $\mcL$. Therefore the
	$\omega_{\alpha\beta}$ is block diagonal in that basis, with the
	blocks corresponding to the individual energies.
	
	\section{Details for the nonreciprocal quasi-lumped circuit example}
	\label{sec:Details_circuit_ex2}
	Here we provide the explicit expressions for the differential operators used in the second circuit example, which consists of three TLs connected through a nonreciprocal linear device. Additionally, the first TL is capacitively coupled on the other end to a Josephson junction, whereas the other two are left open (for the sake of concreteness). Thus, the differential operators used to construct a dressed basis of the linear problem requires extra boundaries, i.e., 
	
	\begin{align}
		\mcL\bmW=&\left(-\mDelta\bW'',\right.\nonumber\\
		&\quad\tilde{\bw}_0=\begin{pmatrix}
			-\mDelta\bU'+\msB_0^{-1}\bU\\
			-\msB_0^{-1}\bV'
		\end{pmatrix}_0,\nonumber\\
		&\quad\left.\tilde{w}_d=\begin{pmatrix}
			-\mDelta\bU_d'\\
		\end{pmatrix}^\parallel\right),\nonumber\\
		\mcl{D}(\mcL)=&\bigg\{\left(\bW\equiv\begin{pmatrix}
			\bU\\ \bV
		\end{pmatrix}(x),\bw_0,w_d\right),\label{eq:Lop_A0B0G0_ad}\\
		&\quad\bW\in \text{AC}^1(\mathbbm{R}^+)\otimes\mathbb{C}_{\mSigma}^{2N},\nonumber\\
		&\quad\bw_0=\begin{pmatrix}
			\msA_0 \bU\\
			\bV-\msA_0\bV'
		\end{pmatrix}_0,w_d=-a_d U_d^1,\nonumber\\
		&\,\quad V_d^1=-a_d(V_d^1)',\, (\mDelta\bU_d')^\perp=(\bV_d)^\perp=0  \bigg\},\nonumber
	\end{align}
	respectively, where $\bU_0\equiv\bU(0)$ and $\bV_0\equiv\bV(0)$, and $U_d^1\equiv U^1(d)$. Here, $(\cdot)^\parallel$ ($(\cdot)^\perp$) refers to the components of the inner vector parallel (orthogonal) with respect to $\bsb{n}=(1,0,0)^T$, e.g., $(\bU_d)^\parallel\equiv U^1(d)$. 
	
	In this case, we have considered the Hilbert space $\mcl{H}=\left[L^2(\mathbbm{R}^+)\otimes\mathbb{C}_{\mSigma}^{2N}\right]\oplus\mathbb{C}_{\msA^{-1}}^{N}\oplus\mathbb{C}_{\msB}^{N}\oplus\mathbb{C}$, with elements
	$\bmW=(\bW,\bw_0,w_d)\in\mcl{H}$ and inner product 
	\begin{align}
		\langle\bmW_{1},\bmW_2\rangle=&\int_{\mathbbm{R}^+}\dd x \bW_1^\dag (x)\mSigma\bW_2(x)\\
		&+(\bw_0)_{11}^\dag\msA_0^{-1}(\bw_0)_{21}+(\bw_0)_{12}^\dag\msB_0(\bw_0)_{22}\nonumber\\
		&\,+a_d^{-1}(w_d)_1 (w_d)_2 .\nonumber
	\end{align}
	On the other hand, the duality operator in this case is extended to 
	\begin{align}
		\mcT \bmW=&\left(\begin{pmatrix}
			-i\bV'\\-i\mDelta\bU'
		\end{pmatrix},\check{\bw}_0=\begin{pmatrix}
			-i\msA_0 \bV'\\
			-i\msB_0^{-1}\bU
		\end{pmatrix}_0,\right.\\
		&\,\qquad\qquad\qquad\left.\check{w}_d= i a_d (V_0^1)'\right).\nonumber
	\end{align}
	Following the steps in App.~\ref{sec:L_op_app} and \ref{sec:T_op_app} it is easy to prove that the new operators $\mcL$ and $\mcT$ are self-adjoint and essentially self-adjoint, respectively.  
	
	\section{Details on the construction of multiport nonreciprocal dissipative responses}
	\label{sec:details_dissip_multiport_NR_circuit}
	Here, we will provide a detailed explanation of how to construct the dissipative multiport impedance matrix (the same approach applies to the admittance matrix) by taking the continuum limit of a series of lossless multiport elements.
	
	Each individual non-zero (nor infinite) pole of the Cauer series is associated to a term of the form
	\begin{equation}
		\label{eq:Zk_s}
		\msZ_k(s)= \frac{1}{s^2+\Omega^2_k}\left(s \msA_k+ \msB_k\right)
	\end{equation}
	in the expansion, where $\msA_k$ ($\msB_k$) is a real symmetric (skew-symmetric) matrix, independent of the complex variable $s$. Therefore, the associated contribution to the boundary distribution, $\tilde{\msZ}_k(\omega)$, is readily computed to be
	\begin{align}
		\label{eq:HOcontrib}
		\begin{aligned}
			\tilde{\msZ}_k(\omega)&= \frac{\pi}{2}\left[\delta\left(\omega-\Omega_k\right)+\delta\left(\omega+\Omega_k\right)\right]\msA_k\\
			&\quad +\frac{i\pi}{2\Omega_k}\left[\delta\left(\omega+\Omega_k\right)-\delta\left(\omega-\Omega_k\right)\right]\msB_k\\
			&\quad +i \mcl{P}\frac{\omega}{\Omega_k^2-\omega^2} \msA_k+\mcl{P}\frac{1}{\Omega_k^2-\omega^2}\msB_k\,.
		\end{aligned}
	\end{align}
	$\mcl{P}$ denotes principal part. Remember that the boundary distribution of these matrix functions of $s$ is (distributionally)
	\begin{align}
		\label{eq:boundarydistribution}
		\tilde{\msZ}_k(\omega)= \lim_{\epsilon\to0^+}\msZ_k(-i\omega+\epsilon)\,,
	\end{align}
	with real $\omega$.
	
	The real part of $\tilde{\msZ}_k(\omega)$ in Eq. \eqref{eq:HOcontrib} is an even distribution for the $\omega$ variable, while the imaginary part is odd. Please observe that the hermitian and antihermitian parts are a mix of the real (even) and imaginary (odd) parts. 
	
	The poles at zero and infinity have a different structure, namely
	\begin{align}
		\label{eq:oinfty}
		\tilde{\msZ}_0(\omega)&= \left[\pi \delta(\omega)-i \mcl{P}\frac{1}{\omega}\right] \msA_0\,,\\
		\tilde{\msZ}_\infty(\omega)&=i \omega \msA_\infty\,.
	\end{align}
	The pole at infinity does not contribute to the hermitian part. As to the pole at zero, its analysis will be analogous to the one-port case, see for instance~\cite{Vool:2017}.
	
	The sum $\sum_k \tilde{\msZ}_k(\omega)$ is to be understood, for our purposes, as a Riemann sum for an integral. To that purpose, we understand the matrices $\msA_k$ and $\msB_k$ in our sequence as associated with matrix functions $\msA(\Omega)$ and $\msB(\Omega)$ by
	\begin{align}
		\label{eq:distocont}
		\msA_k= k \Delta\Omega\,\msA(k\Delta\Omega)
	\end{align}
	and correspondingly for $\msB_k$, where $\Delta\Omega$ is the Riemann step that will be taken to zero. Thus we are led to the question of the representability of $\tilde{\msZ}(\omega)$ in terms of two matrix functions, as follows:\begin{widetext}
		\begin{align}
			\label{eq:tildeZw}
			\begin{aligned}
				\tilde{\msZ}(\omega)&= \frac{\pi}{2}\left[\msA(\omega)+\msA(-\omega)\right]+\mcl{P}\int\frac{\dd\Omega}{\Omega^2-\omega^2}\msB(\Omega)-\frac{i\pi}{2\omega}\left[\msB(\omega)+\msB(-\omega)\right]+i\omega\mcl{P}\int\frac{\dd\Omega}{\Omega^2-\omega^2}\msA(\Omega)\,. 
			\end{aligned}
		\end{align}
		
		\subsection*{Application to the circuit example}
		We have as the final target, for the circuit in Fig.~\ref{fig:NR_2port_CaldeiraLeggett}, the impedance matrix
		\begin{align}
			\label{eq:zs}
			\msZ(s)= \frac{1}{\left(sC+Y_0\right)^2+G^2}\left[(sC+Y_0)\mone- G \msJ\right],
		\end{align}
		recalling $\msJ= i\sigma^y$, and $G=1/R$. Now we require the boundary matrix, $\tilde{\msZ}(\omega)=\msZ(-i\omega+0^+)$, and its expression as in Eq.~(\ref{eq:tildeZw}). Given the simple structure of the matrix, we make the simple ans\"atze for the component matrix functions, namely  $\msA(\omega)= a(\omega)\mone$ and $\msB(\omega)=b(\omega)\msJ$, with $a(\omega)$ and $b(\omega)$ scalar functions, which yield

		\begin{subequations} \label{eq:various}
			\begin{align}
				\frac{\pi}{2}\left[a(\omega)+a(-\omega)\right] &= \frac{Y_0\left(Y_0^2+\omega^2C^2+G^2\right)}{\left(Y_0^2+G^2-\omega^2C^2\right)^2+4\omega^2 C^2 Y_0^2}\,,\label{eq:even_aw}\\
				\mcl{P}\int_{-\infty}^\infty\dd\Omega\frac{a(\Omega)}{\Omega^2-\omega^2} &=\frac{C\left(Y_0^2+\omega^2C^2-G^2\right)}{\left(Y_0^2+G^2-\omega^2C^2\right)^2+4\omega^2 C^2 Y_0^2} \,,\label{eq:variousb}\\
				\frac{\pi}{2}\left[b(\omega)+ b(-\omega)\right]&=\frac{2 \omega^2 Y_0C G}{\left(Y_0^2+G^2-\omega^2C^2\right)^2+4\omega^2 C^2 Y_0^2}\,,\label{eq:even_bw}\\
				\mcl{P}\int_{-\infty}^\infty\dd\Omega\frac{b(\Omega)}{\Omega^2-\omega^2} &=\frac{-G\left(Y_0^2-\omega^2C^2+G^2\right)}{\left(Y_0^2+G^2-\omega^2C^2\right)^2+4\omega^2 C^2 Y_0^2}\label{eq:variousd}\,.
			\end{align}
		\end{subequations}
	\end{widetext}
	In this case, therefore, we can take $a(\omega)$ and $b(\omega)$ as even functions, from inspection of Eqs. \eqref{eq:even_aw} and \eqref{eq:even_bw}. Accordingly, the sequences $\msA_k$ and $\msB_k$ and thus, of capacitors and $C_k$ and gyrators $R_k$ can be delineated.

	\bibliography{bibliography}

\begin{thebibliography}{90}%
\makeatletter
\providecommand \@ifxundefined [1]{%
 \@ifx{#1\undefined}
}%
\providecommand \@ifnum [1]{%
 \ifnum #1\expandafter \@firstoftwo
 \else \expandafter \@secondoftwo
 \fi
}%
\providecommand \@ifx [1]{%
 \ifx #1\expandafter \@firstoftwo
 \else \expandafter \@secondoftwo
 \fi
}%
\providecommand \natexlab [1]{#1}%
\providecommand \enquote  [1]{``#1''}%
\providecommand \bibnamefont  [1]{#1}%
\providecommand \bibfnamefont [1]{#1}%
\providecommand \citenamefont [1]{#1}%
\providecommand \href@noop [0]{\@secondoftwo}%
\providecommand \href [0]{\begingroup \@sanitize@url \@href}%
\providecommand \@href[1]{\@@startlink{#1}\@@href}%
\providecommand \@@href[1]{\endgroup#1\@@endlink}%
\providecommand \@sanitize@url [0]{\catcode `\\12\catcode `\$12\catcode
  `\&12\catcode `\#12\catcode `\^12\catcode `\_12\catcode `\%12\relax}%
\providecommand \@@startlink[1]{}%
\providecommand \@@endlink[0]{}%
\providecommand \url  [0]{\begingroup\@sanitize@url \@url }%
\providecommand \@url [1]{\endgroup\@href {#1}{\urlprefix }}%
\providecommand \urlprefix  [0]{URL }%
\providecommand \Eprint [0]{\href }%
\providecommand \doibase [0]{https://doi.org/}%
\providecommand \selectlanguage [0]{\@gobble}%
\providecommand \bibinfo  [0]{\@secondoftwo}%
\providecommand \bibfield  [0]{\@secondoftwo}%
\providecommand \translation [1]{[#1]}%
\providecommand \BibitemOpen [0]{}%
\providecommand \bibitemStop [0]{}%
\providecommand \bibitemNoStop [0]{.\EOS\space}%
\providecommand \EOS [0]{\spacefactor3000\relax}%
\providecommand \BibitemShut  [1]{\csname bibitem#1\endcsname}%
\let\auto@bib@innerbib\@empty
\bibitem [{\citenamefont {Parra-Rodriguez}\ and\ \citenamefont
  {Egusquiza}(2024)}]{ParraRodriguez:2024a}%
  \BibitemOpen
  \bibfield  {author} {\bibinfo {author} {\bibfnamefont {A.}~\bibnamefont
  {Parra-Rodriguez}}\ and\ \bibinfo {author} {\bibfnamefont {I.~L.}\
  \bibnamefont {Egusquiza}},\ }\bibfield  {title} {\bibinfo {title}
  {Geometrical description and {F}addeev-{J}ackiw quantization of electrical
  networks},\ }\href {https://doi.org/10.22331/q-2024-09-09-1466} {\bibfield
  {journal} {\bibinfo  {journal} {{Quantum}}\ }\textbf {\bibinfo {volume}
  {8}},\ \bibinfo {pages} {1466} (\bibinfo {year} {2024})}\BibitemShut
  {NoStop}%
\bibitem [{\citenamefont {Jackson}(1999)}]{Jackson:1999}%
  \BibitemOpen
  \bibfield  {author} {\bibinfo {author} {\bibfnamefont {J.~D.}\ \bibnamefont
  {Jackson}},\ }\href@noop {} {\emph {\bibinfo {title} {Classical
  electrodynamics}}},\ \bibinfo {edition} {3rd}\ ed.\ (\bibinfo  {publisher}
  {Wiley},\ \bibinfo {address} {New York},\ \bibinfo {year} {1999})\BibitemShut
  {NoStop}%
\bibitem [{\citenamefont {Feynman}\ \emph {et~al.}(2010)\citenamefont
  {Feynman}, \citenamefont {Leighton},\ and\ \citenamefont
  {Sands}}]{Feynman:2010}%
  \BibitemOpen
  \bibfield  {author} {\bibinfo {author} {\bibfnamefont {R.}~\bibnamefont
  {Feynman}}, \bibinfo {author} {\bibfnamefont {R.}~\bibnamefont {Leighton}},\
  and\ \bibinfo {author} {\bibfnamefont {M.}~\bibnamefont {Sands}},\ }\href
  {https://books.google.es/books?id=hlRhwGK40fgC} {\emph {\bibinfo {title} {The
  Feynman Lectures on Physics, Vol. II: Mainly Electromagnetism and Matter}}},\
  \bibinfo {edition} {new millennium}\ ed.\ (\bibinfo  {publisher} {Basic
  Books},\ \bibinfo {address} {New York},\ \bibinfo {year} {2010})\BibitemShut
  {NoStop}%
\bibitem [{\citenamefont {Caldeira}\ and\ \citenamefont
  {Leggett}(1981)}]{CaldeiraLeggett:1981}%
  \BibitemOpen
  \bibfield  {author} {\bibinfo {author} {\bibfnamefont {A.~O.}\ \bibnamefont
  {Caldeira}}\ and\ \bibinfo {author} {\bibfnamefont {A.~J.}\ \bibnamefont
  {Leggett}},\ }\bibfield  {title} {\bibinfo {title} {Influence of dissipation
  on quantum tunneling in macroscopic systems},\ }\href
  {https://doi.org/10.1103/PhysRevLett.46.211} {\bibfield  {journal} {\bibinfo
  {journal} {Physical Review Letters}\ }\textbf {\bibinfo {volume} {46}},\
  \bibinfo {pages} {211} (\bibinfo {year} {1981})}\BibitemShut {NoStop}%
\bibitem [{\citenamefont {Caldeira}\ and\ \citenamefont
  {Leggett}(1983)}]{Caldeira:1983}%
  \BibitemOpen
  \bibfield  {author} {\bibinfo {author} {\bibfnamefont {A.~O.}\ \bibnamefont
  {Caldeira}}\ and\ \bibinfo {author} {\bibfnamefont {A.~J.}\ \bibnamefont
  {Leggett}},\ }\bibfield  {title} {\bibinfo {title} {Quantum tunnelling in a
  dissipative system},\ }\href
  {https://doi.org/https://doi.org/10.1016/0003-4916(83)90202-6} {\bibfield
  {journal} {\bibinfo  {journal} {Annals of Physics}\ }\textbf {\bibinfo
  {volume} {149}},\ \bibinfo {pages} {374} (\bibinfo {year}
  {1983})}\BibitemShut {NoStop}%
\bibitem [{\citenamefont {Yurke}\ and\ \citenamefont
  {Denker}(1984)}]{YurkeDenker:1984}%
  \BibitemOpen
  \bibfield  {author} {\bibinfo {author} {\bibfnamefont {B.}~\bibnamefont
  {Yurke}}\ and\ \bibinfo {author} {\bibfnamefont {J.~S.}\ \bibnamefont
  {Denker}},\ }\bibfield  {title} {\bibinfo {title} {{Quantum network
  theory}},\ }\href {https://doi.org/10.1103/PhysRevA.29.1419} {\bibfield
  {journal} {\bibinfo  {journal} {Physical Review A}\ }\textbf {\bibinfo
  {volume} {29}},\ \bibinfo {pages} {1419} (\bibinfo {year}
  {1984})}\BibitemShut {NoStop}%
\bibitem [{\citenamefont {Devoret}\ and\ \citenamefont
  {Schoelkopf}(2013)}]{Devoret:2013}%
  \BibitemOpen
  \bibfield  {author} {\bibinfo {author} {\bibfnamefont {M.}~\bibnamefont
  {Devoret}}\ and\ \bibinfo {author} {\bibfnamefont {R.}~\bibnamefont
  {Schoelkopf}},\ }\bibfield  {title} {\bibinfo {title} {{Superconducting
  circuits for quantum information: An outlook}},\ }\href
  {https://doi.org/10.1126/science.1231930} {\bibfield  {journal} {\bibinfo
  {journal} {Science}\ }\textbf {\bibinfo {volume} {339}},\ \bibinfo {pages}
  {1169} (\bibinfo {year} {2013})}\BibitemShut {NoStop}%
\bibitem [{\citenamefont {Blais}\ \emph {et~al.}(2021)\citenamefont {Blais},
  \citenamefont {Grimsmo}, \citenamefont {Girvin},\ and\ \citenamefont
  {Wallraff}}]{Blais:2021}%
  \BibitemOpen
  \bibfield  {author} {\bibinfo {author} {\bibfnamefont {A.}~\bibnamefont
  {Blais}}, \bibinfo {author} {\bibfnamefont {A.~L.}\ \bibnamefont {Grimsmo}},
  \bibinfo {author} {\bibfnamefont {S.~M.}\ \bibnamefont {Girvin}},\ and\
  \bibinfo {author} {\bibfnamefont {A.}~\bibnamefont {Wallraff}},\ }\bibfield
  {title} {\bibinfo {title} {Circuit quantum electrodynamics},\ }\href
  {https://link.aps.org/doi/10.1103/RevModPhys.93.025005} {\bibfield  {journal}
  {\bibinfo  {journal} {Review Modern Physics}\ }\textbf {\bibinfo {volume}
  {93}},\ \bibinfo {pages} {025005} (\bibinfo {year} {2021})}\BibitemShut
  {NoStop}%
\bibitem [{\citenamefont {Parra-Rodriguez}(2021)}]{ParraRodriguezPhD:2021}%
  \BibitemOpen
  \bibfield  {author} {\bibinfo {author} {\bibfnamefont {A.}~\bibnamefont
  {Parra-Rodriguez}},\ }\href {http://hdl.handle.net/10810/51132} {\emph
  {\bibinfo {title} {PhD Thesis: Canonical Quantization of Superconducting
  Circuits}}}\ (\bibinfo  {publisher} {Universidad del Pais Vasco/Euskal
  Herriko Unibertsitatea},\ \bibinfo {address} {Leioa},\ \bibinfo {year}
  {2021})\BibitemShut {NoStop}%
\bibitem [{\citenamefont {Janhsen}\ \emph {et~al.}(1992)\citenamefont
  {Janhsen}, \citenamefont {Schiek},\ and\ \citenamefont
  {Hansen}}]{Janhsen:1992}%
  \BibitemOpen
  \bibfield  {author} {\bibinfo {author} {\bibfnamefont {A.}~\bibnamefont
  {Janhsen}}, \bibinfo {author} {\bibfnamefont {B.}~\bibnamefont {Schiek}},\
  and\ \bibinfo {author} {\bibfnamefont {V.}~\bibnamefont {Hansen}},\
  }\bibfield  {title} {\bibinfo {title} {On the definition of quasi lumped
  elements in planar microwave circuits},\ }in\ \href
  {https://doi.org/10.1109/EUMA.1992.335748} {\emph {\bibinfo {booktitle} {1992
  22nd European Microwave Conference}}},\ Vol.~\bibinfo {volume} {1}\ (\bibinfo
  {year} {1992})\ p.\ \bibinfo {pages} {251}\BibitemShut {NoStop}%
\bibitem [{\citenamefont {Blais}\ \emph {et~al.}(2004)\citenamefont {Blais},
  \citenamefont {Huang}, \citenamefont {Wallraff}, \citenamefont {Girvin},\
  and\ \citenamefont {Schoelkopf}}]{Blais:2004}%
  \BibitemOpen
  \bibfield  {author} {\bibinfo {author} {\bibfnamefont {A.}~\bibnamefont
  {Blais}}, \bibinfo {author} {\bibfnamefont {R.-S.}\ \bibnamefont {Huang}},
  \bibinfo {author} {\bibfnamefont {A.}~\bibnamefont {Wallraff}}, \bibinfo
  {author} {\bibfnamefont {S.~M.}\ \bibnamefont {Girvin}},\ and\ \bibinfo
  {author} {\bibfnamefont {R.~J.}\ \bibnamefont {Schoelkopf}},\ }\bibfield
  {title} {\bibinfo {title} {Cavity quantum electrodynamics for superconducting
  electrical circuits: An architecture for quantum computation},\ }\href
  {https://doi.org/10.1103/PhysRevA.69.062320} {\bibfield  {journal} {\bibinfo
  {journal} {Physical ReviewA}\ }\textbf {\bibinfo {volume} {69}},\ \bibinfo
  {pages} {062320} (\bibinfo {year} {2004})}\BibitemShut {NoStop}%
\bibitem [{\citenamefont {Minev}\ \emph
  {et~al.}(2021{\natexlab{a}})\citenamefont {Minev}, \citenamefont {McConkey},
  \citenamefont {Takita}, \citenamefont {Corcoles},\ and\ \citenamefont
  {Gambetta}}]{Minev:2021}%
  \BibitemOpen
  \bibfield  {author} {\bibinfo {author} {\bibfnamefont {Z.~K.}\ \bibnamefont
  {Minev}}, \bibinfo {author} {\bibfnamefont {T.~G.}\ \bibnamefont {McConkey}},
  \bibinfo {author} {\bibfnamefont {M.}~\bibnamefont {Takita}}, \bibinfo
  {author} {\bibfnamefont {A.~D.}\ \bibnamefont {Corcoles}},\ and\ \bibinfo
  {author} {\bibfnamefont {J.~M.}\ \bibnamefont {Gambetta}},\ }\bibfield
  {title} {\bibinfo {title} {Circuit quantum electrodynamics (cqed) with
  modular quasi-lumped models},\ }\Eprint {https://arxiv.org/abs/2103.10344}
  {arXiv:2103.10344 [quant-ph]}  (\bibinfo {year}
  {2021}{\natexlab{a}})\BibitemShut {NoStop}%
\bibitem [{\citenamefont {Tellegen}(1948{\natexlab{a}})}]{Tellegen:1948a}%
  \BibitemOpen
  \bibfield  {author} {\bibinfo {author} {\bibfnamefont {B.~D.~H.}\
  \bibnamefont {Tellegen}},\ }\bibfield  {title} {\bibinfo {title} {The
  gyrator, a new electric network element},\ }\href
  {{https://web.archive.org/web/20140423045739/http://techpreservation.dyndns.org/beitman/abpr/newfiles/The%20Gyrator.pdf}}
  {\bibfield  {journal} {\bibinfo  {journal} {Philips Research Reports}\
  }\textbf {\bibinfo {volume} {3}},\ \bibinfo {pages} {81} (\bibinfo {year}
  {1948}{\natexlab{a}})}\BibitemShut {NoStop}%
\bibitem [{\citenamefont {Pozar}(2009)}]{Pozar:2009}%
  \BibitemOpen
  \bibfield  {author} {\bibinfo {author} {\bibfnamefont {D.~M.}\ \bibnamefont
  {Pozar}},\ }\href
  {https://www.wiley.com/en-us/Microwave+Engineering%2C+4th+Edition-p-9780470631553}
  {\emph {\bibinfo {title} {Microwave Engineering}}},\ \bibinfo {edition}
  {4th}\ ed.\ (\bibinfo  {publisher} {John Wiley \& Sons},\ \bibinfo {address}
  {Hoboken, New York},\ \bibinfo {year} {2009})\BibitemShut {NoStop}%
\bibitem [{\citenamefont {Caloz}\ \emph {et~al.}(2018)\citenamefont {Caloz},
  \citenamefont {Al\`u}, \citenamefont {Tretyakov}, \citenamefont {Sounas},
  \citenamefont {Achouri},\ and\ \citenamefont {Deck-L\'eger}}]{Caloz:2018}%
  \BibitemOpen
  \bibfield  {author} {\bibinfo {author} {\bibfnamefont {C.}~\bibnamefont
  {Caloz}}, \bibinfo {author} {\bibfnamefont {A.}~\bibnamefont {Al\`u}},
  \bibinfo {author} {\bibfnamefont {S.}~\bibnamefont {Tretyakov}}, \bibinfo
  {author} {\bibfnamefont {D.}~\bibnamefont {Sounas}}, \bibinfo {author}
  {\bibfnamefont {K.}~\bibnamefont {Achouri}},\ and\ \bibinfo {author}
  {\bibfnamefont {Z.-L.}\ \bibnamefont {Deck-L\'eger}},\ }\bibfield  {title}
  {\bibinfo {title} {Electromagnetic nonreciprocity},\ }\href
  {https://doi.org/10.1103/PhysRevApplied.10.047001} {\bibfield  {journal}
  {\bibinfo  {journal} {Physical Review Applied}\ }\textbf {\bibinfo {volume}
  {10}},\ \bibinfo {pages} {047001} (\bibinfo {year} {2018})}\BibitemShut
  {NoStop}%
\bibitem [{\citenamefont {Belevitch}(1950)}]{Belevitch:1950}%
  \BibitemOpen
  \bibfield  {author} {\bibinfo {author} {\bibfnamefont {V.}~\bibnamefont
  {Belevitch}},\ }\bibfield  {title} {\bibinfo {title} {{Theory of 2n-terminal
  networks with application to conference telephony}},\ }\href
  {https://worldradiohistory.com/hd2/IDX-Site-Technical/Company-Pubications/Archive-ITT-IDX/IDX/50s/ITT-Vol-27-1950-03-OCR-Page-0064.pdf}
  {\bibfield  {journal} {\bibinfo  {journal} {Electrical Communication}\
  }\textbf {\bibinfo {volume} {27}},\ \bibinfo {pages} {231} (\bibinfo {year}
  {1950})}\BibitemShut {NoStop}%
\bibitem [{\citenamefont {Kamal}\ \emph {et~al.}(2011)\citenamefont {Kamal},
  \citenamefont {Clarke},\ and\ \citenamefont {Devoret}}]{Kamal:2011}%
  \BibitemOpen
  \bibfield  {author} {\bibinfo {author} {\bibfnamefont {A.}~\bibnamefont
  {Kamal}}, \bibinfo {author} {\bibfnamefont {J.}~\bibnamefont {Clarke}},\ and\
  \bibinfo {author} {\bibfnamefont {M.~H.}\ \bibnamefont {Devoret}},\
  }\bibfield  {title} {\bibinfo {title} {Noiseless non-reciprocity in a
  parametric active device},\ }\href {https://doi.org/10.1038/nphys1893}
  {\bibfield  {journal} {\bibinfo  {journal} {Nature Physics}\ }\textbf
  {\bibinfo {volume} {7}},\ \bibinfo {pages} {311} (\bibinfo {year}
  {2011})}\BibitemShut {NoStop}%
\bibitem [{\citenamefont {Viola}\ and\ \citenamefont
  {DiVincenzo}(2014)}]{Viola:2014}%
  \BibitemOpen
  \bibfield  {author} {\bibinfo {author} {\bibfnamefont {G.}~\bibnamefont
  {Viola}}\ and\ \bibinfo {author} {\bibfnamefont {D.~P.}\ \bibnamefont
  {DiVincenzo}},\ }\bibfield  {title} {\bibinfo {title} {Hall effect gyrators
  and circulators},\ }\href {https://doi.org/10.1103/PhysRevX.4.021019}
  {\bibfield  {journal} {\bibinfo  {journal} {Physical Review X}\ }\textbf
  {\bibinfo {volume} {4}},\ \bibinfo {pages} {021019} (\bibinfo {year}
  {2014})}\BibitemShut {NoStop}%
\bibitem [{\citenamefont {Kerckhoff}\ \emph {et~al.}(2015)\citenamefont
  {Kerckhoff}, \citenamefont {Lalumi\`ere}, \citenamefont {Chapman},
  \citenamefont {Blais},\ and\ \citenamefont {Lehnert}}]{Kerckhoff:2015}%
  \BibitemOpen
  \bibfield  {author} {\bibinfo {author} {\bibfnamefont {J.}~\bibnamefont
  {Kerckhoff}}, \bibinfo {author} {\bibfnamefont {K.}~\bibnamefont
  {Lalumi\`ere}}, \bibinfo {author} {\bibfnamefont {B.~J.}\ \bibnamefont
  {Chapman}}, \bibinfo {author} {\bibfnamefont {A.}~\bibnamefont {Blais}},\
  and\ \bibinfo {author} {\bibfnamefont {K.~W.}\ \bibnamefont {Lehnert}},\
  }\bibfield  {title} {\bibinfo {title} {On-chip superconducting microwave
  circulator from synthetic rotation},\ }\href
  {https://doi.org/10.1103/PhysRevApplied.4.034002} {\bibfield  {journal}
  {\bibinfo  {journal} {Physical Review Applied}\ }\textbf {\bibinfo {volume}
  {4}},\ \bibinfo {pages} {034002} (\bibinfo {year} {2015})}\BibitemShut
  {NoStop}%
\bibitem [{\citenamefont {Sliwa}\ \emph {et~al.}(2015)\citenamefont {Sliwa},
  \citenamefont {Hatridge}, \citenamefont {Narla}, \citenamefont {Shankar},
  \citenamefont {Frunzio}, \citenamefont {Schoelkopf},\ and\ \citenamefont
  {Devoret}}]{Sliwa:2015}%
  \BibitemOpen
  \bibfield  {author} {\bibinfo {author} {\bibfnamefont {K.~M.}\ \bibnamefont
  {Sliwa}}, \bibinfo {author} {\bibfnamefont {M.}~\bibnamefont {Hatridge}},
  \bibinfo {author} {\bibfnamefont {A.}~\bibnamefont {Narla}}, \bibinfo
  {author} {\bibfnamefont {S.}~\bibnamefont {Shankar}}, \bibinfo {author}
  {\bibfnamefont {L.}~\bibnamefont {Frunzio}}, \bibinfo {author} {\bibfnamefont
  {R.~J.}\ \bibnamefont {Schoelkopf}},\ and\ \bibinfo {author} {\bibfnamefont
  {M.~H.}\ \bibnamefont {Devoret}},\ }\bibfield  {title} {\bibinfo {title}
  {Reconfigurable josephson circulator/directional amplifier},\ }\href
  {https://doi.org/10.1103/PhysRevX.5.041020} {\bibfield  {journal} {\bibinfo
  {journal} {Physical Review X}\ }\textbf {\bibinfo {volume} {5}},\ \bibinfo
  {pages} {041020} (\bibinfo {year} {2015})}\BibitemShut {NoStop}%
\bibitem [{\citenamefont {Chapman}\ \emph {et~al.}(2017)\citenamefont
  {Chapman}, \citenamefont {Rosenthal}, \citenamefont {Kerckhoff},
  \citenamefont {Moores}, \citenamefont {Vale}, \citenamefont {Mates},
  \citenamefont {Hilton}, \citenamefont {Lalumi\`ere}, \citenamefont {Blais},\
  and\ \citenamefont {Lehnert}}]{Chapman:2017}%
  \BibitemOpen
  \bibfield  {author} {\bibinfo {author} {\bibfnamefont {B.~J.}\ \bibnamefont
  {Chapman}}, \bibinfo {author} {\bibfnamefont {E.~I.}\ \bibnamefont
  {Rosenthal}}, \bibinfo {author} {\bibfnamefont {J.}~\bibnamefont
  {Kerckhoff}}, \bibinfo {author} {\bibfnamefont {B.~A.}\ \bibnamefont
  {Moores}}, \bibinfo {author} {\bibfnamefont {L.~R.}\ \bibnamefont {Vale}},
  \bibinfo {author} {\bibfnamefont {J.~A.~B.}\ \bibnamefont {Mates}}, \bibinfo
  {author} {\bibfnamefont {G.~C.}\ \bibnamefont {Hilton}}, \bibinfo {author}
  {\bibfnamefont {K.}~\bibnamefont {Lalumi\`ere}}, \bibinfo {author}
  {\bibfnamefont {A.}~\bibnamefont {Blais}},\ and\ \bibinfo {author}
  {\bibfnamefont {K.~W.}\ \bibnamefont {Lehnert}},\ }\bibfield  {title}
  {\bibinfo {title} {Widely tunable on-chip microwave circulator for
  superconducting quantum circuits},\ }\href
  {https://doi.org/10.1103/PhysRevX.7.041043} {\bibfield  {journal} {\bibinfo
  {journal} {Physical Review X}\ }\textbf {\bibinfo {volume} {7}},\ \bibinfo
  {pages} {041043} (\bibinfo {year} {2017})}\BibitemShut {NoStop}%
\bibitem [{\citenamefont {Barzanjeh}\ \emph {et~al.}(2017)\citenamefont
  {Barzanjeh}, \citenamefont {Wulf}, \citenamefont {Peruzzo}, \citenamefont
  {Kalaee}, \citenamefont {Dieterle}, \citenamefont {Painter},\ and\
  \citenamefont {Fink}}]{Barzanjeh:2017}%
  \BibitemOpen
  \bibfield  {author} {\bibinfo {author} {\bibfnamefont {S.}~\bibnamefont
  {Barzanjeh}}, \bibinfo {author} {\bibfnamefont {M.}~\bibnamefont {Wulf}},
  \bibinfo {author} {\bibfnamefont {M.}~\bibnamefont {Peruzzo}}, \bibinfo
  {author} {\bibfnamefont {M.}~\bibnamefont {Kalaee}}, \bibinfo {author}
  {\bibfnamefont {P.}~\bibnamefont {Dieterle}}, \bibinfo {author}
  {\bibfnamefont {O.}~\bibnamefont {Painter}},\ and\ \bibinfo {author}
  {\bibfnamefont {J.}~\bibnamefont {Fink}},\ }\bibfield  {title} {\bibinfo
  {title} {{Mechanical on-chip microwave circulator}},\ }\href
  {https://doi.org/10.1038/s41467-017-01304-x} {\bibfield  {journal} {\bibinfo
  {journal} {Nature Communications}\ }\textbf {\bibinfo {volume} {8}},\
  \bibinfo {pages} {953} (\bibinfo {year} {2017})}\BibitemShut {NoStop}%
\bibitem [{\citenamefont {Rosenthal}\ \emph {et~al.}(2017)\citenamefont
  {Rosenthal}, \citenamefont {Chapman}, \citenamefont {Higginbotham},
  \citenamefont {Kerckhoff},\ and\ \citenamefont {Lehnert}}]{Rosenthal:2017}%
  \BibitemOpen
  \bibfield  {author} {\bibinfo {author} {\bibfnamefont {E.~I.}\ \bibnamefont
  {Rosenthal}}, \bibinfo {author} {\bibfnamefont {B.~J.}\ \bibnamefont
  {Chapman}}, \bibinfo {author} {\bibfnamefont {A.~P.}\ \bibnamefont
  {Higginbotham}}, \bibinfo {author} {\bibfnamefont {J.}~\bibnamefont
  {Kerckhoff}},\ and\ \bibinfo {author} {\bibfnamefont {K.~W.}\ \bibnamefont
  {Lehnert}},\ }\bibfield  {title} {\bibinfo {title} {Breaking lorentz
  reciprocity with frequency conversion and delay},\ }\href
  {https://doi.org/10.1103/PhysRevLett.119.147703} {\bibfield  {journal}
  {\bibinfo  {journal} {Physical Review Letters}\ }\textbf {\bibinfo {volume}
  {119}},\ \bibinfo {pages} {147703} (\bibinfo {year} {2017})}\BibitemShut
  {NoStop}%
\bibitem [{\citenamefont {Josephson}(1962)}]{Josephson:1962}%
  \BibitemOpen
  \bibfield  {author} {\bibinfo {author} {\bibfnamefont {B.~D.}\ \bibnamefont
  {Josephson}},\ }\bibfield  {title} {\bibinfo {title} {Possible new effects in
  superconductive tunnelling},\ }\href
  {https://doi.org/https://doi.org/10.1016/0031-9163(62)91369-0} {\bibfield
  {journal} {\bibinfo  {journal} {Physics Letters}\ }\textbf {\bibinfo {volume}
  {1}},\ \bibinfo {pages} {251} (\bibinfo {year} {1962})}\BibitemShut {NoStop}%
\bibitem [{\citenamefont {Mooij}\ and\ \citenamefont
  {Nazarov}(2006)}]{Mooij:2006}%
  \BibitemOpen
  \bibfield  {author} {\bibinfo {author} {\bibfnamefont {J.}~\bibnamefont
  {Mooij}}\ and\ \bibinfo {author} {\bibfnamefont {Y.}~\bibnamefont
  {Nazarov}},\ }\bibfield  {title} {\bibinfo {title} {{Superconducting
  nanowires as quantum phase-slip junctions}},\ }\href
  {https://doi.org/10.1038/nphys234} {\bibfield  {journal} {\bibinfo  {journal}
  {Nature Physics}\ }\textbf {\bibinfo {volume} {2}},\ \bibinfo {pages} {169}
  (\bibinfo {year} {2006})}\BibitemShut {NoStop}%
\bibitem [{\citenamefont {Parra-Rodriguez}\ \emph {et~al.}(2018)\citenamefont
  {Parra-Rodriguez}, \citenamefont {Rico}, \citenamefont {Solano},\ and\
  \citenamefont {Egusquiza}}]{ParraRodriguez:2018}%
  \BibitemOpen
  \bibfield  {author} {\bibinfo {author} {\bibfnamefont {A.}~\bibnamefont
  {Parra-Rodriguez}}, \bibinfo {author} {\bibfnamefont {E.}~\bibnamefont
  {Rico}}, \bibinfo {author} {\bibfnamefont {E.}~\bibnamefont {Solano}},\ and\
  \bibinfo {author} {\bibfnamefont {I.~L.}\ \bibnamefont {Egusquiza}},\
  }\bibfield  {title} {\bibinfo {title} {Quantum networks in divergence-free
  circuit {QED}},\ }\href {https://doi.org/10.1088/2058-9565/aab1ba} {\bibfield
   {journal} {\bibinfo  {journal} {Quantum Science and Technology}\ }\textbf
  {\bibinfo {volume} {3}},\ \bibinfo {pages} {024012} (\bibinfo {year}
  {2018})}\BibitemShut {NoStop}%
\bibitem [{\citenamefont {Paladino}\ \emph {et~al.}(2003)\citenamefont
  {Paladino}, \citenamefont {Taddei}, \citenamefont {Giaquinta},\ and\
  \citenamefont {Falci}}]{Paladino:2003}%
  \BibitemOpen
  \bibfield  {author} {\bibinfo {author} {\bibfnamefont {E.}~\bibnamefont
  {Paladino}}, \bibinfo {author} {\bibfnamefont {F.}~\bibnamefont {Taddei}},
  \bibinfo {author} {\bibfnamefont {G.}~\bibnamefont {Giaquinta}},\ and\
  \bibinfo {author} {\bibfnamefont {G.}~\bibnamefont {Falci}},\ }\bibfield
  {title} {\bibinfo {title} {{Josephson nanocircuit in the presence of linear
  quantum noise}},\ }\href {https://doi.org/10.1016/S1386-9477(02)00948-7}
  {\bibfield  {journal} {\bibinfo  {journal} {Physica E: Low-Dimensional
  Systems and Nanostructures}\ }\textbf {\bibinfo {volume} {18}},\ \bibinfo
  {pages} {39} (\bibinfo {year} {2003})}\BibitemShut {NoStop}%
\bibitem [{\citenamefont {Ao}\ \emph {et~al.}(2023)\citenamefont {Ao},
  \citenamefont {Ashhab}, \citenamefont {Yoshihara}, \citenamefont {Fuse},
  \citenamefont {Kakuyanagi}, \citenamefont {Saito}, \citenamefont {Aoki},\
  and\ \citenamefont {Semba}}]{Ao:2023}%
  \BibitemOpen
  \bibfield  {author} {\bibinfo {author} {\bibfnamefont {Z.}~\bibnamefont
  {Ao}}, \bibinfo {author} {\bibfnamefont {S.}~\bibnamefont {Ashhab}}, \bibinfo
  {author} {\bibfnamefont {F.}~\bibnamefont {Yoshihara}}, \bibinfo {author}
  {\bibfnamefont {T.}~\bibnamefont {Fuse}}, \bibinfo {author} {\bibfnamefont
  {K.}~\bibnamefont {Kakuyanagi}}, \bibinfo {author} {\bibfnamefont
  {S.}~\bibnamefont {Saito}}, \bibinfo {author} {\bibfnamefont
  {T.}~\bibnamefont {Aoki}},\ and\ \bibinfo {author} {\bibfnamefont
  {K.}~\bibnamefont {Semba}},\ }\bibfield  {title} {\bibinfo {title} {Extremely
  large {Lamb} shift in a deep-strongly coupled circuit {QED} system with a
  multimode resonator},\ }\href {https://doi.org/10.1038/s41598-023-36547-w}
  {\bibfield  {journal} {\bibinfo  {journal} {Scientific Reports}\ }\textbf
  {\bibinfo {volume} {13}},\ \bibinfo {pages} {11340} (\bibinfo {year}
  {2023})}\BibitemShut {NoStop}%
\bibitem [{\citenamefont {Parra-Rodriguez}\ and\ \citenamefont
  {Egusquiza}(2022{\natexlab{a}})}]{ParraRodriguez:2022}%
  \BibitemOpen
  \bibfield  {author} {\bibinfo {author} {\bibfnamefont {A.}~\bibnamefont
  {Parra-Rodriguez}}\ and\ \bibinfo {author} {\bibfnamefont {I.~L.}\
  \bibnamefont {Egusquiza}},\ }\bibfield  {title} {\bibinfo {title} {Canonical
  quantisation of telegrapher's equations coupled by ideal nonreciprocal
  elements},\ }\href {https://doi.org/10.22331/q-2022-04-04-681} {\bibfield
  {journal} {\bibinfo  {journal} {{Quantum}}\ }\textbf {\bibinfo {volume}
  {6}},\ \bibinfo {pages} {681} (\bibinfo {year}
  {2022}{\natexlab{a}})}\BibitemShut {NoStop}%
\bibitem [{\citenamefont {Egusquiza}\ and\ \citenamefont
  {Parra-Rodriguez}(2022)}]{Egusquiza:2022}%
  \BibitemOpen
  \bibfield  {author} {\bibinfo {author} {\bibfnamefont {I.~L.}\ \bibnamefont
  {Egusquiza}}\ and\ \bibinfo {author} {\bibfnamefont {A.}~\bibnamefont
  {Parra-Rodriguez}},\ }\bibfield  {title} {\bibinfo {title} {Algebraic
  canonical quantization of lumped superconducting networks},\ }\href
  {https://doi.org/10.1103/PhysRevB.106.024510} {\bibfield  {journal} {\bibinfo
   {journal} {Physical Review B}\ }\textbf {\bibinfo {volume} {106}},\ \bibinfo
  {pages} {024510} (\bibinfo {year} {2022})}\BibitemShut {NoStop}%
\bibitem [{\citenamefont {Devoret}(1995)}]{Devoret:1997}%
  \BibitemOpen
  \bibfield  {author} {\bibinfo {author} {\bibfnamefont {M.~H.}\ \bibnamefont
  {Devoret}},\ }\bibfield  {title} {\bibinfo {title} {Quantum fluctuations in
  electrical circuits},\ }in\ \href@noop {} {\emph {\bibinfo {booktitle}
  {Proceedings of the Les Houches Summer School, Session LXIII}}}\ (\bibinfo
  {publisher} {Elsevier, edited by S. Reynaud, E. Giacobino, and J.
  Zinn-Justin},\ \bibinfo {address} {New York},\ \bibinfo {year}
  {1995})\BibitemShut {NoStop}%
\bibitem [{\citenamefont {Vool}\ and\ \citenamefont
  {Devoret}(2017)}]{Vool:2017}%
  \BibitemOpen
  \bibfield  {author} {\bibinfo {author} {\bibfnamefont {U.}~\bibnamefont
  {Vool}}\ and\ \bibinfo {author} {\bibfnamefont {M.}~\bibnamefont {Devoret}},\
  }\bibfield  {title} {\bibinfo {title} {Introduction to quantum
  electromagnetic circuits},\ }\href
  {https://doi.org/https://doi.org/10.1002/cta.2359} {\bibfield  {journal}
  {\bibinfo  {journal} {International Journal of Circuit Theory and
  Applications}\ }\textbf {\bibinfo {volume} {45}},\ \bibinfo {pages} {897}
  (\bibinfo {year} {2017})}\BibitemShut {NoStop}%
\bibitem [{\citenamefont {Parra-Rodriguez}\ \emph {et~al.}(2019)\citenamefont
  {Parra-Rodriguez}, \citenamefont {Egusquiza}, \citenamefont {DiVincenzo},\
  and\ \citenamefont {Solano}}]{ParraRodriguez:2019}%
  \BibitemOpen
  \bibfield  {author} {\bibinfo {author} {\bibfnamefont {A.}~\bibnamefont
  {Parra-Rodriguez}}, \bibinfo {author} {\bibfnamefont {I.~L.}\ \bibnamefont
  {Egusquiza}}, \bibinfo {author} {\bibfnamefont {D.~P.}\ \bibnamefont
  {DiVincenzo}},\ and\ \bibinfo {author} {\bibfnamefont {E.}~\bibnamefont
  {Solano}},\ }\bibfield  {title} {\bibinfo {title} {Canonical circuit
  quantization with linear nonreciprocal devices},\ }\href
  {https://doi.org/10.1103/PhysRevB.99.014514} {\bibfield  {journal} {\bibinfo
  {journal} {Physical Review B}\ }\textbf {\bibinfo {volume} {99}},\ \bibinfo
  {pages} {014514} (\bibinfo {year} {2019})}\BibitemShut {NoStop}%
\bibitem [{\citenamefont {Parra-Rodriguez}\ and\ \citenamefont
  {Egusquiza}(2022{\natexlab{b}})}]{ParraRodriguez:2022b}%
  \BibitemOpen
  \bibfield  {author} {\bibinfo {author} {\bibfnamefont {A.}~\bibnamefont
  {Parra-Rodriguez}}\ and\ \bibinfo {author} {\bibfnamefont {I.~L.}\
  \bibnamefont {Egusquiza}},\ }\bibfield  {title} {\bibinfo {title} {Quantum
  fluctuations in electrical multiport linear systems},\ }\href
  {https://doi.org/10.1103/PhysRevB.106.054504} {\bibfield  {journal} {\bibinfo
   {journal} {Physical Review B}\ }\textbf {\bibinfo {volume} {106}},\ \bibinfo
  {pages} {054504} (\bibinfo {year} {2022}{\natexlab{b}})}\BibitemShut
  {NoStop}%
\bibitem [{\citenamefont {Guillemin}(1953)}]{Guillemin:1953}%
  \BibitemOpen
  \bibfield  {author} {\bibinfo {author} {\bibfnamefont {E.~A.}\ \bibnamefont
  {Guillemin}},\ }\href@noop {} {\emph {\bibinfo {title} {Introductory circuit
  theory}}}\ (\bibinfo  {publisher} {John Wiley \& Sons},\ \bibinfo {address}
  {New York},\ \bibinfo {year} {1953})\BibitemShut {NoStop}%
\bibitem [{\citenamefont {Devoret}(2021)}]{Devoret:2021}%
  \BibitemOpen
  \bibfield  {author} {\bibinfo {author} {\bibfnamefont {M.~H.}\ \bibnamefont
  {Devoret}},\ }\bibfield  {title} {\bibinfo {title} {Does {Brian}
  {Josephson}'s gauge-invariant phase difference live on a line or a circle?},\
  }\href {https://doi.org/10.1007/s10948-020-05784-9} {\bibfield  {journal}
  {\bibinfo  {journal} {Journal of Superconductivity and Novel Magnetism}\
  }\textbf {\bibinfo {volume} {34}},\ \bibinfo {pages} {1633} (\bibinfo {year}
  {2021})}\BibitemShut {NoStop}%
\bibitem [{\citenamefont {Astafiev}\ \emph {et~al.}(2012)\citenamefont
  {Astafiev}, \citenamefont {Ioffe}, \citenamefont {Kafanov}, \citenamefont
  {Pashkin}, \citenamefont {Arutyunov}, \citenamefont {Shahar}, \citenamefont
  {Cohen},\ and\ \citenamefont {Tsai}}]{Astafiev:2012}%
  \BibitemOpen
  \bibfield  {author} {\bibinfo {author} {\bibfnamefont {O.}~\bibnamefont
  {Astafiev}}, \bibinfo {author} {\bibfnamefont {L.}~\bibnamefont {Ioffe}},
  \bibinfo {author} {\bibfnamefont {S.}~\bibnamefont {Kafanov}}, \bibinfo
  {author} {\bibfnamefont {Y.}~\bibnamefont {Pashkin}}, \bibinfo {author}
  {\bibfnamefont {K.}~\bibnamefont {Arutyunov}}, \bibinfo {author}
  {\bibfnamefont {D.}~\bibnamefont {Shahar}}, \bibinfo {author} {\bibfnamefont
  {O.}~\bibnamefont {Cohen}},\ and\ \bibinfo {author} {\bibfnamefont
  {J.}~\bibnamefont {Tsai}},\ }\bibfield  {title} {\bibinfo {title} {{Coherent
  quantum phase slip}},\ }\href {https://doi.org/10.1038/nature10930}
  {\bibfield  {journal} {\bibinfo  {journal} {Nature}\ }\textbf {\bibinfo
  {volume} {484}},\ \bibinfo {pages} {355} (\bibinfo {year}
  {2012})}\BibitemShut {NoStop}%
\bibitem [{\citenamefont {Foster}(1924)}]{Foster:1924}%
  \BibitemOpen
  \bibfield  {author} {\bibinfo {author} {\bibfnamefont {R.~M.}\ \bibnamefont
  {Foster}},\ }\bibfield  {title} {\bibinfo {title} {{A reactance theorem}},\
  }\href@noop {} {\bibfield  {journal} {\bibinfo  {journal} {Bell Systems
  Technical Journal}\ }\textbf {\bibinfo {volume} {6}},\ \bibinfo {pages} {259}
  (\bibinfo {year} {1924})}\BibitemShut {NoStop}%
\bibitem [{\citenamefont {Cauer}(1926)}]{Cauer:1926}%
  \BibitemOpen
  \bibfield  {author} {\bibinfo {author} {\bibfnamefont {W.}~\bibnamefont
  {Cauer}},\ }\href@noop {} {\emph {\bibinfo {title} {Doktorarbeit: Die
  Verwirklichung der Wechselstrom- widerst\"ande vorgeschriebener
  Frequenzabh\"angigkeit.}}}\ (\bibinfo  {publisher} {TH Berlin},\ \bibinfo
  {address} {Berlin},\ \bibinfo {year} {1926})\BibitemShut {NoStop}%
\bibitem [{\citenamefont {Tellegen}(1948{\natexlab{b}})}]{Tellegen:1948}%
  \BibitemOpen
  \bibfield  {author} {\bibinfo {author} {\bibfnamefont {B.~D.~H.}\
  \bibnamefont {Tellegen}},\ }\bibfield  {title} {\bibinfo {title} {The
  gyrator, a new electric network element},\ }\href@noop {} {\bibfield
  {journal} {\bibinfo  {journal} {Philips Research Reports}\ }\textbf {\bibinfo
  {volume} {3}},\ \bibinfo {pages} {81} (\bibinfo {year}
  {1948}{\natexlab{b}})}\BibitemShut {NoStop}%
\bibitem [{\citenamefont {Newcomb}(1966)}]{Newcomb:1966}%
  \BibitemOpen
  \bibfield  {author} {\bibinfo {author} {\bibfnamefont {R.~W.}\ \bibnamefont
  {Newcomb}},\ }\href@noop {} {\emph {\bibinfo {title} {Linear Multiport
  Synthesis}}}\ (\bibinfo  {publisher} {McGraw-Hill},\ \bibinfo {address} {New
  York},\ \bibinfo {year} {1966})\BibitemShut {NoStop}%
\bibitem [{\citenamefont {Solgun}\ and\ \citenamefont
  {DiVincenzo}(2015)}]{Solgun:2015}%
  \BibitemOpen
  \bibfield  {author} {\bibinfo {author} {\bibfnamefont {F.}~\bibnamefont
  {Solgun}}\ and\ \bibinfo {author} {\bibfnamefont {D.}~\bibnamefont
  {DiVincenzo}},\ }\bibfield  {title} {\bibinfo {title} {{Multiport impedance
  quantization}},\ }\href {https://doi.org/10.1016/j.aop.2015.07.005}
  {\bibfield  {journal} {\bibinfo  {journal} {Annals of Physics}\ }\textbf
  {\bibinfo {volume} {361}},\ \bibinfo {pages} {605} (\bibinfo {year}
  {2015})}\BibitemShut {NoStop}%
\bibitem [{\citenamefont {Nigg}\ \emph {et~al.}(2012)\citenamefont {Nigg},
  \citenamefont {Paik}, \citenamefont {Vlastakis}, \citenamefont {Kirchmair},
  \citenamefont {Shankar}, \citenamefont {Frunzio}, \citenamefont {Devoret},
  \citenamefont {Schoelkopf},\ and\ \citenamefont {Girvin}}]{Nigg:2012}%
  \BibitemOpen
  \bibfield  {author} {\bibinfo {author} {\bibfnamefont {S.~E.}\ \bibnamefont
  {Nigg}}, \bibinfo {author} {\bibfnamefont {H.}~\bibnamefont {Paik}}, \bibinfo
  {author} {\bibfnamefont {B.}~\bibnamefont {Vlastakis}}, \bibinfo {author}
  {\bibfnamefont {G.}~\bibnamefont {Kirchmair}}, \bibinfo {author}
  {\bibfnamefont {S.}~\bibnamefont {Shankar}}, \bibinfo {author} {\bibfnamefont
  {L.}~\bibnamefont {Frunzio}}, \bibinfo {author} {\bibfnamefont {M.~H.}\
  \bibnamefont {Devoret}}, \bibinfo {author} {\bibfnamefont {R.~J.}\
  \bibnamefont {Schoelkopf}},\ and\ \bibinfo {author} {\bibfnamefont {S.~M.}\
  \bibnamefont {Girvin}},\ }\bibfield  {title} {\bibinfo {title} {Black-box
  superconducting circuit quantization},\ }\href
  {https://doi.org/10.1103/PhysRevLett.108.240502} {\bibfield  {journal}
  {\bibinfo  {journal} {Physical Review Letters}\ }\textbf {\bibinfo {volume}
  {108}},\ \bibinfo {pages} {240502} (\bibinfo {year} {2012})}\BibitemShut
  {NoStop}%
\bibitem [{\citenamefont {Solgun}\ \emph {et~al.}(2014)\citenamefont {Solgun},
  \citenamefont {Abraham},\ and\ \citenamefont {DiVincenzo}}]{Solgun:2014}%
  \BibitemOpen
  \bibfield  {author} {\bibinfo {author} {\bibfnamefont {F.}~\bibnamefont
  {Solgun}}, \bibinfo {author} {\bibfnamefont {D.~W.}\ \bibnamefont
  {Abraham}},\ and\ \bibinfo {author} {\bibfnamefont {D.~P.}\ \bibnamefont
  {DiVincenzo}},\ }\bibfield  {title} {\bibinfo {title} {Blackbox quantization
  of superconducting circuits using exact impedance synthesis},\ }\href
  {https://doi.org/10.1103/PhysRevB.90.134504} {\bibfield  {journal} {\bibinfo
  {journal} {Physical Review B}\ }\textbf {\bibinfo {volume} {90}},\ \bibinfo
  {pages} {134504} (\bibinfo {year} {2014})}\BibitemShut {NoStop}%
\bibitem [{\citenamefont {Weyl}(1923)}]{Weyl:1923}%
  \BibitemOpen
  \bibfield  {author} {\bibinfo {author} {\bibfnamefont {H.}~\bibnamefont
  {Weyl}},\ }\bibfield  {title} {\bibinfo {title} {Repartici{\'o}n de corriente
  en una red conductora. ({I}ntroducci{\'o}n al an{\'a}lisis combinatorio)},\
  }\href@noop {} {\bibfield  {journal} {\bibinfo  {journal} {Revista
  Matem{\'a}tica Hispano-Americana}\ }\textbf {\bibinfo {volume} {5}},\
  \bibinfo {pages} {153} (\bibinfo {year} {1923})}\BibitemShut {NoStop}%
\bibitem [{\citenamefont {Rymarz}\ and\ \citenamefont
  {DiVincenzo}(2023)}]{Rymarz:2023}%
  \BibitemOpen
  \bibfield  {author} {\bibinfo {author} {\bibfnamefont {M.}~\bibnamefont
  {Rymarz}}\ and\ \bibinfo {author} {\bibfnamefont {D.~P.}\ \bibnamefont
  {DiVincenzo}},\ }\bibfield  {title} {\bibinfo {title} {Consistent
  quantization of nearly singular superconducting circuits},\ }\href
  {https://doi.org/10.1103/PhysRevX.13.021017} {\bibfield  {journal} {\bibinfo
  {journal} {Physical Review X}\ }\textbf {\bibinfo {volume} {13}},\ \bibinfo
  {pages} {021017} (\bibinfo {year} {2023})}\BibitemShut {NoStop}%
\bibitem [{\citenamefont {Minev}\ \emph
  {et~al.}(2021{\natexlab{b}})\citenamefont {Minev}, \citenamefont {Leghtas},
  \citenamefont {Mundhada}, \citenamefont {Christakis}, \citenamefont {Pop},\
  and\ \citenamefont {Devoret}}]{Minev:2021b}%
  \BibitemOpen
  \bibfield  {author} {\bibinfo {author} {\bibfnamefont {Z.~K.}\ \bibnamefont
  {Minev}}, \bibinfo {author} {\bibfnamefont {Z.}~\bibnamefont {Leghtas}},
  \bibinfo {author} {\bibfnamefont {S.~O.}\ \bibnamefont {Mundhada}}, \bibinfo
  {author} {\bibfnamefont {L.}~\bibnamefont {Christakis}}, \bibinfo {author}
  {\bibfnamefont {I.~M.}\ \bibnamefont {Pop}},\ and\ \bibinfo {author}
  {\bibfnamefont {M.~H.}\ \bibnamefont {Devoret}},\ }\bibfield  {title}
  {\bibinfo {title} {Energy-participation quantization of josephson circuits},\
  }\href {https://doi.org/10.1038/s41534-021-00461-8} {\bibfield  {journal}
  {\bibinfo  {journal} {npj Quantum Information}\ }\textbf {\bibinfo {volume}
  {7}},\ \bibinfo {pages} {1} (\bibinfo {year}
  {2021}{\natexlab{b}})}\BibitemShut {NoStop}%
\bibitem [{\citenamefont {Ciani}\ \emph {et~al.}(2023)\citenamefont {Ciani},
  \citenamefont {DiVincenzo},\ and\ \citenamefont {Terhal}}]{Ciani:2023}%
  \BibitemOpen
  \bibfield  {author} {\bibinfo {author} {\bibfnamefont {A.}~\bibnamefont
  {Ciani}}, \bibinfo {author} {\bibfnamefont {D.~P.}\ \bibnamefont
  {DiVincenzo}},\ and\ \bibinfo {author} {\bibfnamefont {B.~M.}\ \bibnamefont
  {Terhal}},\ }\bibfield  {title} {\bibinfo {title} {Lecture {Notes} on
  {Quantum} {Electrical} {Circuits}},\ }\Eprint
  {https://arxiv.org/abs/arXiv:2312.05329} {arXiv:2312.05329 [quant-ph]}
  (\bibinfo {year} {2023})\BibitemShut {NoStop}%
\bibitem [{\citenamefont {Osborne}\ \emph {et~al.}(2024)\citenamefont
  {Osborne}, \citenamefont {Larson}, \citenamefont {Jones}, \citenamefont
  {Simmonds}, \citenamefont {Gyenis},\ and\ \citenamefont
  {Lucas}}]{Osborne_v2:2024}%
  \BibitemOpen
  \bibfield  {author} {\bibinfo {author} {\bibfnamefont {A.}~\bibnamefont
  {Osborne}}, \bibinfo {author} {\bibfnamefont {T.}~\bibnamefont {Larson}},
  \bibinfo {author} {\bibfnamefont {S.~G.}\ \bibnamefont {Jones}}, \bibinfo
  {author} {\bibfnamefont {R.~W.}\ \bibnamefont {Simmonds}}, \bibinfo {author}
  {\bibfnamefont {A.}~\bibnamefont {Gyenis}},\ and\ \bibinfo {author}
  {\bibfnamefont {A.}~\bibnamefont {Lucas}},\ }\bibfield  {title} {\bibinfo
  {title} {Symplectic geometry and circuit quantization},\ }\href
  {https://doi.org/10.1103/PRXQuantum.5.020309} {\bibfield  {journal} {\bibinfo
   {journal} {PRX Quantum}\ }\textbf {\bibinfo {volume} {5}},\ \bibinfo {pages}
  {020309} (\bibinfo {year} {2024})}\BibitemShut {NoStop}%
\bibitem [{\citenamefont {Faddeev}(1969)}]{Faddeev:1969}%
  \BibitemOpen
  \bibfield  {author} {\bibinfo {author} {\bibfnamefont {L.~D.}\ \bibnamefont
  {Faddeev}},\ }\bibfield  {title} {\bibinfo {title} {The feynman integral for
  singular lagrangians},\ }\href {https://doi.org/10.1007/BF01028566}
  {\bibfield  {journal} {\bibinfo  {journal} {Theoretical and Mathematical
  Physics}\ }\textbf {\bibinfo {volume} {1}},\ \bibinfo {pages} {1} (\bibinfo
  {year} {1969})}\BibitemShut {NoStop}%
\bibitem [{\citenamefont {Faddeev}\ and\ \citenamefont
  {Jackiw}(1988)}]{Faddeev:1988}%
  \BibitemOpen
  \bibfield  {author} {\bibinfo {author} {\bibfnamefont {L.}~\bibnamefont
  {Faddeev}}\ and\ \bibinfo {author} {\bibfnamefont {R.}~\bibnamefont
  {Jackiw}},\ }\bibfield  {title} {\bibinfo {title} {{Hamiltonian Reduction of
  Unconstrained and Constrained Systems}},\ }\href
  {https://doi.org/10.1103/PhysRevLett.60.1692} {\bibfield  {journal} {\bibinfo
   {journal} {Physical Review Letters}\ }\textbf {\bibinfo {volume} {60}},\
  \bibinfo {pages} {1692} (\bibinfo {year} {1988})}\BibitemShut {NoStop}%
\bibitem [{\citenamefont {{Jackiw}}(1993)}]{Jackiw:1993}%
  \BibitemOpen
  \bibfield  {author} {\bibinfo {author} {\bibfnamefont {R.}~\bibnamefont
  {{Jackiw}}},\ }\bibfield  {title} {\bibinfo {title} {{(Constrained)
  Quantization Without Tears}},\ }\Eprint
  {https://arxiv.org/abs/hep-th/9306075} {arXiv:hep-th/9306075 [hep-th]}
  (\bibinfo {year} {1993})\BibitemShut {NoStop}%
\bibitem [{\citenamefont {Toms}(2015)}]{Toms:2015}%
  \BibitemOpen
  \bibfield  {author} {\bibinfo {author} {\bibfnamefont {D.~J.}\ \bibnamefont
  {Toms}},\ }\bibfield  {title} {\bibinfo {title} {Faddeev-jackiw quantization
  and the path integral},\ }\href {https://doi.org/10.1103/PhysRevD.92.105026}
  {\bibfield  {journal} {\bibinfo  {journal} {Physical Review D}\ }\textbf
  {\bibinfo {volume} {92}},\ \bibinfo {pages} {105026} (\bibinfo {year}
  {2015})}\BibitemShut {NoStop}%
\bibitem [{\citenamefont {B\"uttiker}(1987)}]{Buettiker:1987}%
  \BibitemOpen
  \bibfield  {author} {\bibinfo {author} {\bibfnamefont {M.}~\bibnamefont
  {B\"uttiker}},\ }\bibfield  {title} {\bibinfo {title} {Zero-current
  persistent potential drop across small-capacitance josephson junctions},\
  }\href {https://doi.org/10.1103/PhysRevB.36.3548} {\bibfield  {journal}
  {\bibinfo  {journal} {Physical Review B}\ }\textbf {\bibinfo {volume} {36}},\
  \bibinfo {pages} {3548} (\bibinfo {year} {1987})}\BibitemShut {NoStop}%
\bibitem [{\citenamefont {Koch}\ \emph {et~al.}(2007)\citenamefont {Koch},
  \citenamefont {Yu}, \citenamefont {Gambetta}, \citenamefont {Houck},
  \citenamefont {Schuster}, \citenamefont {Majer}, \citenamefont {Blais},
  \citenamefont {Devoret}, \citenamefont {Girvin},\ and\ \citenamefont
  {Schoelkopf}}]{Koch:2007}%
  \BibitemOpen
  \bibfield  {author} {\bibinfo {author} {\bibfnamefont {J.}~\bibnamefont
  {Koch}}, \bibinfo {author} {\bibfnamefont {T.~M.}\ \bibnamefont {Yu}},
  \bibinfo {author} {\bibfnamefont {J.}~\bibnamefont {Gambetta}}, \bibinfo
  {author} {\bibfnamefont {A.~A.}\ \bibnamefont {Houck}}, \bibinfo {author}
  {\bibfnamefont {D.~I.}\ \bibnamefont {Schuster}}, \bibinfo {author}
  {\bibfnamefont {J.}~\bibnamefont {Majer}}, \bibinfo {author} {\bibfnamefont
  {A.}~\bibnamefont {Blais}}, \bibinfo {author} {\bibfnamefont {M.~H.}\
  \bibnamefont {Devoret}}, \bibinfo {author} {\bibfnamefont {S.~M.}\
  \bibnamefont {Girvin}},\ and\ \bibinfo {author} {\bibfnamefont {R.~J.}\
  \bibnamefont {Schoelkopf}},\ }\bibfield  {title} {\bibinfo {title}
  {Charge-insensitive qubit design derived from the cooper pair box},\ }\href
  {https://doi.org/10.1103/PhysRevA.76.042319} {\bibfield  {journal} {\bibinfo
  {journal} {Physical Review A}\ }\textbf {\bibinfo {volume} {76}},\ \bibinfo
  {pages} {042319} (\bibinfo {year} {2007})}\BibitemShut {NoStop}%
\bibitem [{\citenamefont {Rymarz}\ \emph {et~al.}(2021)\citenamefont {Rymarz},
  \citenamefont {Bosco}, \citenamefont {Ciani},\ and\ \citenamefont
  {DiVincenzo}}]{Rymarz:2021}%
  \BibitemOpen
  \bibfield  {author} {\bibinfo {author} {\bibfnamefont {M.}~\bibnamefont
  {Rymarz}}, \bibinfo {author} {\bibfnamefont {S.}~\bibnamefont {Bosco}},
  \bibinfo {author} {\bibfnamefont {A.}~\bibnamefont {Ciani}},\ and\ \bibinfo
  {author} {\bibfnamefont {D.~P.}\ \bibnamefont {DiVincenzo}},\ }\bibfield
  {title} {\bibinfo {title} {Hardware-encoding grid states in a nonreciprocal
  superconducting circuit},\ }\href
  {https://doi.org/10.1103/PhysRevX.11.011032} {\bibfield  {journal} {\bibinfo
  {journal} {Physical Review X}\ }\textbf {\bibinfo {volume} {11}},\ \bibinfo
  {pages} {011032} (\bibinfo {year} {2021})}\BibitemShut {NoStop}%
\bibitem [{\citenamefont {Egusquiza}\ \emph {et~al.}(2022)\citenamefont
  {Egusquiza}, \citenamefont {I\~niguez}, \citenamefont {Rico},\ and\
  \citenamefont {Villarino}}]{Egusquiza:2022b}%
  \BibitemOpen
  \bibfield  {author} {\bibinfo {author} {\bibfnamefont {I.~L.}\ \bibnamefont
  {Egusquiza}}, \bibinfo {author} {\bibfnamefont {A.}~\bibnamefont
  {I\~niguez}}, \bibinfo {author} {\bibfnamefont {E.}~\bibnamefont {Rico}},\
  and\ \bibinfo {author} {\bibfnamefont {A.}~\bibnamefont {Villarino}},\
  }\bibfield  {title} {\bibinfo {title} {Role of anomalous symmetry in
  $0\text{\ensuremath{-}}\ensuremath{\pi}$ qubits},\ }\href
  {https://doi.org/10.1103/PhysRevB.105.L201104} {\bibfield  {journal}
  {\bibinfo  {journal} {Physical Review B}\ }\textbf {\bibinfo {volume}
  {105}},\ \bibinfo {pages} {L201104} (\bibinfo {year} {2022})}\BibitemShut
  {NoStop}%
\bibitem [{\citenamefont {Bouchiat}\ \emph {et~al.}(1998)\citenamefont
  {Bouchiat}, \citenamefont {Vion}, \citenamefont {Joyez}, \citenamefont
  {Esteve},\ and\ \citenamefont {Devoret}}]{Bouchiat:1998}%
  \BibitemOpen
  \bibfield  {author} {\bibinfo {author} {\bibfnamefont {V.}~\bibnamefont
  {Bouchiat}}, \bibinfo {author} {\bibfnamefont {D.}~\bibnamefont {Vion}},
  \bibinfo {author} {\bibfnamefont {P.}~\bibnamefont {Joyez}}, \bibinfo
  {author} {\bibfnamefont {D.}~\bibnamefont {Esteve}},\ and\ \bibinfo {author}
  {\bibfnamefont {M.~H.}\ \bibnamefont {Devoret}},\ }\bibfield  {title}
  {\bibinfo {title} {Quantum coherence with a single cooper pair},\ }\href
  {https://doi.org/10.1238/physica.topical.076a00165} {\bibfield  {journal}
  {\bibinfo  {journal} {Physica Scripta}\ }\textbf {\bibinfo {volume} {T76}},\
  \bibinfo {pages} {165} (\bibinfo {year} {1998})}\BibitemShut {NoStop}%
\bibitem [{\citenamefont {Chua}(1980{\natexlab{a}})}]{Chua:1980}%
  \BibitemOpen
  \bibfield  {author} {\bibinfo {author} {\bibfnamefont {L.}~\bibnamefont
  {Chua}},\ }\bibfield  {title} {\bibinfo {title} {Device modeling via
  nonlinear circuit elements},\ }\href
  {https://doi.org/10.1109/TCS.1980.1084742} {\bibfield  {journal} {\bibinfo
  {journal} {IEEE Transactions on Circuits and Systems}\ }\textbf {\bibinfo
  {volume} {27}},\ \bibinfo {pages} {1014} (\bibinfo {year}
  {1980}{\natexlab{a}})}\BibitemShut {NoStop}%
\bibitem [{\citenamefont {{Parra-Rodriguez}}\ and\ \citenamefont
  {Egusquiza}(2024)}]{ParraRodriguez:2024b}%
  \BibitemOpen
  \bibfield  {author} {\bibinfo {author} {\bibfnamefont {A.}~\bibnamefont
  {{Parra-Rodriguez}}}\ and\ \bibinfo {author} {\bibfnamefont {I.~L.}\
  \bibnamefont {Egusquiza}},\ }\bibfield  {title} {\bibinfo {title}
  {Geometrical description and {{Faddeev-Jackiw}} quantization of electrical
  networks},\ }\href {https://doi.org/10.22331/q-2024-09-09-1466} {\bibfield
  {journal} {\bibinfo  {journal} {Quantum}\ }\textbf {\bibinfo {volume} {8}},\
  \bibinfo {pages} {1466} (\bibinfo {year} {2024})}\BibitemShut {NoStop}%
\bibitem [{\citenamefont {Egusquiza}\ and\ \citenamefont
  {{Parra-Rodriguez}}()}]{PRXComment}%
  \BibitemOpen
  \bibfield  {author} {\bibinfo {author} {\bibfnamefont {I.~L.}\ \bibnamefont
  {Egusquiza}}\ and\ \bibinfo {author} {\bibfnamefont {A.}~\bibnamefont
  {{Parra-Rodriguez}}},\ }\bibfield  {title} {\bibinfo {title} {Comment on
  ``{C}onsistent {Q}uantization of {N}early {S}ingular {S}uperconducting
  {C}ircuits''},\ }\href@noop {} {\bibinfo  {journal} {Physical Review X}\ ,\
  \bibinfo {pages} {to appear}}\BibitemShut {NoStop}%
\bibitem [{\citenamefont {Heaviside}(1971)}]{Heaviside:1971}%
  \BibitemOpen
\bibfield  {journal} {  }\bibfield  {author} {\bibinfo {author} {\bibfnamefont
  {O.}~\bibnamefont {Heaviside}},\ }\href@noop {} {\emph {\bibinfo {title}
  {Electromagnetic theory. V. 1}}},\ \bibinfo {edition} {3rd}\ ed.\ (\bibinfo
  {publisher} {Chelsea},\ \bibinfo {address} {New York},\ \bibinfo {year}
  {1971})\BibitemShut {NoStop}%
\bibitem [{\citenamefont {Veselago}(1968)}]{Veselago:1968}%
  \BibitemOpen
  \bibfield  {author} {\bibinfo {author} {\bibfnamefont {V.~G.}\ \bibnamefont
  {Veselago}},\ }\bibfield  {title} {\bibinfo {title} {The electrodynamics of
  substances with simultaneously negative values of $\epsilon$ and $\mu$},\
  }\href {https://doi.org/10.1070/PU1968v010n04ABEH003699} {\bibfield
  {journal} {\bibinfo  {journal} {Soviet Physics Uspekhi}\ }\textbf {\bibinfo
  {volume} {10}},\ \bibinfo {pages} {509} (\bibinfo {year} {1968})}\BibitemShut
  {NoStop}%
\bibitem [{\citenamefont {Egger}\ and\ \citenamefont
  {Wilhelm}(2013)}]{Egger:2013}%
  \BibitemOpen
  \bibfield  {author} {\bibinfo {author} {\bibfnamefont {D.~J.}\ \bibnamefont
  {Egger}}\ and\ \bibinfo {author} {\bibfnamefont {F.~K.}\ \bibnamefont
  {Wilhelm}},\ }\bibfield  {title} {\bibinfo {title} {Multimode circuit quantum
  electrodynamics with hybrid metamaterial transmission lines},\ }\href
  {https://doi.org/10.1103/PhysRevLett.111.163601} {\bibfield  {journal}
  {\bibinfo  {journal} {Physical Review Lett.}\ }\textbf {\bibinfo {volume}
  {111}},\ \bibinfo {pages} {163601} (\bibinfo {year} {2013})}\BibitemShut
  {NoStop}%
\bibitem [{\citenamefont {Indrajeet}\ \emph {et~al.}(2020)\citenamefont
  {Indrajeet}, \citenamefont {Wang}, \citenamefont {Hutchings}, \citenamefont
  {Taketani}, \citenamefont {Wilhelm}, \citenamefont {LaHaye},\ and\
  \citenamefont {Plourde}}]{Indrajeet:2020}%
  \BibitemOpen
  \bibfield  {author} {\bibinfo {author} {\bibfnamefont {S.}~\bibnamefont
  {Indrajeet}}, \bibinfo {author} {\bibfnamefont {H.}~\bibnamefont {Wang}},
  \bibinfo {author} {\bibfnamefont {M.~D.}\ \bibnamefont {Hutchings}}, \bibinfo
  {author} {\bibfnamefont {B.~G.}\ \bibnamefont {Taketani}}, \bibinfo {author}
  {\bibfnamefont {F.~K.}\ \bibnamefont {Wilhelm}}, \bibinfo {author}
  {\bibfnamefont {M.~D.}\ \bibnamefont {LaHaye}},\ and\ \bibinfo {author}
  {\bibfnamefont {B.~L.~T.}\ \bibnamefont {Plourde}},\ }\bibfield  {title}
  {\bibinfo {title} {Coupling a superconducting qubit to a left-handed
  metamaterial resonator},\ }\href
  {https://doi.org/10.1103/PhysRevApplied.14.064033} {\bibfield  {journal}
  {\bibinfo  {journal} {Physical Review Applied}\ }\textbf {\bibinfo {volume}
  {14}},\ \bibinfo {pages} {064033} (\bibinfo {year} {2020})}\BibitemShut
  {NoStop}%
\bibitem [{\citenamefont {Liberal}\ and\ \citenamefont
  {Ziolkowski}(2021)}]{Liberal:2021}%
  \BibitemOpen
  \bibfield  {author} {\bibinfo {author} {\bibfnamefont {I.}~\bibnamefont
  {Liberal}}\ and\ \bibinfo {author} {\bibfnamefont {R.~W.}\ \bibnamefont
  {Ziolkowski}},\ }\bibfield  {title} {\bibinfo {title} {Nonperturbative decay
  dynamics in metamaterial waveguides},\ }\href
  {https://doi.org/10.1063/5.0044103} {\bibfield  {journal} {\bibinfo
  {journal} {Applied Physics Letters}\ }\textbf {\bibinfo {volume} {118}},\
  \bibinfo {pages} {111103} (\bibinfo {year} {2021})}\BibitemShut {NoStop}%
\bibitem [{\citenamefont {{Solgun}}\ \emph {et~al.}(2019)\citenamefont
  {{Solgun}}, \citenamefont {{DiVincenzo}},\ and\ \citenamefont
  {{Gambetta}}}]{Solgun:2019}%
  \BibitemOpen
  \bibfield  {author} {\bibinfo {author} {\bibfnamefont {F.}~\bibnamefont
  {{Solgun}}}, \bibinfo {author} {\bibfnamefont {D.~P.}\ \bibnamefont
  {{DiVincenzo}}},\ and\ \bibinfo {author} {\bibfnamefont {J.~M.}\ \bibnamefont
  {{Gambetta}}},\ }\bibfield  {title} {\bibinfo {title} {Simple impedance
  response formulas for the dispersive interaction rates in the effective
  hamiltonians of low anharmonicity superconducting qubits},\ }\href
  {https://doi.org/10.1109/TMTT.2019.2893639} {\bibfield  {journal} {\bibinfo
  {journal} {IEEE Transactions on Microwave Theory and Techniques}\ }\textbf
  {\bibinfo {volume} {67}},\ \bibinfo {pages} {928} (\bibinfo {year}
  {2019})}\BibitemShut {NoStop}%
\bibitem [{\citenamefont {Labarca}\ \emph {et~al.}(2023)\citenamefont
  {Labarca}, \citenamefont {Benhayoune-Khadraoui}, \citenamefont {Blais},\ and\
  \citenamefont {Parra-Rodriguez}}]{Labarca:2023}%
  \BibitemOpen
  \bibfield  {author} {\bibinfo {author} {\bibfnamefont {L.}~\bibnamefont
  {Labarca}}, \bibinfo {author} {\bibfnamefont {O.}~\bibnamefont
  {Benhayoune-Khadraoui}}, \bibinfo {author} {\bibfnamefont {A.}~\bibnamefont
  {Blais}},\ and\ \bibinfo {author} {\bibfnamefont {A.}~\bibnamefont
  {Parra-Rodriguez}},\ }\bibfield  {title} {\bibinfo {title} {Toolbox for
  nonreciprocal dispersive models in circuit {QED}},\ }\Eprint
  {https://arxiv.org/abs/2312.08354} {arXiv:2312.08354 [quant-ph]}  (\bibinfo
  {year} {2023})\BibitemShut {NoStop}%
\bibitem [{\citenamefont {Chua}(1980{\natexlab{b}})}]{Chua:1980a}%
  \BibitemOpen
  \bibfield  {author} {\bibinfo {author} {\bibfnamefont {L.}~\bibnamefont
  {Chua}},\ }\bibfield  {title} {\bibinfo {title} {Dynamic nonlinear networks:
  State-of-the-art},\ }\href {https://doi.org/10.1109/TCS.1980.1084745}
  {\bibfield  {journal} {\bibinfo  {journal} {IEEE Transactions on Circuits and
  Systems}\ }\textbf {\bibinfo {volume} {27}},\ \bibinfo {pages} {1059}
  (\bibinfo {year} {1980}{\natexlab{b}})}\BibitemShut {NoStop}%
\bibitem [{\citenamefont {Roska}(1981)}]{Roska:1981}%
  \BibitemOpen
  \bibfield  {author} {\bibinfo {author} {\bibfnamefont {T.}~\bibnamefont
  {Roska}},\ }\bibfield  {title} {\bibinfo {title} {The limits of modeling of
  nonlinear circuits},\ }\href {https://doi.org/10.1109/TCS.1981.1084974}
  {\bibfield  {journal} {\bibinfo  {journal} {IEEE Transactions on Circuits and
  Systems}\ }\textbf {\bibinfo {volume} {28}},\ \bibinfo {pages} {212}
  (\bibinfo {year} {1981})}\BibitemShut {NoStop}%
\bibitem [{\citenamefont {Miano}\ \emph {et~al.}(2023)\citenamefont {Miano},
  \citenamefont {Joshi}, \citenamefont {Liu}, \citenamefont {Dai},
  \citenamefont {Parakh}, \citenamefont {Frunzio},\ and\ \citenamefont
  {Devoret}}]{Miano:2023}%
  \BibitemOpen
  \bibfield  {author} {\bibinfo {author} {\bibfnamefont {A.}~\bibnamefont
  {Miano}}, \bibinfo {author} {\bibfnamefont {V.~R.}\ \bibnamefont {Joshi}},
  \bibinfo {author} {\bibfnamefont {G.}~\bibnamefont {Liu}}, \bibinfo {author}
  {\bibfnamefont {W.}~\bibnamefont {Dai}}, \bibinfo {author} {\bibfnamefont
  {P.~D.}\ \bibnamefont {Parakh}}, \bibinfo {author} {\bibfnamefont
  {L.}~\bibnamefont {Frunzio}},\ and\ \bibinfo {author} {\bibfnamefont {M.~H.}\
  \bibnamefont {Devoret}},\ }\bibfield  {title} {\bibinfo {title} {Hamiltonian
  extrema of an arbitrary flux-biased josephson circuit},\ }\href
  {https://doi.org/10.1103/PRXQuantum.4.030324} {\bibfield  {journal} {\bibinfo
   {journal} {PRX Quantum}\ }\textbf {\bibinfo {volume} {4}},\ \bibinfo {pages}
  {030324} (\bibinfo {year} {2023})}\BibitemShut {NoStop}%
\bibitem [{\citenamefont {Walter}(1973)}]{Walter:1973}%
  \BibitemOpen
  \bibfield  {author} {\bibinfo {author} {\bibfnamefont {J.}~\bibnamefont
  {Walter}},\ }\bibfield  {title} {\bibinfo {title} {Regular eigenvalue
  problems with eigenvalue parameter in the boundary condition},\ }\href
  {https://doi.org/10.1007/BF01177870} {\bibfield  {journal} {\bibinfo
  {journal} {Mathematische Zeitschrift}\ }\textbf {\bibinfo {volume} {133}},\
  \bibinfo {pages} {301} (\bibinfo {year} {1973})}\BibitemShut {NoStop}%
\bibitem [{\citenamefont {Fulton}(1977)}]{Fulton:1977}%
  \BibitemOpen
  \bibfield  {author} {\bibinfo {author} {\bibfnamefont {C.~T.}\ \bibnamefont
  {Fulton}},\ }\bibfield  {title} {\bibinfo {title} {Two-point boundary value
  problems with eigenvalue parameter contained in the boundary conditions},\
  }\href {https://doi.org/10.1017/S030821050002521X} {\bibfield  {journal}
  {\bibinfo  {journal} {Proceedings of the Royal Society of Edinburgh: Section
  A Mathematics}\ }\textbf {\bibinfo {volume} {77}},\ \bibinfo {pages} {293}
  (\bibinfo {year} {1977})}\BibitemShut {NoStop}%
\bibitem [{\citenamefont {Forn-D{\'{i}}az}\ \emph {et~al.}(2017)\citenamefont
  {Forn-D{\'{i}}az}, \citenamefont {Garc{\'{i}}a-Ripoll}, \citenamefont
  {Peropadre}, \citenamefont {Orgiazzi}, \citenamefont {Yurtalan},
  \citenamefont {Belyansky}, \citenamefont {Wilson},\ and\ \citenamefont
  {Lupascu}}]{FornDiaz:2017}%
  \BibitemOpen
  \bibfield  {author} {\bibinfo {author} {\bibfnamefont {P.}~\bibnamefont
  {Forn-D{\'{i}}az}}, \bibinfo {author} {\bibfnamefont {J.}~\bibnamefont
  {Garc{\'{i}}a-Ripoll}}, \bibinfo {author} {\bibfnamefont {B.}~\bibnamefont
  {Peropadre}}, \bibinfo {author} {\bibfnamefont {J.-L.}\ \bibnamefont
  {Orgiazzi}}, \bibinfo {author} {\bibfnamefont {M.}~\bibnamefont {Yurtalan}},
  \bibinfo {author} {\bibfnamefont {R.}~\bibnamefont {Belyansky}}, \bibinfo
  {author} {\bibfnamefont {C.}~\bibnamefont {Wilson}},\ and\ \bibinfo {author}
  {\bibfnamefont {A.}~\bibnamefont {Lupascu}},\ }\bibfield  {title} {\bibinfo
  {title} {{Ultrastrong coupling of a single artificial atom to an
  electromagnetic continuum in the nonperturbative regime}},\ }\href
  {https://doi.org/10.1038/nphys3905} {\bibfield  {journal} {\bibinfo
  {journal} {Nature Physics}\ }\textbf {\bibinfo {volume} {13}},\ \bibinfo
  {pages} {39} (\bibinfo {year} {2017})}\BibitemShut {NoStop}%
\bibitem [{\citenamefont {Frisk~Kockum}\ \emph {et~al.}(2019)\citenamefont
  {Frisk~Kockum}, \citenamefont {Miranowicz}, \citenamefont {De~Liberato},
  \citenamefont {Savasta},\ and\ \citenamefont {Nori}}]{Kockum:2019}%
  \BibitemOpen
  \bibfield  {author} {\bibinfo {author} {\bibfnamefont {A.}~\bibnamefont
  {Frisk~Kockum}}, \bibinfo {author} {\bibfnamefont {A.}~\bibnamefont
  {Miranowicz}}, \bibinfo {author} {\bibfnamefont {S.}~\bibnamefont
  {De~Liberato}}, \bibinfo {author} {\bibfnamefont {S.}~\bibnamefont
  {Savasta}},\ and\ \bibinfo {author} {\bibfnamefont {F.}~\bibnamefont
  {Nori}},\ }\bibfield  {title} {\bibinfo {title} {Ultrastrong coupling between
  light and matter},\ }\href {https://doi.org/10.1038/s42254-018-0006-2}
  {\bibfield  {journal} {\bibinfo  {journal} {Nature Reviews Physics}\ }\textbf
  {\bibinfo {volume} {1}},\ \bibinfo {pages} {19} (\bibinfo {year}
  {2019})}\BibitemShut {NoStop}%
\bibitem [{\citenamefont {Mehta}\ \emph {et~al.}(2022)\citenamefont {Mehta},
  \citenamefont {Ciuti}, \citenamefont {Kuzmin},\ and\ \citenamefont
  {Manucharyan}}]{Mehta:2022}%
  \BibitemOpen
  \bibfield  {author} {\bibinfo {author} {\bibfnamefont {N.}~\bibnamefont
  {Mehta}}, \bibinfo {author} {\bibfnamefont {C.}~\bibnamefont {Ciuti}},
  \bibinfo {author} {\bibfnamefont {R.}~\bibnamefont {Kuzmin}},\ and\ \bibinfo
  {author} {\bibfnamefont {V.~E.}\ \bibnamefont {Manucharyan}},\ }\bibfield
  {title} {\bibinfo {title} {Theory of strong down-conversion in multi-mode
  cavity and circuit {QED}},\ }\Eprint {https://arxiv.org/abs/arXiv:2210.14681}
  {arXiv:2210.14681 [quant-ph]}  (\bibinfo {year} {2022})\BibitemShut {NoStop}%
\bibitem [{\citenamefont {Feynman}\ and\ \citenamefont
  {Vernon}(1963)}]{Feynman:1963}%
  \BibitemOpen
  \bibfield  {author} {\bibinfo {author} {\bibfnamefont {R.}~\bibnamefont
  {Feynman}}\ and\ \bibinfo {author} {\bibfnamefont {F.}~\bibnamefont
  {Vernon}},\ }\bibfield  {title} {\bibinfo {title} {The theory of a general
  quantum system interacting with a linear dissipative system},\ }\href
  {https://doi.org/https://doi.org/10.1016/0003-4916(63)90068-X} {\bibfield
  {journal} {\bibinfo  {journal} {Annals of Physics}\ }\textbf {\bibinfo
  {volume} {24}},\ \bibinfo {pages} {118} (\bibinfo {year} {1963})}\BibitemShut
  {NoStop}%
\bibitem [{\citenamefont {Cattaneo}\ and\ \citenamefont
  {Paraoanu}(2021)}]{Cattaneo:2021}%
  \BibitemOpen
  \bibfield  {author} {\bibinfo {author} {\bibfnamefont {M.}~\bibnamefont
  {Cattaneo}}\ and\ \bibinfo {author} {\bibfnamefont {G.~S.}\ \bibnamefont
  {Paraoanu}},\ }\bibfield  {title} {\bibinfo {title} {Engineering dissipation
  with resistive elements in circuit quantum electrodynamics},\ }\href
  {https://doi.org/https://doi.org/10.1002/qute.202100054} {\bibfield
  {journal} {\bibinfo  {journal} {Advanced Quantum Technologies}\ }\textbf
  {\bibinfo {volume} {4}},\ \bibinfo {pages} {2100054} (\bibinfo {year}
  {2021})}\BibitemShut {NoStop}%
\bibitem [{\citenamefont {Ashida}\ \emph {et~al.}(2022)\citenamefont {Ashida},
  \citenamefont {Yokota}, \citenamefont {\.{I}mamo\u{g}lu},\ and\ \citenamefont
  {Demler}}]{Ashida:2022}%
  \BibitemOpen
  \bibfield  {author} {\bibinfo {author} {\bibfnamefont {Y.}~\bibnamefont
  {Ashida}}, \bibinfo {author} {\bibfnamefont {T.}~\bibnamefont {Yokota}},
  \bibinfo {author} {\bibfnamefont {A.}~\bibnamefont {\.{I}mamo\u{g}lu}},\ and\
  \bibinfo {author} {\bibfnamefont {E.}~\bibnamefont {Demler}},\ }\bibfield
  {title} {\bibinfo {title} {Nonperturbative waveguide quantum
  electrodynamics},\ }\href {https://doi.org/10.1103/PhysRevResearch.4.023194}
  {\bibfield  {journal} {\bibinfo  {journal} {Physical Review Research}\
  }\textbf {\bibinfo {volume} {4}},\ \bibinfo {pages} {023194} (\bibinfo {year}
  {2022})}\BibitemShut {NoStop}%
\bibitem [{\citenamefont {Paulisch}\ \emph {et~al.}(2016)\citenamefont
  {Paulisch}, \citenamefont {Kimble},\ and\ \citenamefont
  {González-Tudela}}]{Paulisch:2016}%
  \BibitemOpen
  \bibfield  {author} {\bibinfo {author} {\bibfnamefont {V.}~\bibnamefont
  {Paulisch}}, \bibinfo {author} {\bibfnamefont {H.~J.}\ \bibnamefont
  {Kimble}},\ and\ \bibinfo {author} {\bibfnamefont {A.}~\bibnamefont
  {González-Tudela}},\ }\bibfield  {title} {\bibinfo {title} {Universal
  quantum computation in waveguide {QED} using decoherence free subspaces},\
  }\href {https://doi.org/10.1088/1367-2630/18/4/043041} {\bibfield  {journal}
  {\bibinfo  {journal} {New Journal of Physics}\ }\textbf {\bibinfo {volume}
  {18}},\ \bibinfo {pages} {043041} (\bibinfo {year} {2016})}\BibitemShut
  {NoStop}%
\bibitem [{\citenamefont {Mirhosseini}\ \emph {et~al.}(2019)\citenamefont
  {Mirhosseini}, \citenamefont {Kim}, \citenamefont {Zhang}, \citenamefont
  {Sipahigil}, \citenamefont {Dieterle}, \citenamefont {Keller}, \citenamefont
  {Asenjo-Garcia}, \citenamefont {Chang},\ and\ \citenamefont
  {Painter}}]{Mirhosseini:2019}%
  \BibitemOpen
  \bibfield  {author} {\bibinfo {author} {\bibfnamefont {M.}~\bibnamefont
  {Mirhosseini}}, \bibinfo {author} {\bibfnamefont {E.}~\bibnamefont {Kim}},
  \bibinfo {author} {\bibfnamefont {X.}~\bibnamefont {Zhang}}, \bibinfo
  {author} {\bibfnamefont {A.}~\bibnamefont {Sipahigil}}, \bibinfo {author}
  {\bibfnamefont {P.~B.}\ \bibnamefont {Dieterle}}, \bibinfo {author}
  {\bibfnamefont {A.~J.}\ \bibnamefont {Keller}}, \bibinfo {author}
  {\bibfnamefont {A.}~\bibnamefont {Asenjo-Garcia}}, \bibinfo {author}
  {\bibfnamefont {D.~E.}\ \bibnamefont {Chang}},\ and\ \bibinfo {author}
  {\bibfnamefont {O.}~\bibnamefont {Painter}},\ }\bibfield  {title} {\bibinfo
  {title} {Cavity quantum electrodynamics with atom-like mirrors},\ }\href
  {https://doi.org/10.1038/s41586-019-1196-1} {\bibfield  {journal} {\bibinfo
  {journal} {Nature}\ }\textbf {\bibinfo {volume} {569}},\ \bibinfo {pages}
  {692} (\bibinfo {year} {2019})}\BibitemShut {NoStop}%
\bibitem [{\citenamefont {Kannan}\ \emph {et~al.}(2020)\citenamefont {Kannan},
  \citenamefont {Ruckriegel}, \citenamefont {Campbell}, \citenamefont
  {Frisk~Kockum}, \citenamefont {Braumüller}, \citenamefont {Kim},
  \citenamefont {Kjaergaard}, \citenamefont {Krantz}, \citenamefont {Melville},
  \citenamefont {Niedzielski}, \citenamefont {Vepsäläinen}, \citenamefont
  {Winik}, \citenamefont {Yoder}, \citenamefont {Nori}, \citenamefont
  {Orlando}, \citenamefont {Gustavsson},\ and\ \citenamefont
  {Oliver}}]{Kannan:2020}%
  \BibitemOpen
  \bibfield  {author} {\bibinfo {author} {\bibfnamefont {B.}~\bibnamefont
  {Kannan}}, \bibinfo {author} {\bibfnamefont {M.~J.}\ \bibnamefont
  {Ruckriegel}}, \bibinfo {author} {\bibfnamefont {D.~L.}\ \bibnamefont
  {Campbell}}, \bibinfo {author} {\bibfnamefont {A.}~\bibnamefont
  {Frisk~Kockum}}, \bibinfo {author} {\bibfnamefont {J.}~\bibnamefont
  {Braumüller}}, \bibinfo {author} {\bibfnamefont {D.~K.}\ \bibnamefont
  {Kim}}, \bibinfo {author} {\bibfnamefont {M.}~\bibnamefont {Kjaergaard}},
  \bibinfo {author} {\bibfnamefont {P.}~\bibnamefont {Krantz}}, \bibinfo
  {author} {\bibfnamefont {A.}~\bibnamefont {Melville}}, \bibinfo {author}
  {\bibfnamefont {B.~M.}\ \bibnamefont {Niedzielski}}, \bibinfo {author}
  {\bibfnamefont {A.}~\bibnamefont {Vepsäläinen}}, \bibinfo {author}
  {\bibfnamefont {R.}~\bibnamefont {Winik}}, \bibinfo {author} {\bibfnamefont
  {J.~L.}\ \bibnamefont {Yoder}}, \bibinfo {author} {\bibfnamefont
  {F.}~\bibnamefont {Nori}}, \bibinfo {author} {\bibfnamefont {T.~P.}\
  \bibnamefont {Orlando}}, \bibinfo {author} {\bibfnamefont {S.}~\bibnamefont
  {Gustavsson}},\ and\ \bibinfo {author} {\bibfnamefont {W.~D.}\ \bibnamefont
  {Oliver}},\ }\bibfield  {title} {\bibinfo {title} {Waveguide quantum
  electrodynamics with superconducting artificial giant atoms},\ }\href
  {https://doi.org/10.1038/s41586-020-2529-9} {\bibfield  {journal} {\bibinfo
  {journal} {Nature}\ }\textbf {\bibinfo {volume} {583}},\ \bibinfo {pages}
  {775} (\bibinfo {year} {2020})}\BibitemShut {NoStop}%
\bibitem [{\citenamefont {Sheremet}\ \emph {et~al.}(2023)\citenamefont
  {Sheremet}, \citenamefont {Petrov}, \citenamefont {Iorsh}, \citenamefont
  {Poshakinskiy},\ and\ \citenamefont {Poddubny}}]{Sheremet:2023}%
  \BibitemOpen
  \bibfield  {author} {\bibinfo {author} {\bibfnamefont {A.~S.}\ \bibnamefont
  {Sheremet}}, \bibinfo {author} {\bibfnamefont {M.~I.}\ \bibnamefont
  {Petrov}}, \bibinfo {author} {\bibfnamefont {I.~V.}\ \bibnamefont {Iorsh}},
  \bibinfo {author} {\bibfnamefont {A.~V.}\ \bibnamefont {Poshakinskiy}},\ and\
  \bibinfo {author} {\bibfnamefont {A.~N.}\ \bibnamefont {Poddubny}},\
  }\bibfield  {title} {\bibinfo {title} {Waveguide quantum electrodynamics:
  Collective radiance and photon-photon correlations},\ }\href
  {https://doi.org/10.1103/RevModPhys.95.015002} {\bibfield  {journal}
  {\bibinfo  {journal} {Review Modern Physics}\ }\textbf {\bibinfo {volume}
  {95}},\ \bibinfo {pages} {015002} (\bibinfo {year} {2023})}\BibitemShut
  {NoStop}%
\bibitem [{\citenamefont {Pechenezhskiy}\ \emph {et~al.}(2020)\citenamefont
  {Pechenezhskiy}, \citenamefont {Mencia}, \citenamefont {Nguyen},
  \citenamefont {Lin},\ and\ \citenamefont {Manucharyan}}]{Pechenezhskiy:2020}%
  \BibitemOpen
  \bibfield  {author} {\bibinfo {author} {\bibfnamefont {I.~V.}\ \bibnamefont
  {Pechenezhskiy}}, \bibinfo {author} {\bibfnamefont {R.~A.}\ \bibnamefont
  {Mencia}}, \bibinfo {author} {\bibfnamefont {L.~B.}\ \bibnamefont {Nguyen}},
  \bibinfo {author} {\bibfnamefont {Y.-H.}\ \bibnamefont {Lin}},\ and\ \bibinfo
  {author} {\bibfnamefont {V.~E.}\ \bibnamefont {Manucharyan}},\ }\bibfield
  {title} {\bibinfo {title} {The superconducting quasicharge qubit},\ }\href
  {https://doi.org/10.1038/s41586-020-2687-9} {\bibfield  {journal} {\bibinfo
  {journal} {Nature}\ }\textbf {\bibinfo {volume} {585}},\ \bibinfo {pages}
  {368} (\bibinfo {year} {2020})}\BibitemShut {NoStop}%
\bibitem [{\citenamefont {Crescini}\ \emph {et~al.}(2023)\citenamefont
  {Crescini}, \citenamefont {Cailleaux}, \citenamefont {Guichard},
  \citenamefont {Naud}, \citenamefont {Buisson}, \citenamefont {W.~Murch},\
  and\ \citenamefont {Roch}}]{Crescini:2023}%
  \BibitemOpen
  \bibfield  {author} {\bibinfo {author} {\bibfnamefont {N.}~\bibnamefont
  {Crescini}}, \bibinfo {author} {\bibfnamefont {S.}~\bibnamefont {Cailleaux}},
  \bibinfo {author} {\bibfnamefont {W.}~\bibnamefont {Guichard}}, \bibinfo
  {author} {\bibfnamefont {C.}~\bibnamefont {Naud}}, \bibinfo {author}
  {\bibfnamefont {O.}~\bibnamefont {Buisson}}, \bibinfo {author} {\bibfnamefont
  {K.}~\bibnamefont {W.~Murch}},\ and\ \bibinfo {author} {\bibfnamefont
  {N.}~\bibnamefont {Roch}},\ }\bibfield  {title} {\bibinfo {title} {Evidence
  of dual {Shapiro} steps in a {Josephson} junction array},\ }\href
  {https://doi.org/10.1038/s41567-023-01961-4} {\bibfield  {journal} {\bibinfo
  {journal} {Nature Physics}\ }\textbf {\bibinfo {volume} {19}},\ \bibinfo
  {pages} {851} (\bibinfo {year} {2023})}\BibitemShut {NoStop}%
\bibitem [{\citenamefont {Ardati}\ \emph {et~al.}(2024)\citenamefont {Ardati},
  \citenamefont {L\'eger}, \citenamefont {Kumar}, \citenamefont {Suresh},
  \citenamefont {Nicolas}, \citenamefont {Mori}, \citenamefont {D'Esposito},
  \citenamefont {Vakhtel}, \citenamefont {Buisson}, \citenamefont {Ficheux},\
  and\ \citenamefont {Roch}}]{Ardati:2024}%
  \BibitemOpen
  \bibfield  {author} {\bibinfo {author} {\bibfnamefont {W.}~\bibnamefont
  {Ardati}}, \bibinfo {author} {\bibfnamefont {S.}~\bibnamefont {L\'eger}},
  \bibinfo {author} {\bibfnamefont {S.}~\bibnamefont {Kumar}}, \bibinfo
  {author} {\bibfnamefont {V.~N.}\ \bibnamefont {Suresh}}, \bibinfo {author}
  {\bibfnamefont {D.}~\bibnamefont {Nicolas}}, \bibinfo {author} {\bibfnamefont
  {C.}~\bibnamefont {Mori}}, \bibinfo {author} {\bibfnamefont {F.}~\bibnamefont
  {D'Esposito}}, \bibinfo {author} {\bibfnamefont {T.}~\bibnamefont {Vakhtel}},
  \bibinfo {author} {\bibfnamefont {O.}~\bibnamefont {Buisson}}, \bibinfo
  {author} {\bibfnamefont {Q.}~\bibnamefont {Ficheux}},\ and\ \bibinfo {author}
  {\bibfnamefont {N.}~\bibnamefont {Roch}},\ }\bibfield  {title} {\bibinfo
  {title} {Using bifluxon tunneling to protect the fluxonium qubit},\ }\href
  {https://doi.org/10.1103/PhysRevX.14.041014} {\bibfield  {journal} {\bibinfo
  {journal} {Physical Review X}\ }\textbf {\bibinfo {volume} {14}},\ \bibinfo
  {pages} {041014} (\bibinfo {year} {2024})}\BibitemShut {NoStop}%
\bibitem [{\citenamefont {Kuzmin}\ \emph {et~al.}(2024)\citenamefont {Kuzmin},
  \citenamefont {Mehta}, \citenamefont {Grabon}, \citenamefont {Mencia},
  \citenamefont {Burshtein}, \citenamefont {Goldstein},\ and\ \citenamefont
  {Manucharyan}}]{Kuzmin:2024}%
  \BibitemOpen
  \bibfield  {author} {\bibinfo {author} {\bibfnamefont {R.}~\bibnamefont
  {Kuzmin}}, \bibinfo {author} {\bibfnamefont {N.}~\bibnamefont {Mehta}},
  \bibinfo {author} {\bibfnamefont {N.}~\bibnamefont {Grabon}}, \bibinfo
  {author} {\bibfnamefont {R.~A.}\ \bibnamefont {Mencia}}, \bibinfo {author}
  {\bibfnamefont {A.}~\bibnamefont {Burshtein}}, \bibinfo {author}
  {\bibfnamefont {M.}~\bibnamefont {Goldstein}},\ and\ \bibinfo {author}
  {\bibfnamefont {V.~E.}\ \bibnamefont {Manucharyan}},\ }\bibfield  {title}
  {\bibinfo {title} {Observation of the {Schmid}–{Bulgadaev} dissipative
  quantum phase transition},\ }\href
  {https://doi.org/10.1038/s41567-024-02695-7} {\bibfield  {journal} {\bibinfo
  {journal} {Nature Physics}\ ,\ \bibinfo {pages} {1}} (\bibinfo {year}
  {2024})},\ \bibinfo {note} {publisher: Nature Publishing Group}\BibitemShut
  {NoStop}%
\bibitem [{\citenamefont {Labarca}\ \emph {et~al.}(2024)\citenamefont
  {Labarca}, \citenamefont {Benhayoune-Khadraoui}, \citenamefont {Blais},\ and\
  \citenamefont {Parra-Rodriguez}}]{Labarca:2024}%
  \BibitemOpen
  \bibfield  {author} {\bibinfo {author} {\bibfnamefont {L.}~\bibnamefont
  {Labarca}}, \bibinfo {author} {\bibfnamefont {O.}~\bibnamefont
  {Benhayoune-Khadraoui}}, \bibinfo {author} {\bibfnamefont {A.}~\bibnamefont
  {Blais}},\ and\ \bibinfo {author} {\bibfnamefont {A.}~\bibnamefont
  {Parra-Rodriguez}},\ }\bibfield  {title} {\bibinfo {title} {Toolbox for
  nonreciprocal dispersive models in circuit quantum electrodynamics},\ }\href
  {https://doi.org/10.1103/PhysRevApplied.22.034038} {\bibfield  {journal}
  {\bibinfo  {journal} {Physical Review Applied}\ }\textbf {\bibinfo {volume}
  {22}},\ \bibinfo {pages} {034038} (\bibinfo {year} {2024})}\BibitemShut
  {NoStop}%
\bibitem [{\citenamefont {Flanders}(1963)}]{Flanders:1963}%
  \BibitemOpen
  \bibfield  {author} {\bibinfo {author} {\bibfnamefont {H.}~\bibnamefont
  {Flanders}},\ }\href {https://books.google.es/books?id=pG0PllIO08kC} {\emph
  {\bibinfo {title} {Differential Forms with Applications to the Physical
  Sciences}}},\ Dover books on advanced mathematics\ (\bibinfo  {publisher}
  {Academic Press},\ \bibinfo {year} {1963})\BibitemShut {NoStop}%
\bibitem [{\citenamefont {Galindo}\ and\ \citenamefont
  {Pascual}(2012)}]{Galindo:2012}%
  \BibitemOpen
  \bibfield  {author} {\bibinfo {author} {\bibfnamefont {A.}~\bibnamefont
  {Galindo}}\ and\ \bibinfo {author} {\bibfnamefont {P.}~\bibnamefont
  {Pascual}},\ }\href {https://doi.org/10.1007/978-3-642-83854-5} {\emph
  {\bibinfo {title} {Quantum Mechanics I}}},\ Theoretical and Mathematical
  Physics\ (\bibinfo  {publisher} {Springer Berlin Heidelberg},\ \bibinfo
  {year} {2012})\BibitemShut {NoStop}%
\end{thebibliography}%
\end{document}